\documentstyle[prb,aps]{revtex}
\def\f{$f$}
\def\rr{{\bf r}}

\def\HH{{\cal H}}
\def\dd{{\rm d}}

\begin{document}
\date{\today}

\title{Theory of Strongly Correlated Electron Systems. \\
II.Intersite Coulomb Interaction and the Approximation of Renormalized 
Fermions in Total Energy Calculations.}

\author{I. Sandalov$^{1,2}$, U. Lundin$^1$, O. Eriksson$^1$}
\address{$^1$Condensed Matter Theory group, Uppsala University, Box 530,
SE-751 21 Uppsala, Sweden\\
$^2$Dept.\ of Physics, Link\"oping University,
SE-581 83 Link\"{o}ping, Sweden}

\maketitle

\begin{abstract}
The strong-coupling perturbation theory (SCPT) for correlated electron 
systems is extended to the case of full Coulomb interaction. 
The Coulomb mechanism of the orbital polarization 
is discussed and attention is paid to the importance of spectral weight
transfer
between the localized and delocalized subsystems of electrons. 
A one-to-one correspondence between subsets of Feynman graphs of SCPT
(which we name the approximation of renormalized Fermions (ARF)) 
and weak-coupling perturbation theory (WCPT) is established.
The comparison of the Galitskii and Migdal expression for the total energy
and the Sham equation (which connects the self-energy and the 
exchange-correlation
potential in density functional theory) for WCPT with the ones for 
the systems with strong
electron correlations is used for the
formulation of a
simple theory for extending the local density approximation to density 
functional theory (DFT) 
to include explicitly correlations. The extension 
requires the inclusion of the many-electron spectral weights in 
the definition of the charge density and the renormalization 
of the mixing and hopping matrix elements caused by many body effects.
\end{abstract}

\section{Introduction}
The standard model of the lanthanides successfully explains most of magnetic 
properties and, 
within {\it ab
initio} calculations, also the equilibrium lattice parameter, bulk moduli and 
cohesive energy;
actually, the picture of localized $f$-electrons works much better for
compounds with rare earths than the picture of fully delocalized 
$f$-electrons. Nevertheless, it is not sufficient for an explanation of 
all the 
complex physical and chemical properties. A recent example is given in the 
neutron scattering
experiment on Pr metal~\cite{clausen94}. 
The mechanism which can provide the \f-localization, has
been recognized long ago as the intra shell Coulomb interaction, which forms
large energy gaps between occupied and unoccupied $d$- or $f$-orbitals. It has 
been explained within
Hartree-Fock approximation by Brandow~\cite{brandow} in connection with
physics of Mott insulators and explicitly exploited in such phenomenological
theories, as LDA+U~\cite{AZA}, self-interaction corrected LDA DFT~\cite%
{temmerman,svane1}. It is clear that this picture is a bit oversimplified,
since in many cases it is difficult to describe a local subsystem of materials
within a single-electron approach. LDA-DFT is not a single-electron
theory since it goes beyond the Hartree-Fock
approximation at least up to the random phase approximation (RPA). 
However, it is also well-known that
without formation of an energy gap with the help of SIC or LDA+U between $%
f(d)$-orbitals it is difficult to obtain electron localization in practical 
implementations of
DFT.

However, in total energy calculations within the local density approximation to
density functional theory, it is possible to treat the $f$-electrons either
as core (or valence) electrons, obtaining localized states. Irrespective of
if the \f-states are localized or delocalized, there is a problem of taking
into account strong correlations between $f$-electrons of non-filled shells,
since the expression for the exchange-correlation
potential usually is based on the expression derived from the theory
for \emph{homogeneous} electron gas. The treatment of $f$-electrons as 
core states also 
requires a prescription of how many of them
should be forced to be localized in the core. Even though the total energy may 
may be minimized 
with respect to the \f-occupation~\cite{delin97} 
this approach must rely on experimental data of 
the \f-shell and it has problems of including fluctuations 
or non-integral occupations of the \f-states. 
A theoretical method that attempts to describe all the intricate properties 
of \f- (and $d$-) 
electron materials, such as heavy Fermion behavior, Kondo effects, Mott 
insulators 
(with a transition from a metal to insulator), unconventional 
super-conductivity, hybridization gaps etc., must be build on a many body 
framework, and 
it seems that a fruitful way to do this is to join many body corrections 
and electronic 
structure methods. The main motivation for this combination is that 
the many body theory described accurately the correlated \f- (sometimes 
{\it d}-) states, 
whereas electronic structure methods treat the diffuse, more extended 
states well. 

Here we suggest to look at the problem using a theory of
strongly correlated electrons (SCE). We would like to combine 
field-theoretical methods with DFT. For the electrons which are 
localized the approach starting from the atomic limit seems to be
appropriate. The
derivation of the connection between DFT and many-body theory for the total
energy, given by Sham~\cite{sham} uses the expression for the total 
energy~\cite{ward_luttinger} which contains the sum of all skeleton graphs. 
The theory
of SCE from the atomic limit is based mainly on the 
Pauli-principle 
(e.g.\ for destruction operators $c_{fa\sigma }^{2}=0$).
Since some infinite sequence of graphs of perturbation theory from
the weak-coupling side (say, for the self-energy for the Fermion Green function
(GF)) corresponds to the expression where the relation like $\hat{n}%
_{fa\sigma }^{2}=\hat{n}_{fa\sigma }$ has been used within the
strong-coupling perturbation theory, it is not easy to find this sequence and
provide a one-to-one correspondence between these perturbation theories. It is
possible, however, to express the Fermion GFs in terms of GFs for
many-electron operators. Here we will exploit this connection.

The description within the framework of the LDA to DFT, when the $f$-electrons
are treated as core electrons with only a certain number of $f$-orbitals
occupied may be provided within the 
many-body approach in the following way. Let us consider an ion which has $n$
$f$-electrons in
the ground state number. Then, only the transitions $\Gamma
_{n}\rightarrow \Gamma_{n\pm 1}$ will be involved in the formation of
the spectrum of the \emph{single-electron} excitations while all other
transitions, like $\Gamma_{n}\rightarrow \Gamma_{n\pm 2},\Gamma_{n\pm 3}$, 
involving larger number of electrons will be strongly suppressed by a large
energy separation between these states. If we, in such theory, take the limit 
when the energy of the atomic-like transition $\Delta _{2}\equiv E_{\Gamma
}^{(n+1)}-E_{\Gamma ^{\prime }}^{(n)}$ between any $n+1$ and 
$n$-electron states $\Gamma $ and $\Gamma^{\prime }$ of the $f$-ion is much
higher than the Fermi energy, $\varepsilon _{F}$, the number of $f$-electrons
in the ion will be fixed. Indeed, in this limit this upper ''single-electron''
level $\Delta_{2}$ is empty. In the rare earth elements the populated part of 
the $f$-spectral density corresponding to the transitions $\Delta_{1}\equiv
E_{\Gamma^{\prime\prime}}^{(n)}-E_{\Gamma^{^{\prime\prime\prime
}}}^{(n-1)}$ is much below $\varepsilon_{F}$. It can also be (much) below
the bottom of the conduction-electron bands. This mechanism leads to a 
self-consistent switching off of the mixing interaction (see figures 4,5,6 
for LDA+Hubbard-Kanamori corrections in Ref.~\cite{olof}) and 
removes the overlap
between these core-like levels and the conduction electrons. Thus, this physical
picture exactly corresponds to a type of \emph{ab initio} calculation 
with a fixed number of localized electrons. The 
photo-electron spectroscopy experiments, show that even in the rare
earth elements, for which this picture seems to be most appropriate, the
level $\Delta _{2}$ is sometimes only slightly above Fermi energy.
Therefore, due to mixing interaction it should contribute to the cohesive
energy~\cite{urbansqrt}. The total energy of the system and Fermion GFs can
be calculated in a simple approximation by making use of the theory developed
in Ref.~\cite{DT}.
This allows to construct the Sham equation for the exchange-correlation
potential and to derive corrections to the standard LDA expression.

The paper is organized as follows. 
In Section~\ref{CIaGF:sect}, 
in order to provide an opportunity to directly compare the 
contributions from Coulomb interaction to the equations of motion for GFs
in WCPT and SCPT, the equations of motion 
are written exactly via functional derivatives of the GFs in WCPT 
(see Section~\ref{CIaGFa:sect}) 
and in SCPT (see Section~\ref{CIaGFb:sect}) 
within a real-space non-orthogonal representation. 
Here we introduce the approximation 
of renormalized Fermions (ARF), which allows us to establish a one-to-one 
correspondence 
between certain subsequences of graphs in WCPT and SCPT. This fact is used 
further for the formulation of an 
recipe for a correlation corrected form of LDA on the basis of the analysis 
by Sham~\cite{sham} for the exchange-correlation potential for a normal and 
a correlated system. 
In Section~\ref{standard:sect} 
we discuss the standard model for the lanthanides and compare the
self-energies within WCPT and SCPT in the ARF. Since the graphs
which are not taken into account by this approximation can be written
explicitly, this allows us to establish the domain of validity of the standard
approach to the case when the system has (quasi)localized electrons, and,
also, suggests a possible recipe to overcome the difficulties existing in
the LDA in DFT. 
Section~\ref{disc:sect} contains the conclusions and a discussion. 
The contributions to the equations of motion for the GFs are rewritten via
functional derivatives of GFs. This allows to construct a regular SCPT via an 
iteration procedure for the equations for the GFs. 
In Appendix~\ref{3-level:app} we discuss the transfer of spectral weight 
between the low and high energy regions and 
show the role played by spectral weights in the formation 
of the orbital polarization in the simple example of 3-orbital atoms 
with 2 strongly interacting electrons.

\section{Coulomb Interaction and Green Functions}
\label{CIaGF:sect}
\subsection{WCPT for an Non-Orthogonal Set}
\label{CIaGFa:sect}
Perturbation theory from the atomic limit should, on one hand, be 
constructed in real space in order to be able to treat the single-site 
Coulomb interactions better than within the LDA. The wave functions centered 
on different atoms are not supposed to be orthogonal to each other. 
On the other hand, we have to compare the perturbation series for the GFs 
within weak coupling and strong coupling.  For this reason we have 
to understand how the equations reflect the non-orthogonality of the
basis set. Here, we shortly repeat the well-known 
Kadanoff and Baym derivation for
the exact expression for the self-energy and obtain the first
corrections which we will compare with the ones obtained from the SCPT 
approach later. First we have to introduce the $S$-matrix with an external 
field $U_{23}(t)a_{2}^{\dagger }a_{3}$~\cite{kadanoff_baym}\footnote{The
external field $U_{23}$ should not be confused with the Hubbard $U$}. 
Then, for the Hamiltonian 
\begin{equation}
{\cal H}=h_{23}^{0}a_{2}^{\dagger }a^{\vphantom{\dagger}}_{3}+
\frac{1}{2}v_{2345}a_{2}^{\dagger
}a_{3}^{\dagger }a^{\vphantom{\dagger}}_{4}a^{\vphantom{\dagger}}_{5},
\end{equation}
where $h_{23}^{0}=(2|\frac{p^{2}}{2m}+v(r)|3)$ and $v_{2345}$ is a matrix
element of the Coulomb interaction, the equation of motion for the GF $%
F_{11^{\prime }}(tt^{\prime })\equiv 1/i \langle {\cal T}
a^{\vphantom{\dagger}}_{1}(t)
a_{1^{\prime}}^{\dagger }(t^{\prime })\rangle $ has the form: 
(the letter $F$ will be used for \emph{Fermion}
GFs below keeping $G$ for the GFs involving Hubbard operators) 
\begin{eqnarray}
\{ &&\delta^{\vphantom{\dagger}}_{13}i\partial^{\vphantom{\dagger}}_{t}-
O_{12}^{-1}[\tilde{h}_{23}^{0}(t)-v_{\sigma
,23}^{el}]\}F^{\vphantom{\dagger}}_{31^{\prime }}
(t,t^{\prime }|U)=i\delta (t-t^{\prime
})O_{11^{\prime }}^{-1}  \nonumber \\
+ &&O_{12}^{-1}v^{\vphantom{\dagger}}_{[23]45}\langle {\cal T}a_{3}^{\dagger
}(t)a^{\vphantom{\dagger}}_{4}(t)a^{\vphantom{\dagger}}_{5}(t)
a_{1^{\prime }}^{\dagger }(t^{\prime })\rangle _{U},
\end{eqnarray}
where $v_{[23]45}\equiv \frac{1}{2}(v_{2345}-v_{3245}),$ $\tilde{h}%
_{23}^{0}=h_{23}^{0}+U_{23}(t)$ and the notation $U$ in the GF $%
F_{31^{\prime }}(tt^{\prime }|U)$ means that it satisfies the equation 
of motion in
the external field $U$. The boundary conditions 
for GFs are based on cyclic permutations of operators under 
sign of trace and, therefore, are not changed by the
non-orthogonality.

The Dyson equation for the full GF, 
\begin{eqnarray}
&&\hspace*{-5mm}
\{\delta _{13}i\partial _{t}-O_{12}^{-1}\int \dd t_{1}[(\tilde{h}%
_{23}^{0}(t) \nonumber \\ 
&&-v_{\sigma ,23}^{el})\delta (t-t_{1})-\Sigma
_{23}(t,t_{1}|U)]\}F_{31^{\prime }}(t_{1},t^{\prime }|U)\nonumber \\
&&=i\delta (t-t^{\prime
})O_{11^{\prime }}^{-1},
\end{eqnarray}
can be written if we write the double-electron GF via a functional derivative  
\begin{eqnarray}
&&\hspace*{-5mm}
\langle {\cal T}a_{3}^{\dagger }(t)a_{4}(t)a_{5}(t)a_{1^{\prime }}^{\dagger
}(t^{\prime })\rangle _{U}=\left[ F_{34}(t,t^{+}|U) \right. \nonumber \\
&&\left. +\frac{\delta }{\delta
U_{34}(t^{+})}\right] F_{51^{\prime }}(t,t^{\prime }|U),
\end{eqnarray}
and define the self-energy as  
\begin{eqnarray}
\Sigma _{21^{\prime }}(t,t^{\prime }|U)=\{ v_{[23]45}\left[
F_{34}(t,t^{+}|U)+   \right . \nonumber \\ \left .
\frac{\delta }{\delta U_{34}(t^{+})}\right]
F_{56}(t,t_{1}|U)\} F_{61^{\prime }}^{-1}(t_{1},t^{\prime }|U).
\end{eqnarray}
Here, the inverse of the GF is defined by the equations  
\begin{eqnarray}
&&\hspace*{-5mm}
F_{56}(t,t_{1}|U)F_{61^{\prime }}^{-1}(t_{1},t^{\prime
}|U)=F_{56}^{-1}(t,t_{1}|U)F_{61^{\prime }}(t_{1},t^{\prime }|U) \nonumber \\
&&=\delta
(t-t^{\prime })\delta _{11^{\prime }}. 
\end{eqnarray}
The Hartree correction appears if we neglect the functional derivative 
in the self-energy
\begin{equation}
\Sigma_{21^{\prime}}^{H}(t,t_{1}|U)=v_{[23]41^{\prime }}F_{34}(t,t^{+}|U),
\end{equation}
and the Fock correction comes from the derivative $\delta F^{0}/\delta U$: 
\begin{equation}
\Sigma _{21^{\prime }}^{F}(t,t_{1}|U)=-v_{[23]1^{\prime }5}F_{53}(t,t^{+}|U).
\end{equation}
As we see, the first-order corrections to the self-energy do not contain 
the overlap
matrix. The next corrections does not contain it either. 
We will also need the RPA-screened Coulomb interaction. The
difference between the standard procedure for Fermions and the one in this 
representation
consists only in the necessity to write the representation indices. Introducing
the effective field 
\begin{equation}
U_{21^{\prime }}^{eff}(t)=U_{21^{\prime }}(t)+\Sigma _{21^{\prime }}^{H}(t)
\end{equation}
and $\Sigma ^{\prime }=\Sigma -\Sigma ^{H}$ we can rewrite $\Sigma ^{\prime
} $ in terms of $U^{eff}$ as follows: 
\begin{eqnarray}
\Sigma _{21^{\prime }}^{\prime }(t,t^{\prime }|U) &=&v_{[23]45}\frac{\delta
U_{78}^{eff}(t_{2})}{\delta U_{34}(t^{+})}\frac{\delta F_{56}(t,t_{1}|U)}{%
\delta U_{78}^{eff}(t_{2})}F_{61^{\prime }}^{-1}(t_{1},t^{\prime }|U) 
\nonumber \\
&=&\tilde{v}_{2785}(t_{2},t^{+})\frac{\delta F_{56}(t,t_{1}|U)}{\delta
U_{78}^{eff}(t_{2})}F_{61^{\prime }}^{-1}(t_{1},t^{\prime }|U)  \nonumber \\
&=&\tilde{v}_{2785}(t_{2},t^{+})F_{59}(t,t_{9}|U)\frac{\delta F_{91^{\prime
}}^{-1}(t_{9},t^{\prime }|U)}{\delta U_{78}^{eff}(t_{2})}.
\end{eqnarray}
The effective interaction, $\tilde{v}$, is 
\begin{equation}
\tilde{v}_{2785}(t_{2},t^{+})\equiv v_{[23]45}\frac{\delta
U_{78}^{eff}(t_{2})}{\delta U_{34}(t^{+})}\equiv v_{[23]45}\,\varepsilon
_{78,34}^{-1}(t_{2},t^{+}).
\end{equation}
Using the zero approximation for the vertex 
($F^{-1}\rightarrow F_{0}^{-1}$) 
\begin{equation}
\frac{\delta F_{91^{\prime }}^{-1}(t_{9},t^{\prime }|U)}{\delta
U_{78}^{eff}(t_{2})}\simeq -\delta (t_{9}-t^{\prime })\delta
(t_{2}-t^{\prime })\delta _{97}\delta _{1^{\prime }8}
\end{equation}
we obtain the desired expression for self-energy in RPA: 
\begin{equation}
\Sigma _{21^{\prime }}^{{\rm RPA}}(t,t^{\prime }|U)=-\tilde{v}_{271^{\prime
}5}(t^{\prime },t^{+})F_{57}(t,t^{\prime }|U).
\end{equation}
The inverse of the dielectric permeability, $\varepsilon^{-1}$, entering
the effective interaction is determined by the equation: 
\begin{eqnarray}
\label{epsilon_wc:eq}
&&\hspace{-5mm}
\varepsilon _{71^{\prime },34}^{-1}(t^{\prime },t^{+}) =\delta
(t^{+}-t^{\prime })\delta _{37}\delta _{1^{\prime }4} \nonumber \\
&&+v_{[73]41^{\prime }}F_{35}(t,t_{5}|U)F_{64}(t_{5},t^{+}|U)\varepsilon
_{56,34}^{-1}(t^{\prime },t^{+}).
\end{eqnarray}
Putting the external fields $U=0$ and making a Fourier transformation with
respect to time we find the equation in terms of Matsubara frequencies 
(below we use $i\omega$ for Fermionic and $i\Omega$ for Bosonic frequencies) 
\begin{equation}
\label{dielectrrr:eq}
\varepsilon _{71^{\prime },34}^{-1}(i\Omega )=\delta _{37}\delta _{1^{\prime
}4}+v_{[73]41^{\prime }}\Pi _{35,64}^{(0)}(i\Omega )\varepsilon
_{56,34}^{-1}(i\Omega ),
\end{equation}
where 
\begin{equation}
\Pi _{35,64}^{(0)}(i\Omega )\equiv T\sum_{\omega }F_{35}(i\omega
)F_{64}(i\omega +i\Omega ).
\end{equation}
Then, the self-energy ($U=0$) is  
\begin{equation}
\label{sigma_RPA:eq}
\Sigma _{21^{\prime }}^{{\rm RPA}}(i\omega )=-v_{[23]45}T\sum_{\omega
_{1}}\varepsilon _{71^{\prime },34}^{-1}(i\omega -i\omega
_{1})F_{57}(i\omega _{1}).
\end{equation}
Let us write it as sum of ''static'' and ''dynamic'' parts: 
\begin{equation}
\Sigma _{21^{\prime }}^{{\rm RPA}}(i\omega )=\Sigma _{st,21^{\prime
}}^{{\rm RPA}}+\Sigma _{{\rm dyn},21^{\prime }}^{{\rm RPA}}(i\omega ),
\end{equation}
where 
\begin{equation}
\Sigma _{st,21^{\prime }}^{{\rm RPA}}=-v_{[23]45}\varepsilon _{st;71^{\prime
},34}^{-1}T\sum_{\omega _{1}}F_{57}(i\omega _{1})=v_{21^{\prime }}^{(ex)},
\end{equation}
\begin{eqnarray}
&&\hspace{-5mm} 
\Sigma _{{\rm dyn},21^{\prime}}^{{\rm RPA}}(i\omega)=-v_{[23]45}T\sum_{\omega
_{1}}[\varepsilon _{71^{\prime},34}^{-1}(i\omega -i\omega _{1}) 
\nonumber \\ 
&&-\varepsilon_{st;71^{\prime },34}^{-1}]F_{57}(i\omega _{1}).
\end{eqnarray}
Here we separated the frequency-independent part of the dielectric
permeability $\varepsilon _{st;71^{\prime },34}^{-1}$. Its contribution
to the self-energy, $\Sigma _{st,21^{\prime }}^{{\rm RPA}}$, which is the 
frequency-independent  part of the screened exchange interaction, is
denoted $v_{21^{\prime }}^{(ex)}$. The formulas given above have to be
compared with the corresponding expressions in SCPT. We will see below that in
SCPT $f$- and non-$f$-electrons are renormalized differently, for this
reason we will need these formulas in terms of $f$- and non-$f$ indices ($c$%
) separately. Each of the corrections, say, $\Sigma _{21^{\prime
}}^{H},\Sigma _{21^{\prime }}^{{\rm RPA}}$ contains four terms since each index
takes two values: $1=(c,f)$, where $c=(jL)$ and $f=(j\mu )$  
\begin{eqnarray}
&&\hspace*{-5mm}
\Sigma_{jL,j^{\prime }L^{\prime }}^{H}(tt^{+}|U) \nonumber \\
&=& v_{[jL,j_{3}L_{3}]j_{4}L_{4},\,j^{\prime }L^{\prime
}}F_{j_{3}L_{3},j_{4}L_{4}}(tt^{+}|U) \\
&&+v_{[jL,j_{3}\mu _{3}]j_{4}L_{4},\,j^{\prime }L^{\prime }}F_{j_{3}\mu
_{3},j_{4}L_{4}}(tt^{+}|U) \\
&&+v_{[jL,j_{3}L_{3}]j_{4}\mu _{4},\,j^{\prime }L^{\prime
}}F_{j_{3}L_{3},j_{4}\mu _{4}}(tt^{+}|U) \\
&&+v_{[jL,j_{3}\mu _{3}]j_{4}\mu _{4},\,j^{\prime }L^{\prime }}F_{j_{3}\mu
_{3},j_{4}\mu _{4}}(tt^{+}|U),
\end{eqnarray}
\begin{eqnarray}
&&\hspace*{-5mm}
\Sigma _{jL,j^{\prime }L^{\prime }}^{{\rm RPA}}(tt^{\prime }|U)  \nonumber \\
&=&-[\tilde{v}%
_{jL,j_{7}L_{7},j^{\prime }L^{\prime },\,\,j_{5}L_{5}}(t^{\prime
}t^{+})F_{j_{5}L_{5},j_{7}L_{7}}(tt^{\prime }|U) \\
&&+\tilde{v}_{jL,j_{7}\mu _{7},j^{\prime }L^{\prime
},\,\,j_{5}L_{5}}(t^{\prime }t^{+})F_{,j_{5}L_{5},j_{7}\mu _{7}}(tt^{\prime
}|U) \\
&&+\tilde{v}_{jL,j_{7}L_{7},j^{\prime }L^{\prime },\,\,j_{5}\mu
_{5}}(t^{\prime }t^{+})F_{,j_{5}\mu _{5},j_{7}L_{7}}(tt^{\prime }|U) \\
&&+\tilde{v}_{jL,j_{7}\mu _{7},j^{\prime }L^{\prime },\,\,j_{5}\mu
_{5}}(t^{\prime }t^{+})F_{j_{5}\mu _{5}j_{7}\mu _{7}}(tt^{\prime }|U)],
\end{eqnarray}
and the same for the other contributions $\Sigma _{j\mu ,j^{\prime }L^{\prime
}}^{H}(tt^{+}|U),\Sigma _{j\mu ,j^{\prime }L^{\prime }}^{{\rm RPA}}
(tt'|U)$, 
etc.

Thus, we can conclude, that within WCPT the overlap matrixes enter only the
equations via the definition of the zero GF. As we shall soon see, 
the situation is more complex within SCPT, since in this case, the 
overlap matrixes enter in a
form combining the inverse of the overlap matrix with many-electron
population numbers. In the case when we take into account only the static part
of the screening, the formula for the total energy can still be written in a 
simple
form. Let us write it on the basis of Eq.(\ref{tote:eq}). Let us denote the
Hartree part of the Hamiltonian as 
\begin{equation}
h_{12}^{H}=(1|h^{H}({\bf r})|2)=(1|\frac{p^2}{2m}+v_{ne}(%
{\bf r})+v^{H}({\bf r})|2).
\end{equation}
Then, representing the overlap matrix in the Cholesky form, 
$O_{12}=Z_{1\gamma }\bar{Z}_{\gamma 2}$, and introducing the eigenvectors, 
$u_{\gamma}^{\nu}$, and eigenstates ,$\varepsilon^{\nu}$, by the equation 
\begin{equation}
\sum_{12\delta }Z_{\gamma 1}^{-1}[h_{12}^{H}+v_{12}^{(ex)}]\bar{Z}_{2\delta
}^{-1}\cdot u_{\delta }^{\nu }=\varepsilon ^{\nu }u_{\gamma }^{\nu }.
\end{equation}
We see that the solution to the equation for the GF in this approximation, 
\begin{equation}
\{O_{12}\omega -[h_{12}^{H}+v_{12}^{(ex)}]\}F_{21^{\prime }}(\omega )=\delta
_{11^{\prime }},
\end{equation}
can be written in the following form  
\begin{equation}
\label{eff_12:eq}
F_{12}=\langle \langle a_{1}|a_{2}^{\dagger }\rangle \rangle _{\omega
}^{(ret)}=\sum_{\alpha \gamma \nu }\frac{\bar{Z}_{1\gamma }^{-1}u_{\delta
}^{\nu }u_{\alpha }^{\ast \nu }Z_{\alpha 1}^{-1}}{\omega -\varepsilon ^{\nu
}+i\delta }.
\end{equation}
Then, the expectation value $\langle a_{2}^{\dagger }a_{1}\rangle $ is given
by 
\begin{equation}
\langle a_{2}^{\dagger }a_{1}\rangle =\sum_{\alpha \gamma \nu }\bar{Z}%
_{1\gamma }^{-1}u_{\delta }^{\nu }f(\varepsilon ^{\nu })u_{\alpha }^{\nu
}Z_{\alpha 1}^{-1}.
\end{equation}
The charge density $\rho({\bf r})$ can then be written as follows  
\begin{eqnarray}
\rho (x) &=&-i\lim_{x^{\prime }\rightarrow x}\sum_{12}\phi
_{1}(x)[\sum_{\omega }e^{i\omega 0^{+}}F_{12}(i\omega )]\phi _{2}^{\ast
}(x^{\prime }) \nonumber \\
&\equiv& \sum_{12}\phi _{2}^{\ast }(x^{\prime })\phi
_{1}(x)\langle a_{2}^{\dagger }a^{\vphantom{\dagger}}_{1}\rangle \\
&=&-i\lim_{x^{\prime }\rightarrow x}\sum_{12}\phi _{1}(x)[\sum_{\omega
}e^{i\omega 0^{+}}F_{12}^{DFT}(i\omega )]\phi _{2}^{\ast }(x^{\prime
}) \nonumber \\
&\equiv& \sum_{12}\phi _{2}^{\ast }(x^{\prime })\phi _{1}(x)\langle
a_{2}^{\dagger }a^{\vphantom{\dagger}}_{1}\rangle ^{DFT}.
\label{charge_dens:eq}
\end{eqnarray}
Further, we can use the Migdal-Galitskii expression for the total energy, 
written in terms of, say, retarded Fermion Green functions,
\begin{eqnarray}
&&\hspace{-5mm}E_{tot}=\frac{1}{2}\int \dd\omega \lbrack O_{12}\omega
+(h_{12}^{0}-v_{12}^{el})]f(\omega )(-\frac{1}{\pi }) \nonumber \\
&& \times Im F_{21}^{(ret)}(\omega +i\delta ).
\label{tote:eq}
\end{eqnarray}
In order to avoid possible confusion we note that $F_{12}(i\omega )$ are the
\emph{coefficients of expansion} of the GF in a series of the functions $\phi
_{jL}({\bf r})$,
\begin{eqnarray}
&&\hspace{-5mm}
F({\bf r},t;{\bf r}^{\prime },t^{\prime })=\frac{1}{i}\langle {\cal T}\psi (%
{\bf r},t)|\psi ^{\dagger }({\bf r},t^{\prime })\rangle = \nonumber \\
&&\sum_{12}\phi
_{1}(x)F_{12}(i\omega )\phi _{2}^{\ast }(x^{\prime })
\end{eqnarray}
and they do not coincide with the matrix elements
$\tilde{F}_{12}(t,t^{\prime })=\int \dd\rr\dd\rr^{\prime }\phi _{1}^{\ast }(%
{\bf r})F^{(ret)}({\bf r},t;{\bf r}^{\prime },t^{\prime })\phi _{2}(%
{\bf r})$ calculated on this functions.

Inserting the imaginary part of the GF, 
Eq.(\ref{eff_12:eq}), into the expression for
total energy, Eq.(\ref{tote:eq}), we find within the static RPA  
\begin{eqnarray}
&&\hspace*{-5mm}
E_{tot}^{\HH} =\frac{1}{2}\sum_{12\gamma \alpha }\bar{Z}_{2\gamma
}^{-1}u_{\gamma }^{\nu }[O_{12}\varepsilon ^{\nu }+(1|\frac{p^{2}}{%
2m}+v_{ne}(r)|2)] \nonumber \\ 
&\times& f(\varepsilon ^{\nu })u_{\alpha }^{\ast \nu
}Z_{\alpha 1}^{-1} \\
&=&\frac{1}{2}[(1|\frac{p^{2}}{2m}+v_{ne}(r)+v^{H}({\bf %
r})|2)-v_{12}^{(ex)} \nonumber \\
&+&(1|\frac{p^2}{2m}+v_{ne}({\bf r})|2)%
]\langle a_{1}^{\dagger }a^{\vphantom{\dagger}}_{2}\rangle \\
&=&\int \dd\rr\langle \psi _{\sigma }^{\dagger }({\bf r})[\frac{p%
^{2}}{2m}+v_{ne}({\bf r})]\psi _{\sigma }( { \bf r})\rangle  \nonumber \\
&&+\frac{1}{2}\left[ \int \dd\rr\int \dd\rr^{\prime }\frac{\rho ({\bf r}%
)\rho ({\bf r}^{\prime })}{|{\bf r}-{\bf r}^{\prime }|}%
-v_{12}^{(ex)}\langle a_{1}^{\dagger }a^{\vphantom{\dagger}}_{2}\rangle 
\right],
\end{eqnarray}
as expected. 
As follows from the comparison
of the standard Kohn-Sham expression for total energy used in DFT 
calculations, 
the term $v_{12}^{(ex)}\langle a_{1}^{\dagger }a_{2}\rangle $ should be
identified with the term $\int\rho({\bf r})v_{x}({\bf r})$ in DFT, i.e.\ 
with the contribution to the total energy from the screened exchange 
potential.
According to Kotani~\cite{kotani} this expression in a static approximation 
($\omega=0$) should give
(at least, for normal metals) values close to what is usually obtained
within standard LDA DFT calculations. 

One can write the expression for $E_{tot}$ also for the case 
of frequency-dependent screening. Then the self-energy also depends on 
frequency. For this case one can apply the 
idea of Migdal~\cite{migdal} to introduce energy dependent 
eigenfunctions in order to diagonalize the GF  
\begin{equation}
\sum_{12\delta }Z_{\gamma 1}^{-1}[h_{12}^{\HH}+Re \Sigma
_{12}^{{\rm RPA}}(\omega )]\bar{Z}_{2\delta }^{-1}\cdot u_{\delta }^{\nu }
(\omega)=E^{\nu }(\omega )u_{\gamma }^{\nu }(\omega ).
\end{equation}
Then, the Fermion GF is  
\begin{equation}
F_{12}^{(ret)}=\langle \langle a_{1}|a_{2}^{\dagger }\rangle \rangle
_{\omega }^{(ret)}=\sum_{\alpha \gamma \nu }\frac{\bar{Z}_{1\gamma
}^{-1}u_{\delta }^{\nu }(\omega )u_{\alpha }^{\ast \nu }(\omega )Z_{\alpha
1}^{-1}}{\omega -E^{\nu }(\omega )+i\delta }.
\end{equation}
The spectrum of the single-particle excitations is then determined by the 
poles of the GF  
\begin{equation}
\varepsilon^{\nu}=E^{\nu}(\varepsilon^{\nu }),
\end{equation}
where 
\begin{equation}
E^{\nu }(\varepsilon ^{\nu })=
u_{\gamma }^{\ast \nu }(\varepsilon ^{\nu } )Z_{\gamma
1}^{-1}[h_{12}^{H}+Re \Sigma _{12}^{{\rm RPA}}(\varepsilon ^{\nu })]\bar{Z}%
_{2\delta }^{-1}u_{\delta }^{\nu }(\varepsilon ^{\nu } ).
\end{equation}
In turn, the total energy acquires the form  
\begin{eqnarray}
&&\hspace*{-5mm}
E_{tot}=\frac{1}{2}\sum_{12\gamma \alpha }\bar{Z}_{2\gamma }^{-1}u_{\gamma
}^{\nu }(\varepsilon ^{\nu })\frac{[O_{12}\varepsilon ^{\nu }+(1|\frac{
p^2}{2m}+v_{ne}({\bf r})|2)]}{1-\left( \frac{\partial E^{\nu
}(\omega )}{\partial \omega }\right) _{\varepsilon ^{\nu }}} \nonumber \\
&&\times f(\varepsilon
^{\nu })u_{\alpha }^{\ast \nu }(\varepsilon ^{\nu })Z_{\alpha 1}^{-1}.
\end{eqnarray}
Thus, the eigenvectors $u_{\gamma }^{\nu }(\varepsilon ^{\nu })$ are needed
only at $\omega =\varepsilon ^{\nu }$ and the renormalization factor $\left[
1-\left( \frac{\partial E^{\nu }(\omega )}{\partial \omega }\right)
_{\varepsilon ^{\nu }}\right] ^{-1}$ takes into account the correlation
induced deviation of the distribution function from a purely Fermionic one.

At last, we have to note that in the case when one takes into account also
the 
decay of the quasi-particles, the total energy can be calculated directly from
Eq.(\ref{tote:eq}). 

The facts which we are going to exploit further are: 1) the Galitskii-Migdal
formula, Eq.(\ref{tote:eq}), 
gives the total energy in the approximation chosen for \emph{Fermionic} GFs; 
2) the Fermionic GFs can be expressed in the form of a linear
combination of the GFs for the Hubbard operators; 3) the lower intra atomic
Fermi-like transitions give only core-like contribution to the energy. The
latter statement will be considered in the next section. In the region of
strong intra atomic coupling, the self-energies $\Sigma _{j\mu ,j^{\prime }\mu
^{\prime }},\Sigma _{j\mu ,j^{\prime }L^{\prime }}$ for Fermionic GFs also
can be found for many cases by making use of the SCPT where a non-perturbative
calculation should be used for the interaction between electrons in each 
ion, but
perturbative ones with respect to the interactions connecting different
ions. How to calculate the GFs for Hubbard operators is discussed in next
subsection and the details are given in Appendix \ref{coul_eq_of_mot:app}.

\subsection{SCPT and the Approximation 
of Renormalized Fermions}
\label{CIaGFb:sect}
If the Hubbard intra atomic repulsion is strong enough the perturbation
theory from the atomic limit generates the following physical picture. The
partially filled $d$- (or $f$-) shell is separated by a large energy gap to two
subs hells. One of them can be viewed as core-like states, while the other one 
can, 
in a certain approximation, be described in terms of effective,
renormalized, Fermions~\cite{rice}. These quasi-Fermions behave 
very similar to the Fermions in the weak-coupling regime. However, this
simple picture arises only within the lowest approximation where the upper
Hubbard sub bands are similar to the corresponding sub band of the 
so-called Hubbard-I approximation in the Hubbard models
\footnote{Although Hubbard did not consider in his works the inter-site Coulomb
interactions, the approximation we use has the same underlying physics and,
therefore, we will use the same name ''Hubbard-I'' for this approximation.}
and the decay of the quasi-particles is only due to normal
scattering of these quasi Fermions caused by Coulomb intersite interaction. The
scattering on collective excitations like spin waves, etc., as well
as the scattering caused by kinematic interactions, is not taken into
account. In this case the
Fermi-liquid type of decay ($\sim T^{2}$) still holds~\cite
{sand_ovch_fl,urbannfl} at low temperature and excitation energy. 
Our aim will be to show that, in the \emph{ab initio} 
calculations, for materials where the strong correlations are not 
developed very much due to the suppression by the a strong 
localization of a 
part of a shell, one can use the standard expression for the
exchange-correlation potential even in the case of strong electron
correlations (SEC) with the difference that a modified expression for the
charge density should be used and, also, hopping, mixing and overlap
matrixes should be slightly renormalized. In general case 
the renormalization constants, however, cannot be found 
without the system of equations for the GFs derived
within SCPT.
After some simplifying assumptions only one constant 
enters the equations and, therefore, only one 
equation should be added (see next paper).

The idea of the derivation performed consists of the following. The
density functional theory is based on an self-consistent solution of the
Kohn-Sham equations for the charge density and potential. The key question
is, of course, from where does one take the analytical form for this
potential. Usually it is taken from the theory of the \emph{homogeneous}
electron gas and phenomenologically extended to the case of non-homogeneous
systems.
Sham~\cite{sham}, 
using the Hohenberg and Kohn theorem~\cite{hoenberg_kohn}, has suggested a
theory which connects the exact exchange-correlation potential with the
exact self-energy. Although it is far from obvious that the same holds
for the case of an approximate self-energy, we will use this assumption.
Sham and Schl\"{u}ter~\cite{sham_schluter} managed to solve this equation for
the case of semiconductors within the approximation of spherical charge
spheres and found that the analytical form in this approximation is the same
as in the case of an phenomenological extension of the theory of homogeneous
gas. At last, Kotani~\cite{kotani} has shown that the results of the standard
DFT-LDA calculations for the total energy are well reproduced within the
static random phase approximation (RPA). These facts can be used as follows.
If we manage to 

1. establish a one-to-one mapping between a subset of the Feynman graphs 
for the self-energy within SCPT and WCPT (at least, in RPA) and

2. show that the Sham equation for the exchange-correlation potential
within this subset of graphs has the same form as its analogue in WCPT and

3. derive the expression for the charge density $\rho _{P}(r)$ within the
approximation in the SCPT, which is the analogue of the RPA in WCPT,

then, the analytical form for the exchange-correlation potential in this
approximation should be the same as in the standard case, but the usual
expression for the charge density $\rho (r)$ should be changed to $\rho
_{P}(r)$: $v_{xc}[\rho (r)]\rightarrow v_{xc}[\rho _{P}(r)]$, as well as the
matrix elements of different interactions should be renormalized as it is
required by the mapping. Since the LDA for the 
exchange-correlation potential works well for the case of delocalized
electrons, we, thus, have to show that any graph in WCPT describing the
Coulomb interaction between the electrons which belongs to one site
and the other either to another site, or to the delocalized states, has
its counter-part in the SCPT. As follows from the formulas for total energy
and the Sham equation, both require knowledge of the full \emph{Fermionic} GFs.
Therefore, we have to derive the equations for the GFs, $G$, 
constructed in $X$-operators and, then, express the Fermionic GF, 
$F$, in terms of GFs $G$.

In the SCPT the $f$-electron operator can be expressed in terms of 
Fermion-like intra-atomic excitation. In the approximation Hubbard-I
(HIA) the latter can be described in terms of renormalized electrons. The
mapping looks as follows 
\begin{eqnarray}
&&\hspace*{-5mm}
v^{Coul}\cdot \hat{f}_{\mu }=v^{Coul}\cdot f_{\mu }^{a}\hat{X}%
^{a}\Rightarrow (v^{Coul}\sqrt{P^{a}})\cdot f_{\mu }^{a}(\hat{X}^{a}/
\sqrt{P^{a}})\nonumber \\ 
&&\equiv v_{a}^{Coul}\cdot \tilde{f}^{a},
\end{eqnarray}
where $\tilde{f}^{a}$ is the operator of an effective Fermion, and 
$P^a$ is the spectral weight. Formally,
this approximation can be obtained by a simple exchange of the commutation
relations  
\begin{eqnarray}
\{X^{a},X^{\bar{b}}\} &=&\varepsilon _{\xi }^{a\bar{b}}Z^{\xi }=\varepsilon
_{\xi }^{a\bar{b}}\langle Z^{\xi }\rangle +[\varepsilon _{\xi }^{a\bar{b}%
}Z^{\xi }-\varepsilon _{\xi }^{a\bar{b}}\langle Z^{\xi }\rangle ]
\nonumber  \\ 
&\simeq&
\varepsilon _{\xi }^{a\bar{b}}\langle Z^{\xi }\rangle , \\
\{c_{1},X_{j_{2}}^{\bar{b}}\} &\simeq &O_{12}^{-1}f_{2}^{a}\varepsilon _{\xi
}^{a\bar{b}}\langle Z^{\xi }\rangle .
\end{eqnarray}
Within the orbital representation where $\varepsilon _{\xi }^{a\bar{b}%
}=\delta ^{a\bar{b}}\delta (\xi -[\Gamma ,\Gamma ])$, i.e.\, it can only take 
the values $1,0$, and the matrix elements $f_{2}^{a}=1,0,-1$. The
expectation value is non-zero only when it is one of the population
numbers, $\langle Z^{\xi }\rangle =\delta _{\xi ,[\Gamma ,\Gamma ]}\langle
h^{\Gamma }\rangle =\delta _{\xi ,[\Gamma ,\Gamma ]}N_{\Gamma }$; then 
\begin{eqnarray}
&&\hspace*{-5mm}
\varepsilon _{\xi }^{a\bar{b}}\langle Z^{\xi }\rangle =\varepsilon _{\xi }^{a%
\bar{b}}(\delta _{\xi ,[\Gamma ,\Gamma ]}+\delta _{\xi ,[\gamma ,\gamma
]})\delta (a-b)P^{a}\nonumber \\
&\equiv& \delta (a-b)[N_{\Gamma }+N_{\gamma }]
\end{eqnarray}
for the transition $a=[\gamma ,\Gamma ]$ and, therefore, 
\begin{equation}
\{c_{1},X_{j_{2}}^{\bar{b}}\}\simeq
O_{1_{c}2_{f}}^{-1}f_{2}^{a}P_{j_{2}}^{a}\delta (a-b).
\end{equation}
Then, the $X$-operator can be written as $X=(X^{a}/\sqrt{P^{a}})\sqrt{P^{a}}=%
\tilde{f}^{a}\sqrt{P^{a}}$. Then, the interaction is renormalized by the
square root of $P$, which is brought by the $X$- (or, due to
non-orthogonality of the basis set, $c$-) operator into the equation of
motion for the GF and in the lowest approximations plays the role of the 
spectral weight
of the pole, corresponding to the transition $a$. If the fluctuations
described by the GFs, $\langle T[\varepsilon _{\xi }^{a\bar{b}}Z^{\xi
}(t)-\varepsilon _{\xi }^{a\bar{b}}\langle Z^{\xi }(t)\rangle ][\varepsilon
_{\xi }^{a\bar{b}}Z^{\xi }(t^{\prime })-\varepsilon _{\xi }^{a\bar{b}%
}\langle Z^{\xi }(t^{\prime })\rangle ]\rangle$, are neglected, the
perturbation theory contains only Fermi-like Feynman graphs. The latter
means that any Feynman graph obtained within a weak-coupling expansion has its
counter-partner in the strong-coupling perturbation theory. Attention
has to be paid to the fact that the number of renormalized Fermions does not
coincide with the number of ordinary Fermions: the number of the latter is
equal to the number of \emph{orbitals} involved, while the number of the
former is equal to the number of \emph{Fermi-like transitions} involved.
Although it is not difficult in principle to answer the question, how to
derive this approximation and what remains beyond this approximation, it
requires quite lengthy calculations.

Different ways may be used for the calculation of the Fermionic GFs within the
SCPT: 1) directly, either via a chain of equations of motion, until the
chain becomes closed on one site via\ the operator relations $f^{2}=0,[\hat{n%
}_{\mu }^{(f)}]^{2}=\hat{n}_{\mu }^{(f)},$ or, equivalently, in terms of
chain of high-order GFs; 2) using the representation of Hubbard operators,
which diagonalize the on-site interactions, and the diagram technique for
them. Then we have to find the Fermion-like GFs via the Hubbard $X$%
-operators. Here we use the second way, since the SCPT for the GFs are needed
and some approximations are considered in a paper by Sandalov 
{\it et al}.~\cite{DT}
Let us inspect how
the first-order corrections to GFs generated by the same types
of the matrix elements of the Coulomb interaction look like 
(one should remember that
the inverse of the overlap matrix is not included into the definition of
self-energy).

The term of the Hamiltonian describing the Coulomb interaction between
electrons can be written in the $jL$-representation in the form 
\begin{eqnarray}
&&\hspace*{-5mm}
H_{Coul}=\frac{1}{2}v_{2345}[c_{2}^{\dagger }+(f_{2}^{\dagger })^{\bar{a}%
_{2}}X_{2}^{\bar{a}_{2}}][c_{3}^{\dagger }+(f_{3}^{\dagger })^{\bar{a}%
_{3}}X_{3}^{\bar{a}%
_{3}}] \nonumber \\
&\times& [c_{4}+f_{4}^{a_{4}}X_{4}^{a_{4}}][c_{5}+f_{5}^{a_{5}}X_{5}^{a_{5}}].
\end{eqnarray}
The interactions with the core electrons are included into the definition
of the $X$-operators here. Now we have to consider all these sixteen terms 
in order
to take into account single-site correlations via multiplication rules for
the Hubbard operators. It is
convenient to consider the terms which contain different number of $f$%
-operators separately. Thus, we form classes according to the
number of $f$-operators contained in the interaction term. In order to
reduce the number of terms in the equations of motion we collect similar
terms by making use of the commutation relations and a proper symmetrization
of the interaction. We also use the short-hand notations for indices: for
conduction electrons $1_{c}=(j_{1},L_{1})\equiv (j_{1},l_{1}\neq
3,m_{1l},s_{1}=1/2,\sigma _{1})$ and for $f$-electrons $\;1_{f}=(j_{1},\mu
_{1})\equiv (j_{1},l_{1}=3,m_{1l},s_{1}=1/2,\sigma _{1})$ , where it does
not lead to a confusion. Also, it is convenient to introduce the group
constants which are the coefficients in the multiplication rules for $X$%
-operators: 
\begin{equation}
\label{multiplication:eq}
X_{j}^{a}\cdot X_{j}^{\bar{b}}=\kappa _{\xi }^{a\bar{b}}Z_{j}^{\xi
},\;X_{j}^{a}\cdot Z_{j}^{\xi }=\kappa _{b}^{a\xi }X_{j}^{b},\;etc.
\end{equation}
Obviously, 
\begin{equation}
\varepsilon _{\xi }^{a\bar{b}}=\kappa _{\xi }^{a\bar{b}}+\kappa _{\xi }^{%
\bar{b}a}
\end{equation}
and so on.

The $f$-operator in $X$-representation is given by a sum of $X$-operators,
and we are not able to find the Fermionic GF 
$F^{(ff)}=\langle {\cal T}f_{1}(\tau
_{1})f_{1^{\prime }}^{\dagger }(\tau ^{\prime })\rangle $ without finding
the GFs $G^{Xc},G^{cX},G^{XX}$, involving Hubbard $X$-operators.
Therefore, we need the equations for these GFs. For the Hamiltonian which
includes hopping and mixing interactions (Hubbard-Anderson model) these
equations are given in a paper by Sandalov {\it et al}.~\cite{DT}
Therefore, we have to add the terms generated by
Coulomb interaction. If we write the equations in terms of functional
derivatives with respect to the external field, all mixed terms arise when
we iterate these equations. However, as seen from the form of the
contributions from the Coulomb interactions to the equations of motion for the
operators $c$ and $X$ (see Appendix~\ref{coul_eq_of_mot:app}), 
the fields which are introduced in Ref.~\cite{DT}, are not sufficient. 
Now instead of the $S_{{\rm ext}}$-matrix,
describing external fields we have to use the following 
$S_{{\rm ext}}$%
\begin{equation}
\label{action:eq}
S_{{\rm ext}}(-i\beta ,0)=\exp \left\{ -i\int_{0}^{-i\beta }dt{\cal L}%
_{{\rm ext}}(t)\right\} ,
\end{equation}
where
\begin{eqnarray}
\label{lagrange:eq}
{\cal L}_{{\rm ext}}(t) &=&\sum_{j\xi }U_{j\xi }^{Z}(t)Z_{j}^{\xi
}(t)+\sum_{jL,j^{\prime }L^{\prime }}c_{jL}^{\dagger }(t)U_{jL,j^{\prime
}L^{\prime }}^{cc}(t)c^{\vphantom{\dagger}}_{j^{\prime }L^{\prime }}(t) 
\nonumber \\
&+&\sum_{jL,j^{\prime }a^{\prime }}c_{jL}^{\dagger }(t)U_{jL,j^{\prime
}a^{\prime }}^{cX}(t)X_{j^{\prime }}^{a^{\prime }}(t)\nonumber \\
&+&\sum_{ja,j^{\prime
}L^{\prime }}X_{j}^{\bar{a}}(t)U_{j\bar{a},j^{\prime }L^{\prime
}}^{Xc}(t)c^{\vphantom{\dagger}}_{j^{\prime }L^{\prime }}(t)  \nonumber \\
&+&\sum_{ja,j^{\prime }L^{\prime }}X_{j}^{\bar{a}}(t)U_{j\bar{a},j^{\prime
}a^{\prime }}^{XX}(t)X_{j^{\prime }}^{a^{\prime }}(t)(1-\delta _{jj^{\prime
}}).
\end{eqnarray}
As seen, all terms except the one proportional to $U_{j\xi }^{Z}$ renormalize,
the $c$-$c$-hopping of the conduction electrons, mixing and $f$%
-$f$-hopping: 
\begin{eqnarray}
\HH^{cc} &\rightarrow &\tilde{\HH}^{cc}=\HH^{cc}+U^{cc},  \nonumber \\
W &\rightarrow &\tilde{W}=W+U^{cX},  \nonumber \\
t &\rightarrow &\tilde{t}=t+U^{cc},
\end{eqnarray}
therefore, we consider that all these external fields are included
into hopping and mixing matrix elements below. 
These fields  generate
the terms already considered in 
a paper by Sandalov {\it et al}.~\cite{DT}, 
in perturbation theory, and the problem
has a non-linear nature from the very beginning. Now, using the the
contributions into the equations of motion from the Coulomb interaction, given
in Appendix~\ref{coul_eq_of_mot:app}, 
we will express the new complex GFs in terms of the simple ones. It is
convenient to consider these contributions 
class by class; the number of the class corresponds to the number
of $f$-operators in the term of Coulomb interaction. Before doing this
it useful to note that contrary to the weak-coupling expansion the
representation which we use here is not unique. However, after one is chosen, 
it generates the system of graphs of the diagram technique which corresponds
to this choice of the closed form of the equations for the GFs since we 
obtain the
graphs by means of iterations of these equations. As a consequence, special 
care
is required in comparison with some other techniques: one should not expect
graph-to-graph correspondence between different diagram expansions for the
GFs. Our convention here is the following: in the first step we separate the
first Bose-like operator which stands in the left-time product 
and then, if necessary, 
make it again.
This sequence simplifies the calculation of the necessary time limits. The
analogue of the Hubbard-I approximation (HIA) is obtained if one fully
neglects all the contributions coming from the functional derivatives.

The approximation of ''renormalized Fermions'' can be introduced if

a) we will show that, from a physical point of view, the system of
correlated electrons has the Fermion-like excitations, which may differ 
very little from the 
ones in an weakly interacting Fermions gas, at least in some
region of parameters (strength of interactions, dimensionality, temperature,
etc.); 

b) from the point of view of mathematics we will show that even
in the strong-coupling regime a subsequence of Feynman graphs in SCPT exists
which describes these excitations. Then, the remaining graphs will 
determine the region of parameters where this approximation is valid.

We will now search for a solution to the equations of motion for the GFs 
in the form  
\begin{equation}
{\bf G}={\bf {\cal D}P},
\end{equation}
where the bold letters denote matrixes with respect to all indices and times
as well as their product implies matrix multiplication with respect to all
indices. Here the GF is 
\begin{equation}
G_{\alpha \beta }=\frac{1}{i}\langle {\cal T}
\eta^{\vphantom{\dagger}} _{\alpha }(t)\eta _{\beta
}^{\dagger }(t^{\prime })\rangle_{U},\;
\eta^{\vphantom{\dagger}}_{\alpha}(t)=
c^{\vphantom{\dagger}}_{jL}(t),X_{j}^{a}(t).
\end{equation}
The $cc$-component here $G_{11}=G^{(cc)}\equiv F^{(cc)}$ is a Fermion GF. In
terms of the operators $\eta$, the interaction can be written as 
\begin{equation}
\bar{V}_{Coul}=\frac{1}{2}
\bar{v}^{\vphantom{\dagger}}_{2345}\eta_{2}^{\dagger }\eta
_{3}^{\dagger}\eta^{\vphantom{\dagger}}_{4}\eta^{\vphantom{\dagger}}_{5},
\end{equation}
with 
\begin{eqnarray}
\bar{v}_{2345}^{cccc} 
&=&v^{\vphantom{\dagger}}_{j_{2}L_{2},j_{3}L_{3},j_{4}L_{4},j_{5}L_{5}}; 
\nonumber \\ 
\bar{v}_{2345}^{cfcc}&=&v^{\vphantom{\dagger}}_{j_{2}L_{2},j_{3}\mu
_{3},j_{4}L_{4},j_{5}L_{5}}(f_{\mu _{3}}^{\dagger })^{a_{3}};  \nonumber \\
\bar{v}_{2345}^{cffc} 
&=&v^{\vphantom{\dagger}}_{j_{2}L_{2},j_{3}\mu_{3},j_{4}\mu
_{4},j_{5}L_{5}}(f_{\mu _{3}}^{\dagger })^{a_{3}}(f_{\mu
_{4}})^{a_{4}},\;{\rm etc.}
\label{coul_elem:eq}
\end{eqnarray}
The external fields in Eq.(\ref{action:eq}) and Eq.(\ref{lagrange:eq})
can also be written shortly as $\eta
_{3}^{\dagger }U_{34}\eta _{4}$. Of course, the operators belonging to the
same site should be multiplied according to the rules for the $X$-operators, 
Eq.(\ref{multiplication:eq}). We shall consider these terms,
later. Since the anticommutators $\{c,X^{\dagger
}\},\{X,X^{\dagger }\}$ give an operator, not $c$-number, we have to write 
all equations 
for a general case $\{\eta _{1},\eta _{2}^{\dagger }\}=\varepsilon
_{3_{b}}^{12}Z^{3_{b}}$, where $Z$ is Bose-like operator and, therefore, $%
\bar{V}_{Coul}$ gives the following contribution into the equations of motion 
\begin{equation}
\lbrack\eta_{1},\bar{V}_{Coul}]=
\bar{v}^{\vphantom{\dagger}}_{[23]45}\varepsilon
_{3_{b}}^{12}Z^{3_{b}}\eta_{3}^{\dagger}\eta^{\vphantom{\dagger}}_{4}
\eta^{\vphantom{\dagger}}_{5}.
\end{equation}
In the case when $\eta _{1}=c_{1}$ and $\eta_{2}^{\dagger }=c_{2}^{\dagger }
$, the operator $Z^{3_{b}}=1$ and $\varepsilon
_{3_{b}}^{12}=O_{1_{c}2_{c}}^{-1}.$ In the equation for the GF the term
generated by $[\eta _{1},\bar{V}_{Coul}]$ can be written as follows  
\begin{eqnarray}
&&\hspace*{-5mm}
\bar{v}^{\vphantom{\dagger}}_{[23]45}\varepsilon _{3_{b}}^{12}\left[ \langle
Z^{3_{b}}(t^{+}\rangle +i\frac{\delta }{\delta U^{3_{b}}(t^{+})}\right] 
\nonumber \\
&&\times\langle {\cal T}\eta_{3}^{\dagger}(t)\eta^{\vphantom{\dagger}}_{4}(t)
\eta^{\vphantom{\dagger}}_{5}(t)\eta_{1^{\prime
}}^{\dagger }(t^{\prime })\rangle _{U} \nonumber \\
&=&\bar{v}_{[23]45}\varepsilon _{3_{b}}^{12}\left[ \langle
Z^{3_{b}}(t^{++})\rangle +i\frac{\delta }{\delta U^{3_{b}}(t^{++})}\right]
\nonumber \\
&&\times\left[ \langle {\cal T}\eta _{3}^{\dagger }(t^{+})
\eta^{\vphantom{\dagger}}_{4}(t^{+})\rangle +i\frac{%
\delta }{\delta U_{34}(t^{+})}\right] G_{51^{\prime }}^{\eta \eta
}(t,t^{\prime }|U).
\end{eqnarray}
Here $t^{++}$ and $t^{+}$ denote the limits $%
\lim_{t_{1}\rightarrow t+0}\lim_{t_{2}\rightarrow t_{1}+0}$ respectively. 
Recall that $%
[F_{0}^{cc}(t,t^{\prime }|U)^{-1}]_{51^{\prime }}$ does not depend on the
fields $U^{3_{b}}$ and, therefore, $\delta F_{51^{\prime }}^{cc}(t,t^{\prime
}|U)/\delta U^{3_{b}}(t^{\prime \prime })$ can give non-zero
contributions only via the mixed GFs 
$\propto \langle {\cal T}cX^{\dagger }\rangle$.
Therefore, we can write the equation for the GF in the following form  
\begin{eqnarray}
&&\hspace*{-5mm}
G_{1\bar{1}^{\prime }}={\cal D}_{11^{\prime \prime }}^{0}P_{1^{\prime \prime }%
\bar{1}^{\prime }}+{\cal D}_{11^{\prime \prime }}^{0}\varepsilon
_{4_{b}}^{1^{\prime \prime }\bar{2}}\left[ \langle 
Z^{4_{b}}\rangle +i\frac{%
\delta }{\delta U_{4_{b}}}\right] \nonumber \\ 
&&\times[V_{\bar{2}3}+U_{\bar{2}3}]G_{3\bar{1}%
^{\prime }}  \nonumber \\
&&+{\cal D}_{11^{\prime \prime }}^{0}\varepsilon_{4_{b}}^{1^{\prime\prime}
\bar{2}%
}\left[ \langle 
Z^{4_{b}}\rangle +i\frac{\delta }{\delta U_{4_{b}}}\right] \nonumber \\ 
&&\times\bar{v}_{[\bar{2}\bar{3}]45}\left[ -G_{4\bar{3}}
(t_{3}^{+},t_{3})+i\frac{%
\delta }{\delta U_{34}(t_{3}^{+})}\right] G_{5\bar{1}^{\prime }},
\label{iteration:eq}
\end{eqnarray}
where for brevity we write the time argument only in those places where a
confusion may arise. The mixing and hopping are denoted by $V$ following
the definitions in I.
We see that there is an essential difference with the
standard WCPT: here we have to calculate the functional derivatives twice,
while in WCPT only second square bracket is present. Now let us, in order
to give the reader a feeling of the SCPT, make the first iterations and
give both the analytical and graphical expressions for the corrections to the 
GF. In
zero order of SCPT $G_{1\bar{1}^{\prime }}\Rightarrow G_{1\bar{1}^{\prime
}}^{0}={\cal D}_{11^{\prime \prime }}^{0}
P_{1^{\prime \prime }\bar{1}^{\prime }}^{0}
$. Let us now insert $G_{1\bar{1}^{\prime }}\Rightarrow G_{1\bar{1}^{\prime
}}^{0}$ into the right-hand side. We have the terms where 
$\delta /\delta U=0$, 
\begin{eqnarray}
\label{start_comp:eq}
\delta G_{1\bar{1}^{\prime }}^{(1)} &=&{\cal D}_{11^{\prime \prime
}}^{0}\varepsilon _{4_{b}}^{1^{\prime \prime }\bar{2}}\langle
Z^{4_{b}}\rangle \lbrack V_{\bar{2}3}+U_{\bar{2}3}]G_{3\bar{1}^{\prime
}}^{0}  \nonumber \\
&&-{\cal D}_{11^{\prime \prime }}^{0}
\varepsilon _{4_{b}}^{1^{\prime \prime }\bar{2}%
}\langle Z^{4_{b}}\rangle \bar{v}_{[\bar{2}\bar{3}]45}G_{4\bar{3}%
}(t_{3}^{+},t_{3})G_{5\bar{1}^{\prime }}^{0}.
\end{eqnarray}
Since $\varepsilon _{4_{b}}^{1^{\prime \prime }\bar{2}}\langle
Z^{4_{b}}\rangle =P^{1^{\prime \prime }\bar{2}}$ and 
$G={\cal D}P$, we have simply 
\begin{eqnarray}
&&\hspace*{-5mm}
\delta G_{1\bar{1}^{\prime }}^{(1)}=G_{1\bar{2}}^{0}[V_{\bar{2}3}+U_{\bar{2}%
3}]G_{3\bar{1}^{\prime }}^{0} \nonumber \\
&&-G_{1\bar{2}}^{0} P^{1^{\prime \prime }\bar{2}}
\bar{v}_{[\bar{2}\bar{3}]45}G_{4\bar{3}}^{0}(t_{3}^{+},
t_{3})G_{5\bar{1}^{\prime}}^{0}
\label{firstcorr:eq}
\end{eqnarray}
We denote the pseudolocator ${\cal D}$ by a 
solid line with arrow in the Feynman graphs, the end-factor $P$ with an open 
circle, for the mixing-hopping 
plus external field, $V+U$, a wavy line is used 
and for the Coulomb
interaction we use a dashed line. Below, we will also need a notation for
the Bose-like correlation function $K^{1_{b}2_{b}}(t,t^{\prime })=\delta
\langle Z^{1_{b}}(t)\rangle /$ $\delta U_{2_{b}}(t^{\prime })$; 
we will use
the curly line for it. The graphs Fig.~\ref{A:fig} and Fig.~\ref{a:fig}  
correspond  to the analytical expressions in Eq.(\ref{firstcorr:eq}). 
Since $\delta G_{1%
\bar{2}}^{0}/$ $\delta U_{\bar{3}4}=0$ and this derivative removes one of the 
interactions $[V_{\bar{2}3}+U_{\bar{2}3}]$ in any expression, in order
to obtain the other graphs of first order with respect to Coulomb
interaction, we have to insert $G_{5\bar{1}^{\prime
}}\Rightarrow \delta _{V}^{(1)}G_{5\bar{1}^{\prime }}=G_{5\bar{2}}^{0}[V_{%
\bar{2}3}+U_{\bar{2}3}]G_{3\bar{1}^{\prime }}^{0}$ into the generating 
equation. Then, the term with the 
derivative $\delta \lbrack \delta _{V}^{(1)}G_{5\bar{1}^{\prime }}]/$ $%
\delta U_{\bar{3}4}$  produces the exchange graph (see Fig.~\ref{b:fig}): 
\begin{equation}
\label{exc_contr:eq}
G_{1\bar{2}}^{0}\bar{v}_{[\bar{2}\bar{3}]45}G_{5\bar{3}}^{0}G_{4\bar{1}%
^{\prime }}^{0}.
\end{equation}

Next, 
we find four terms generated by the derivative $\delta /\delta U_{4_{b}}$: 
\begin{eqnarray}
&&{\cal D}_{11^{\prime \prime }}^{0}
\varepsilon _{4_{b}}^{1^{\prime \prime}\bar{2}}%
\bar{v}^{\vphantom{\dagger}}_{[\bar{2}\bar{3}]45}
{\cal D}_{46}^{0}(t_{3}^{+},t_{6})\Gamma
_{67,4_{b}}^{0}G_{7\bar{3}}^{0}(t_{7},t_{3})G_{5\bar{1}^{\prime}}^{0} 
\label{c-graph:eq} \\
&-&{\cal D}_{11^{\prime \prime }}^{0}
\varepsilon _{4_{b}}^{1^{\prime \prime}\bar{2}%
}\bar{v}^{\vphantom{\dagger}}_{[\bar{2}\bar{3}]45}
{\cal D}_{46}^{0}\varepsilon _{6_{b}}^{6\bar{3}%
}K^{\vphantom{\dagger}}_{6_{b}4_{b}}G_{5\bar{1}^{\prime }}^{0} 
\label{d-graph:eq} \\
&+&{\cal D}_{11^{\prime \prime }}^{0}
\varepsilon _{4_{b}}^{1^{\prime \prime}\bar{2}}%
\bar{v}^{\vphantom{\dagger}}_{[\bar{2}\bar{3}]45}
G_{4\bar{3}}^{0}(t_{3}^{+},t_{3}){\cal D}_{56}^{0}%
\Gamma _{67,4_{b}}^{0}G_{7\bar{1}^{\prime }}^{0} \label{e-graph:eq}\\
&-&{\cal D}_{11^{\prime \prime }}^{0}
\varepsilon _{4_{b}}^{1^{\prime \prime}\bar{2}%
}\bar{v}^{\vphantom{\dagger}}_{[\bar{2}\bar{3}]45}G_{4\bar{3}%
}^{0}(t_{3}^{+},t_{3}){\cal D}_{56}^{0}
K^{\vphantom{\dagger}}_{6_{b}4_{b}}\varepsilon _{6_{b}}^{6\bar{1}^{\prime}}.
\label{f-graph:eq}
\end{eqnarray}
The corresponding graphs are shown in the Figures \ref{c:fig}, \ref{d:fig}, 
\ref{e:fig} and \ref{f:fig} respectively. 
Continuing these iterations we find that there are a subsequence
given by the graphs in the Figures 
\ref{g:fig}, \ref{h:fig}, \ref{k:fig} and \ref{l:fig}, 
which exactly have the structure of
the WCPT: the only difference is that instead of Fermion GFs, $F$, in WCPT
here we are dealing with the GFs, $G$. 
Obviously, each one of these GFs, $G$, can be
dressed with hopping and mixing, as shown in Fig. \ref{n:fig},
if the corresponding transition is in the
energy region where these interactions are not equal to zero (see next
sections). 

Then, these GFs describe delocalized transitions, or, in other
words, Hubbard type of bands. The graphs \ref{c:fig}, \ref{d:fig}, 
\ref{e:fig} and \ref{f:fig} do not appear in
WCPT for the single-electron GFs and describe contributions from \emph{%
kinematic} interactions. Let us now return to the
definition , Eq.(\ref{coul_elem:eq}), 
of the Coulomb matrix elements, $\bar{v}_{[\bar{2}\bar{3}]4Fig. 5}$, and
consider, for example, the graph \ref{c:fig}. In the case when one of the 
indices, in Eq.(\ref{c-graph:eq}), 
say, $5$, describes a $c$-electron, there are no additional factors, 
$f^{a_{5}}$. If the index 
$5$ correspond to an \f-electron, then there is a factor 
$f^{a_{5}}$. Such
factors are automatically provided by the matrix elements 
$\bar{v}_{[\bar{2}\bar{3}]45}$, Eq.(\ref{coul_elem:eq}),  
for all inner lines, but not for the external ends. Therefore, if we
multiply these type of graphs by the external factors, say, for $c$, and use
the expansion of the Fermi operator in terms of the Hubbard operators, we
find that  
\begin{equation}
\label{end_comp:eq}
f_{\nu _{1}}^{a_{1}}G_{1\bar{2}}^{0}
\bar{v}^{\vphantom{\dagger}}_{[\bar{2}\bar{3}]45}G_{5\bar{3}%
}^{0}G_{4\bar{1}^{\prime }}^{0}(f_{\nu _{1^{\prime }}}^{\dagger
})^{a_{1^{\prime }}}=F_{1\bar{2}}^{0}
v^{\vphantom{\dagger}}_{[\bar{2}\bar{3}]45}F_{5\bar{3}%
}^{0}F_{4\bar{1}^{\prime }}^{0},
\end{equation}
i.e.\ these graphs describe the standard Coulomb scattering of electrons, but 
\emph{the Fermion GFs should be found within SCPT}. 
Besides, as seen from these iterations, 
any application of the derivative of 
the sort 
$\delta /\delta U^{3_{b}}(t^{\prime\prime})$ to the expression containing
at least one GF $G(t,t^{\prime }|U)$ leads to the graph describing the effect
of kinematic interaction and, therefore, destroys the one-to-one correspondence
between the WCPT and SCPT in the ARF series. 
Thus, in order to obtain the equation 
for the GF in ARF we have to neglect in  the first bracket of the 
Eq.(\ref{iteration:eq}) 
$\delta G(t,t^{\prime }|U)/\delta U^{3_{b}}(t^{\prime \prime})$. 
Then, the interaction is renormalized by the expectation value of the
Bose-like operator 
\begin{equation}
\label{general1:eq}
\bar{v}_{2345}\rightarrow 
\bar{v}^{\vphantom{\dagger}}_{2345}\varepsilon _{3_{b}}^{12}\langle
Z^{3_{b}}(t^{++})\rangle =\bar{v}^{\vphantom{\dagger}}_{2345}P^{12}.
\end{equation}
This condition removes not only the kinematic interactions, but 
also any of the
graphs containing the correlation functions $K$ (see the figures 
\ref{d:fig} and \ref{f:fig}). 
It is worth to note that some of these correlators describe spin waves
in the magnetic ordered media, therefore, this type of scattering of 
carriers is beyond the ARF.  In spirit the ARF corresponds 
to the Hubbard-I approximation (in our definition). 

This remarkable similarity can be achieved in the level of equations of
motion for the GFs. Although this equation is approximate  it  is still a
functional equation,  therefore, {\em the similarity holds in any orders of the
PT}, which can be generated by iterations of these equations. Therefore,  the
analogue of  the RPA in WCPT can be constructed too. This is
obvious from the graphs in the figures \ref{l:fig} and \ref{m:fig} 
and further graphs containing an 
increasing number of loops. In matrix notation
the definition of this approximation arises from the ''reduced''
differentiation of the GF. Namely, the higher correlation functions arising
in the equation of motion are expressed in terms of functional derivatives
of the GFs with  respect to the bosonic fields which are needed to
reproduce these higher order GFs. 
Thus, we will call the described approximation  the 
''\emph{approximation of renormalized Fermions}''.  The factor 
${\bf P,}$ is specific for each transition constant, 
renormalizing the hopping, Coulomb and mixing interactions. One can
expect from this comparison that those abnormal features of some 
compounds which
are different from the normal Fermi liquid have to be described by some of the 
remaining graphs which are not included into the set defined above as ARF. 
The 
analogy can be continued, for example, one can  introduce the self
energy in SCPT within the ARF in the form 
\begin{equation}
\label{sigma_scpt:eq}
\Sigma ^{SCPT}=[\bar{v}P(\langle \eta ^{\dagger }\eta \rangle +i\frac{\delta 
}{\delta U})G]G^{-1},
\end{equation}
which is in complete analogue to the standard definition used in the WCPT, 
\begin{equation}
\label{sigma_wcpt:eq}
\Sigma^{WCPT}=[v(\langle \rho \rangle +i\frac{\delta }{\delta U})F]F^{-1}.
\end{equation}
Thus, we have established a one-to-one correspondence between the 
perturbation theory for {\em single}-electron GFs, $F$,  
(weak-coupling expansion)  and a sub-series in 
the perturbation theory for the {\em many}-electron GFs, $G$. 
This is the central result of the present paper.
One of our targets formulated in the introduction was to build a 
bridge between the standard DFT-LDA calculations and SCPT and to find an
extension of LDA which is reasonably simple and takes into account some of
the features of SCE. The  form of the
exchange-correlation potential $v_{xc}^{h.e.g.}$ which is used in DFT-LDA is 
derived from the theory of the \emph{homogeneous} electron gas, i.e.\ for
the case of fully delocalized electrons. Therefore, the potential $v^{DFT}$,
should be the solution to the Sham equation\cite{sham},
\begin{equation}
0=\sum_{\omega }e^{i\omega 0^{+}}\{F_{12}^{DFT}(i\omega )[\tilde{\Sigma}%
_{23}(i\omega )-v_{23}^{DFT}]F_{31^{\prime }}(i\omega )\}.
\label{sham:eq}
\end{equation}
 is expected to have an analytical form
close to $v_{xc}^{h.e.g.}$ 
only for the delocalized Bloch electrons. 
The connection between the matrix elements $\tilde{\Sigma}%
_{25}(i\omega )$ and the coefficients of expansion $\Sigma _{34}(i\omega )$
is given by the relation
\begin{eqnarray}
&&\hspace*{-5mm}
\tilde{\Sigma}_{25}(i\omega )=\int \dd x\dd x^{\prime}\phi _{2}(x_{1})\Sigma
(x,x^{\prime },i\omega )\phi_{5}^{\ast}(x_{2}) \nonumber \\
&& \equiv O_{23}\Sigma_{34}(i\omega)O_{45},
\end{eqnarray}
In turn, this means that the one-to-one
correspondence, equations \ref{sigma_wcpt:eq} and \ref{sigma_scpt:eq}, 
found between the WCPT and SCPT series of graphs can
be constructively used for the delocalized transitions. 
The recipe follows from comparison of these two equations: the following 
replacements should be made in the expression for the self-energy: 
$F \rightarrow \langle \eta \eta ^{\dagger }  \rangle $ and the matrix element 
of Coulomb interaction $v \rightarrow \tilde{v}P $. Besides, the charge density 
should be rewritten in the form, where each $f$-operator is represented in 
terms of $X$-operators: $f_{\mu }= f_{\mu }^a X^a $. As follows from I, 
and will be shown in details in III,
the matrix elements of mixing interaction and overlap matrixes 
are also renormalized by the spectral weights.
Let us now compare the equations for the dielectric permeability within WCPT,
Eq.(\ref{epsilon_wc:eq}),
and SCPT. On one hand, according to our comparison of graphs,
Eq.(\ref{start_comp:eq})-(\ref{end_comp:eq}),
and the more general observation,
Eq.(\ref{general1:eq})-(\ref{sigma_wcpt:eq}),
in order to transform a graph of
WCPT into the one of SCPT we have to replace each bare $f$-GF as follows
\begin{equation}
F_{\nu }^{(0)}=(\omega -\varepsilon _{0}^{f})^{-1}\rightarrow \sum_{a_{2}}%
\frac{|(f_{\nu _{e}})^{a}|^{2}P^{a}}{i\omega -\Delta _{\bar{a}}}.
\end{equation}
On the other hand, the low-energy transitions does almost not form bands
at all.
Let us write down the elementary loop (the
graph beyond exchange) and separate the lower transitions in it .
Each index in the loop
\begin{equation}
\label{rpa_looop:eq}
\Pi _{12,34}^{(0)}(i\Omega )=T\sum_{i\omega }F_{12}(i\omega )F_{34}(i\omega
+i\Omega )
\end{equation}
can take the values $1=1_{c},1_{f}^{e},1_{f}^{f}$ , i.e.\ the index of a
conduction electron, $j_{1}L_{1}$, or one of values for an $f$-electron, $%
j_{1}\nu _{1}^{e}$ or $j_{1}\nu _{1}^{f}$, where $e$ denotes an empty
(delocalized) and $f$ stands for a filled (localized) orbitals. If we
neglect, for a first step, the contributions from the Coulomb matrix elements
with an odd number of f-operators\footnote{We can do it since 
even differential overlap between localized $f$-orbitals and delocalized 
$c$-ones is negligible}, then the lower pole contributes only in
the loops,
 which contain one GF of the $ff$-type, $F_{1_{f}2_{f}}$or $%
F_{3_{f}4_{f}}$. All other loops are standard ones with the only
difference that the standard $f$-GFs, $F_{1_{e}2_{e}}^{(0)}$, when the
f-electrons are treated as valent, however they acquire the factor $|(f_{\nu
_{e}})^{a_{2}}|^{2}P^{a_{2}}$  where the sum is taken over the upper
transitions $a_{2}$, and $\varepsilon _{0}^{f}$ should be replaced by $%
\Delta _{\bar{a}_{2}}$.  Since in the loop
\begin{equation}
\Pi _{1_{f}^{f}2_{f}^{f},3_{c}4_{c}}^{(0)}(i\Omega )\sim
\sum_{k}B_{1_{f}^{f}2_{f}^{f}}B_{3_{c}4_{c}}(k)\frac{f(\Delta _{\bar{a}%
_{1}})-f(\varepsilon _{k\lambda }^{\sigma }-\mu )}{\Delta _{\bar{a}%
_{1}}-\varepsilon _{k\lambda }^{\sigma }+i\Omega }
\end{equation}
$f(\Delta _{\bar{a}_{1}})$ is always equal to 1 and the Fermi
function $f(\varepsilon
_{k\lambda }^{\sigma }-\mu )=0$ only if   $\varepsilon _{k\lambda }^{\sigma
}>\mu $ , the loops $\Pi _{1_{f}^{f}2_{f}^{f},3_{c}4_{c}}^{(0)}(i\Omega )$ have
large denominators in the region where the numerator is not equal to zero
and, therefore, they are small compared to $\Pi
_{1_{c}2_{c},3_{c}4c}^{(0)}(i\Omega )$. 
Here, the coefficients $B$ transform the
GFs from the orbital representation to the bands $k\lambda$. Thus, as it
is expected, the main contribution to the  screening comes from the 
conduction
electrons including the delocalized upper transitions. Within this
approximation the formulas: Eq.(\ref{dielectrrr:eq}) for the
dielectric permeability $\varepsilon
^{-1}$, Eq.(\ref{sigma_RPA:eq}) for the self-energy
$\Sigma^{{\rm RPA}}$  remains valid,
however, with the replacement of the full GF by only the 
upper-pole part. There are,
of course, also  contributions from the electrostatic interactions between
electrons of different sites, which are treated in a standard way with the
difference that each transition enters with the its spectral weight.

Let us turn now to the Galitskii-Migdal expression
Eq.(\ref{tote:eq})
for the total
energy. In the last term of
Eq.(\ref{tote:eq})
we have a summation over the transitions. For
the lower transitions $O_{j_{1}j_{2}}=\delta _{j_{1}j_{2}}$ , $%
t_{j_{1}j_{2}}=\delta _{j_{1}j_{2}}\varepsilon _{0}^{f}$ and the local terms
give a contribution to the energy from the core-like levels.
The remaining expression, within the ARF,
coincides with the contribution to the energy from delocalized electrons
within the WCPT if we make the replacement in the $f$-GFs discussed above.
Exactly the same arguments can be applied to the Sham equation,
Eq.(\ref{sham:eq}),
connecting
the many-body GF and the self-energy with the DFT GF and the
exchange-correlation potential.
Therefore, we
can conclude that the analytical form of the exchange-correlation potential
used for the description of delocalized electrons can be applied
also to the case of correlated  electrons if we will use the expression for
the charge density Eq.(\ref{charge_dens:eq} in the form 
(\ref{charge_dens100:eq}) (the derivation is given in next section).
   Note that in the limit when the lower
spectral weight for the  localized electrons $P=1$ and $\Delta
_{a_{2}}\rightarrow \infty $ we are back to the standard model for the
lanthanides (see the next section), 
since in this case we have $n$ $f$-electrons localized and the
upper transitions do not contribute to the energy. Then, the
possibility to describe photo-electron spectroscopy disappear. When $P<1,
$ we have to solve  a system of equations for the spectral weights
additionally to a standard DFT calculation and bring these two calculations
to self-consistency. The remaining
important question,  the magnitudes of $\Delta_1$ and
$\Delta_2$ will be discussed in last section. 

Since only a part
of the transitions is delocalizing, some difference arises between
electrons and quasi Fermionic excitations, which we call renormalized
Fermions.
 \emph{First}, the numbers of
Fermions and renormalized Fermions are different, since the latter is
determined by the number of Fermi-like intra-atomic transitions, involved
in the physics under consideration. \emph{Second}, the WCPT generates the
graphs describing all intra-core and \f-\f interactions, relating to the
same ion, while they are already taken into account via the algebraic
construction of the many-electron state in the GFs, $G$. The 
intra-atomic Coulomb interactions are
included into the WCPT theory separately; as will be seen later, in SCPT the
transitions, describing deep intra-ionic transitions are naturally separated 
by a big energy gap from the delocalized electrons and actually the
discussed mapping involves only the delocalized states, providing, 
(if necessary) a description of the many-electron structure of the ion
including the Hund rules. \emph{Third} 
In order to make the two series, WCPT and SCPT, fully analogous in the Coulomb 
interaction, Eq.(\ref{coul_elem:eq}), and to derive the ARF, we had to treat 
all terms in the way as if they belonged to different sites. Therefore, we 
had to add the graphs, which treat the correction to the terms 
in Eq.(\ref{coul_elem:eq}) with a few \f-operators belonging to the same 
site, and subtract those, which within the ARF, are treated incorrectly. 
This difference should be included in the error of the ARF. 
Therefore, the accuracy of this approximation (or an
applicability to the physics in question) is determined not only by those
graphs which describe the different manifestations of the kinematic
interactions, but also by the difference between the correlated 
and decoupled single-ion terms in the ARF. 
In Appendix~\ref{coul_eq_of_mot:app}, 
we have separated all Coulomb terms into classes
containing different number of $f(d)$-operators (or core-operators): the
terms belonging to the class 0 do not contain them at all, the ones
belonging to the class 1 contain one $f$-operators etc. Due to the large
number of terms this consideration is quite lengthy. The terms generated by
Coulomb interaction in the equations of motion for 
Green functions operators are given in Appendix~\ref{coul_eq_of_mot:app}.

In order to estimate the error, due to this difference between the correlated 
and decoupled single ion terms, 
we have to take all the graphs which contain some of the sites, 
or all of them, coinciding and subtract the graphs with the same value of
the matrix element of Coulomb interaction but with the analytical expression
for the set of the GFs corresponding to certain choice of links between sites. 
The sites are defined as linked if they belong to the same site. 
These different
choices are considered in Appendix~\ref{coul_eq_of_mot:app}.
The interaction of the
density-density type is expected to give the main contribution, therefore, let
us consider the graph $b$ as a example. It takes into account the Hartree
contributions from the terms of the Coulomb Hamiltonian: $V_{1}^{C}=\bar{v}%
_{[j_{2}\bar{a}_{2},j_{3}\bar{a}_{3}]j_{4}a_{4},j_{5}a_{5}}X_{j_{2}}^{\bar{a}%
_{2}}X_{j_{3}}^{\bar{a}_{3}}X_{j_{4}}^{a_{4}}X_{j_{5}}^{a_{5}}$, $V_{2}^{C}=%
\bar{v}_{[j_{2}\bar{a}_{2},j_{3}L_{3}]j_{4}L_{4},j_{5}a_{5}}X_{j_{2}}^{\bar{a%
}_{2}}c_{j_{3}L_{3}}^{\dagger }c_{j_{4}L_{4}}X_{j_{5}}^{a_{5}}$ and $%
V_{3}^{C}=v_{[j_{2}L_{2},j_{3}L_{3}]j_{4}L_{4},j_{5}L_{5}}c_{j_{2}L_{2}}^{%
\dagger }c_{j_{3}L_{3}}^{\dagger }c_{j_{4}L_{4}}c_{j_{5}L_{5}},$ where all
the sites $j_{i}$ are different. There are also terms of the $V_{1}^{C}$
type where the sites $j_{3}=j_{4}$ or/and $j_{2}=j_{5}$. In terms of
Appendix~\ref{coul_eq_of_mot:app}
the contributions from these interactions belong to
the class 4 and are described by equations 
(\ref{cl4dgff:eq}) to (\ref{cl4dgff2:eq}). We have to put 
$\delta /\delta U^{Z}=0$ in these terms. The term $\bar{v}_{j_{2}\bar{a}%
_{2},j_{3}\bar{a}_{3}j_{3}a_{4},j_{5}a_{5}}X_{j_{2}}^{\bar{a}_{2}}X_{j_{3}}^{%
\bar{a}_{3}}X_{j_{3}}^{a_{4}}X_{j_{5}}^{a_{5}}=\bar{v}_{j_{2}\bar{a}%
_{2},j_{3}\bar{a}_{3}j_{3}a_{4},j_{5}a_{5}}\kappa _{\xi }^{\bar{a}%
_{3}a_{4}}X_{j_{2}}^{\bar{a}_{2}}Z_{j_{3}}^{\xi }X_{j_{5}}^{a_{5}}$
generates the following contributions to the self-energy $\Sigma
_{ja,j_{5}a_{5}}$%
\begin{eqnarray}
\bar{v}_{j_{2}\bar{a}_{2},j_{3}\bar{a}_{3}j_{3}a_{4},j_{5}a_{5}}[\delta
_{jj_{2}}\kappa _{\xi }^{\bar{a}_{3}a_{4}}\varepsilon _{\xi _{1}}^{a\bar{a}%
_{2}}\langle Z_{j}^{\xi _{1}}\rangle \langle Z_{j_{3}}^{\xi }\rangle 
\nonumber \\ -\delta
_{jj_{3}}\kappa _{\xi }^{\bar{a}_{3}a_{4}}\varepsilon _{b}^{a\xi }\langle
X_{j_{2}}^{\bar{a}_{2}}X_{j}^{b}\rangle ],
\end{eqnarray}
whereas the term which has been added for inclusion into the ARF and, 
therefore,
subtracted, is
\begin{eqnarray}
\bar{v}_{j_{2}\bar{a}_{2},j_{3}\bar{a}_{3}j_{3}a_{4},j_{5}a_{5}}[\delta
_{jj_{2}}\kappa _{\xi }^{\bar{a}_{3}a_{4}}\varepsilon _{\xi _{1}}^{a\bar{a}%
_{2}}\langle Z_{j}^{\xi _{1}}\rangle \langle Z_{j_{3}}^{\xi }\rangle 
\nonumber \\ -\delta
_{jj_{3}}\varepsilon _{\xi _{1}}^{a\bar{a}_{3}}\langle Z_{j}^{\xi
_{1}}\rangle \langle X_{j_{2}}^{\bar{a}_{2}}X_{j}^{a_{4}}\rangle ].
\end{eqnarray}
The difference is
\begin{equation}
\bar{v}_{j_{2}\bar{a}_{2},j_{3}\bar{a}_{3}j_{3}a_{4},j_{5}a_{5}}\delta
_{jj_{3}}[-\kappa _{\xi }^{\bar{a}_{3}a_{4}}\varepsilon _{b}^{a\xi
}+\varepsilon _{\xi }^{a\bar{a}_{3}}\delta ^{a_{4}b}\langle Z_{j}^{\xi
}\rangle ]\,\langle X_{j_{2}}^{\bar{a}_{2}}X_{j}^{b}\rangle.
\label{vv72:eq}
\end{equation}
It is convenient to use the equality
\begin{equation}
\kappa _{\xi }^{\bar{a}_{3}a_{4}}\varepsilon _{b}^{a\xi }=\varepsilon _{\xi
}^{a\bar{a}_{3}}\kappa _{b}^{\xi a_{4}},
\end{equation}
which follows from the commutation relations between Hubbard operators.
Then, the difference between the terms is proportional to
\begin{equation}
(-\langle Z_{j}^{\xi }\rangle \delta ^{a_{4}b}+\kappa _{b}^{\xi
a_{4}})\varepsilon _{\xi }^{a\bar{a}_{3}}
\end{equation}
for diagonal $\xi =[\Gamma ,\Gamma ]$ this is $\propto (1-P^{a_{4}})$.
Now we will use an information about the spectral weights which is 
considered in
details in the next sections. Both, localized and delocalized $f$-orbitals have
upper ($\Delta _{a}^{u}\sim \Delta _{1}$) and lower ($\Delta _{a}^{l}\sim
\Delta _{2}$) transitions . For localized orbitals and for lower transitions
$\bar{a}_2$ or/and $b$ in Eq.(\ref{vv72:eq}), 
the expectation values, $\langle X_{j_{2}}^{\bar{a}%
}X_{j}^{b}\rangle$, are small ($\ll 1$) due to the 
suppressed hopping and mixing
in the region of energies at energies $\omega \sim \Delta _{a_{1}}$ and big
energy gap $\varepsilon _{F}-\Delta _{a_{1}}$; the latter suppresses
admixture across the gap. 
For the upper transitions the localized orbitals
themselves have in this region of energies small spectral weight and again $%
\langle X_{j_{2}}^{\bar{a}}X_{j}^{b}\rangle \ll 1$. 
The weight $P^{a_{4}}$ is close to unity for the delocalized
orbitals near the upper transitions ($\Delta_{2}$), therefore, the small
constant decreasing the difference in Eq.(\ref{vv72:eq}) is 
$\propto(1-P^{a_{4}})$. 


It is also worth to discuss a possible confusion: it may seem that these
speculations are wrong just for the simple reason that the equation for the 
GF
in the WCPT is \emph{exact} while both representations (single-electron
Fermionic and many-electron) are equivalent; therefore, WCPT must generate
some graphs which should not appear in the SCPT for an
 approximate equation for the GF
in SCPT. Indeed, all this consideration might be performed within the WCPT,
however, in this case one has to consider the theory, not for the
single-electron GFs, $F$, in WCPT but, for the many-electron ones, $%
G_{F}(x_{1},x_{2},x_{3},...,x_{n})$, (here $x_{i}=(r_{i},t_{i})$ ) which
correspond to the functions $G_{X}$. The functions $%
G_{F}(x_{1},x_{2},x_{3},...,x_{n};R_{f})$ describe the 
relative motion of all $n$
electrons in an ion and, therefore, contain much more information than we
actually need. This means that to find them require much more effort 
than for the simplified functions $G_{X}$, which are
constructed on many-electron operators, $X(t)$, belonging to the same 
time, $t$. Of course, the same argument shows the insufficiency of the
approach discussed in the cases when a knowledge of the 
dynamics of each electron is
essential. This situation arises when the magnitude of the perturbation is so
large, that the complexes are destroyed completely.
In Appendix~\ref{3-level:app} we consider an simple example,
illustrating the difference in the descriptions already in the level of 
the ARF. Now we shall consider how the standard model for the lanthanides, 
widely used in {\it ab initio} methods,  arises in our approach. 

\section{Standard Model for the Lanthanides and its Extension}
\label{standard:sect}
It is instructive to consider the limiting case corresponding to the 
so-called \emph{standard model} for the rare earths first, 
where the {\it f}-electrons are assumed to be completely localized and 
atomic like.
We will show that 
this model 
sometimes needs to be modified, in order to explain all experimental data 
of the lanthanides.
Let us first have a look at the structure of the zero Fermion GF in the atomic
limit, which is given by  
\begin{eqnarray}
F_{\nu }^{(at)}(i\omega ) &\equiv &\langle {\cal T}f_{\nu }f_{\nu }^{\dagger
}\rangle_{i\omega }=\sum_{a}\frac{|f_{\nu }^{a}|^{2}P^{a}}{i\omega -\Delta
_{\bar{a}}}  \nonumber \\
&&\hspace{-1cm}\equiv 
\sum_{\{\Gamma _{n}\}}\frac{|\langle \Gamma _{n}|f_{\nu }|\Gamma
_{n+1}\rangle |^{2}(N_{\Gamma _{n}}+N_{\Gamma _{n+1}})}{i\omega -(E_{\Gamma
_{n+1}}-E_{\Gamma _{n}})}.
\end{eqnarray}
The differences of the energies of the ion states $E_{\Gamma _{n+1}}-E_{\Gamma
_{n}}\sim E_{\Gamma _{n}}-E_{\Gamma _{n-1}}\sim U$, i.e.\ of the order of
value of Hubbard $U$; here $n$ is the number of localized electrons of the
element in question in a given compound. Due to the large value of $U$ the
transition $E_{\Gamma _{n}}-E_{\Gamma _{n-1}}\equiv \Delta _{\Gamma
_{n-1}\Gamma _{n}}$ in many cases (particularly, in rare earths) is much
below the bottom of the valence bands, while the transition 
$E_{\Gamma _{n+1}}-E_{\Gamma
_{n}}\equiv \Delta _{\Gamma _{n+1}\Gamma _{n}}$ is above the Fermi level (as
seen from spectroscopical data, in many rare earths it is not very high
above $\varepsilon _{F}$). The strength of the hopping and mixing depend on the
energy of the states involved since the width of the inter-atomic barrier for
electron tunneling is increasing with a decrease of the energy and, also,
the intra-atomic states of an isolated ion belonging to different energies are
orthogonal to each other; for the deep states of a non-isolated atoms this
picture is changed only by a small amount\footnote{%
It should be noted that this statement is representation dependent
and relates to the methods where the intra-atomic wave functions in the basis
set are taken as atomic-like.}.  
On one hand, as discussed earlier (see Fig.5,6 in Ref.~\cite{olof}), 
the mixing interaction and hopping gradually, but quite fast, are
switching off when, say, the $f$- or $d$-level moves deeper to the energy 
region below the bottom of a band. 
In our language this corresponds to the lower
transitions, $\Delta _{\Gamma _{n-1}\Gamma _{n}}$. Therefore, only the upper
transitions, which are in the proximity of the Fermi level, can participate in
the formation of bands. However, if the upper transition is much higher 
than $\varepsilon_{F}$, the bands formed are not occupied and do not
contribute to the cohesive energy. From the physical point of view this
picture is fully analogous to the standard model: in both cases the
localized, occupied, $f$-bands contribute to the total energy as core states
while the contribution to the cohesive energy in both cases is absent. 

The arguments given in the simple example in 
Appendix~\ref{3-level:app} can be easily 
extended to the general case of $n$ orbitals.
Let us show how this separation (or, in another language,
''orbital polarization'') arises. The immediate objection often arising is
that from the point of view of simple models of strongly correlated
electrons, say, Hubbard-like models, the switching off of mixing and hopping 
seems strange. Indeed, the transformation of the Hamiltonian from the
Fermion representation to the representation of Hubbard operators is
exact, 
\begin{eqnarray}
&&\hspace*{-7mm}
{\cal H}_{Hubb} =\sum_{\langle ij\rangle \sigma }(\varepsilon _{\lambda
}^{0}\delta _{ij}\delta ^{\lambda \nu }+t_{ij}^{\lambda \nu })f_{i\lambda
}^{\dagger }f_{j\nu }+\sum_{i\sigma }U\hat{n}_{i\lambda }\hat{n}_{i\nu } \\
&=&\sum_{i\sigma }\varepsilon _{i\lambda }^{0}X_{i}^{[\lambda ,\lambda
]}+\sum_{i,\lambda \neq \nu }(\varepsilon _{\lambda }^{0}+\varepsilon _{\nu
}^{0}+U)X_{i}^{[\lambda \nu ,\lambda \nu ]}+...  \nonumber \\
&&+\sum_{\langle ij\rangle \lambda \mu ...}t_{ij}^{\lambda \nu }[\langle
\lambda _{1}|f_{i\lambda }^{\dagger }|0\rangle X_{i}^{[\lambda
_{1},0]} \nonumber \\
&+&\langle \lambda _{1}\mu |f_{i\lambda }^{\dagger }|\mu \rangle
X_{i}^{[\lambda _{1}\mu ,\mu ]}+...] \nonumber \\
&&\times[(\langle 0|f_{i\nu }|\lambda _{2}\rangle X_{j}^{0\lambda _{2}}+\langle
\mu |f_{i\nu }|\lambda _{2}\mu \rangle X_{i}^{[\mu ,\lambda _{2}\mu ]}+...],
\end{eqnarray}
therefore, \emph{if} one writes \emph{the same} magnitude of hopping for all
orbitals, $t_{ij}^{\lambda \nu }=t_{ij},$ its magnitude for the lower
transitions coincides with the one for the upper transitions.  Here we have
written a truncated single-site Coulomb interaction in order to simplify the
discussion; the Greek indices denote the orbitals: 
$\lambda =(l,m_{l},\sigma )$,
the vacuum state $|0\rangle $ includes all filled shells of ion, $|\lambda
\rangle =f_{i\lambda }^{\dagger }|0\rangle $, $|\lambda \mu \rangle =f_{i\mu
}^{\dagger }f_{i\lambda }^{\dagger }|0\rangle $, \emph{etc}., and the
orbital $\lambda $ belongs to an open shell. However, this is an 
oversimplification which makes the physics different from what happens in
reality  the hopping is always orbital dependent. Let us denote the occupied
orbitals of the shell by $\nu _{f}$, and unoccupied ones by $\nu _{e}$. In
lowest order one expects that the Hubbard sub bands will be generated by
the poles of the Fermionic GF. Indeed, the ''band structure'' in the Hubbard-I
approximation for the model $t_{ij}^{\lambda \nu }=\delta ^{\lambda \nu
}t_{ij}^{\lambda }$ is given by the equation: 
\begin{equation}
F_{\nu }^{-1}(\omega )=\left[ \sum_{a}\frac{|f_{\nu }^{a}|^{2}P^{a}}{\omega
-[\Delta _{\bar{a}}+i\Gamma _{\bar{a}}]}\right] ^{-1}-t_{k}^{\nu }=0.
\end{equation}
However, as discussed above, if the orbital $\nu _{f}$ is deep in the
potential well, the hopping $t_{k}^{\nu _{f}}$ becomes so small
(exponentially) that the bandwidth becomes of the order of the width of the
transition, $\Gamma _{\bar{a}}$ (the width is due to interaction with
phonons, electro-magnetic fluctuations etc.). In this region of
parameters the coherence is destroyed and bands are not formed. Besides, the
bare hopping entering this equation, actually, does not determine the real
band structure: the equation for it contains the self-energy which 
also should be 
found self-consistently (in both, WCPT and SCPT); the hopping itself in the
Hubbard representation is treated as an interaction and is dressed either
with the vertex corrections within field-theory methods 
or, via
changes of the wave functions within \emph{ab initio} band structure methods.
Second, as follows from the observation that any element shows a valence which
is either integer, or, for the compounds with intermediate valence, between
to nearest integers in a given surrounding and pressure, only three groups
of the transitions  
\begin{eqnarray}
&&\Delta_{\bar{a}_{1}}=\Delta_{\bar{a}}^{(n,n-1)}=
       E_{\Gamma_{n}}-E_{\Gamma _{n-1}} \nonumber \\
&&\Delta_{\bar{a}_{2}}=\Delta_{\bar{a}}^{(n+1,n)}=
       E_{\Gamma _{n+1}}-E_{\Gamma _{n}} \nonumber \\
&&\Delta_{\bar{a}_{3}}=\Delta_{\bar{a}}^{(n+1,n)}=
       E_{\Gamma_{n+2}}-E_{\Gamma _{n+1}}, 
\end{eqnarray}
can contribute to the Fermionic GF. Here $n$ is
the number of localized electrons in the ion. In order to simplify the
discussion we take $|\Gamma _{n}\rangle $ in the simple form: $|\Gamma
_{n}\rangle =f_{1}^{\dagger }f_{2}^{\dagger }...f_{n}^{\dagger }|0\rangle ,$
put $\Delta _{\bar{a}_{i}}=\Delta _{i}$, $i=1,2,3$ and omit the widths of
the levels $\Gamma _{\bar{a}_{i}}^{i}$. Then the matrix elements $|f_{\nu
}^{a_{l}}|^{2}=|\langle 0|f_{n}f_{n-1}...f_{\nu -1}f_{\nu
+1}...f_{n-1}|f_{\nu }|f_{1}^{\dagger }f_{2}^{\dagger }...f_{\nu -1}f_{\nu
}f_{\nu +1}...f_{n}^{\dagger }|0\rangle |^{2}=1$ and are equal to zero if
the orbital $\nu $ is not present in the state $|\Gamma _{n}\rangle $ and
the equations for the spectrum become 
\begin{eqnarray}
&&\hspace*{-5mm}
F_{\nu _{f}}^{-1}=[F_{\nu _{f}}^{0}]^{-1}-t_{k}^{\nu _{f}}  \nonumber \\
&&=[\frac{P_{\nu
_{0}}^{a_{1}}}{\omega -\Delta _{1}}+\frac{P_{\nu _{0}}^{a_{2}}}{\omega
-\Delta _{2}}+\frac{P_{\nu _{0}}^{a_{3}}}{\omega -\Delta _{3}}%
]^{-1}-t_{k}^{\nu _{0}} =0; \\
&&\hspace*{-5mm}
F_{\nu _{e}}^{-1}=[F_{\nu _{e}}^{0}]^{-1}-t_{k}^{\nu _{e}} \nonumber \\
&&= [\frac{P_{\nu
_{e}}^{a_{1}}}{\omega -\Delta _{1}}+\frac{P_{\nu _{e}}^{a_{2}}}{\omega
-\Delta _{2}}+\frac{P_{\nu _{e}}^{a_{3}}}{\omega -\Delta _{3}}%
]^{-1}-t_{k}^{\nu _{e}} =0,
\end{eqnarray}
where 
\begin{equation}
\sum_{a_{i}}|f_{\nu _{0}}^{a_{l}}|^{2}P^{a_{i}}=P_{\nu
_{0}}^{a_{i}},\;\sum_{a_{i}}|f_{\nu _{e}}^{a_{l}}|^{2}P^{a_{i}}=P_{\nu
_{e}}^{a_{i}}.
\end{equation}
Let us consider the combinations of the population numbers $%
P^{a}=P^{a(\Gamma ,\Gamma ^{\prime })}$, which are the spectral weights for 
the 
non-perturbed Fermionic GFs. For simplicity we also take
the sum rule for the population numbers in the form $\sum_{\Gamma }N_{\Gamma
}=1$ (recall that this sum rule is representation dependent; it follows from
the commutation relation $\{f_{j\nu },f_{j\nu }^{\dagger }\}=1,$ but this is
valid only for the orthogonal basis set of wave functions; in the case of
non-orthogonal set $\{f_{j\nu },f_{j\lambda }^{\dagger }\}=O_{\nu \lambda
}^{-1}$ and the sum rule is also different). According to our assumption the
ion has $n$ electrons localized, \emph{i.e.}, the states $\sum C_{\nu
_{1}\nu _{2}...\nu _{n}}^{(i)}f_{\nu _{1}}^{\dagger }f_{\nu _{2}}^{\dagger
}...f_{\nu _{n}}^{\dagger }|0\rangle =|\Gamma _{n}^{(i)}\rangle $ are filled
in the non-interacting ion. If the interactions which remove the degeneracy of
these states are absent (not taken into account, or the temperature is higher
than the corresponding splitting), then, there is no physical reason for the
population numbers of these states to be different. Let us calculate the
population numbers of the states involved for one chosen orbital and then
use the symmetry between the orbitals. Let us choose the orbital $\nu =1$.
Then any of the transitions $f_{1}|1,\nu _{1},\nu _{2}...\nu
_{n-1}\rangle \rightarrow |\nu _{1},\nu _{2}...\nu _{n-1}\rangle $ for $\nu
_{i}$ from the set $\nu =2,3,...,N$ gives a non-zero matrix element $|f_{\nu
_{0}}^{a_{1}}|^{2}$; therefore    
\begin{equation}
N_{\Gamma _{n}^{(i)}}^{0}=\left( 
\begin{array}{c}
N \\ 
n
\end{array}
\right)^{-1} \equiv C^n_N, 
\end{equation}
and all other population numbers are zero. Then 
\begin{equation}
P_{\nu _{0}}^{a_{1}}=\sum_{a_{1}}|f_{\nu _{0}}^{a_{1}}|^{2}\sum_{\Gamma
}\varepsilon _{\lbrack \Gamma ,\Gamma ]}^{a_{1},\bar{a}_{1}}N_{\Gamma
}^{0}=N_{\Gamma _{n-1}}^{0}+N_{\Gamma _{n}}^{0}=N_{\Gamma _{n}}
\end{equation}
and 
\begin{equation}
P_{\nu _{e}}^{a_{1}}=\sum_{a_{1}}|f_{\nu _{e}}^{a_{1}}|^{2}\sum_{\Gamma
}\varepsilon _{\lbrack \Gamma ,\Gamma ]}^{a_{1},\bar{a}_{1}}N_{\Gamma
}=N_{\Gamma _{n}^{\prime }}+N_{\Gamma _{n+1}}=N_{\Gamma _{n}^{\prime }}.
\end{equation}
Therefore, the zero Fermion GF for the particular orbital $\nu _{0}^{1}$ is 
\begin{eqnarray}
F_{\nu _{0}^{1}\nu _{0}^{1}} &&=
\frac{C_{N-1}^{n-1} N_{\Gamma _{n}}}{\omega -\Delta _{1}}+
\frac{C_{N-1}^{n} N_{\Gamma _{n}^{\prime }}}{\omega -\Delta _{2}}
=\frac{C_{N-1}^{n-1} C_{N}^{n} 
}{\omega -\Delta _{1}}+\frac{C_{N-1}^{n}
C_{N}^{n}
}{\omega -\Delta _{2}} \nonumber \\
&&=\frac{n/N}{\omega -\Delta _{1}}+\frac{(N-n)/N}{\omega -\Delta _{2}}.
\label{upp_low:eq}
\end{eqnarray}
The total spectral weight is equal to one as it should be. 
However, at $\omega=\Delta_1$ 
we have the weight $n/N<1$, while the
weight of the low-energy transition from all $N$ orbitals is $n$, as it is
expected within a single-electron picture. 
Thus, within the non-polarized picture one can expect from 
Eq.(\ref{upp_low:eq}) that the 
bandwidth 
of the lower Hubbard sub bands will be proportional to $n/N$ and the upper 
one to $(N-n)/N$.
\footnote{Within the diagrammatic field-theory methods this problem 
in the case of degenerate states is not solved 
yet and we postpone this for a future work.}  
This paramagnetic picture is, however, too rough. First, this 
speculation does not reflect 
the fact, established in the {\em ab initio} calculations, that
if the energy $\Delta _{a_1}$ is
below the bottom of the lowest conduction band, any matrix element of 
hopping or
mixing, calculated within self-consistent scheme becomes 
negligible~\cite{olof}. 
Second, the degeneracy is certainly overestimated: the first
Hund rule is caused by the intra-atomic exchange integrals, but the splitting 
between different orbitals may be a few such integrals (say, the transitions 
between 
three-electron states and two electron ones cover the range from $2U-2J$ to 
$2U-9J$, 
where $J$ is exchange integral); the next-order Coulomb interactions, which 
are responsible for the second  Hund rule, are also much greater than room 
temperature;
at last, the spin-orbit coupling, for example, in rare earths provides a 
splitting of eV-order, between the multiplets. 
Thus, if the atomic-like Hund rules are not
fully destroyed we find in a zero approximation that only one of the 
population numbers 
$N_{\Gamma _{2}}^{0}\simeq 1,\;$and, therefore, due to the 
sum rule, $\sum_{\Gamma
}N_{\Gamma }^{0}=1$, all remaining $N_{\Gamma _{1}}^{0}=0$. The GFs 
\begin{eqnarray}
&&G_{a_{1}}^{(0)}(\omega)=\frac{P_{\nu_{0}}^{a_{1}}}{\omega-\Delta_{\bar{a}%
_{1}}}\approx \frac{1}{\omega -\Delta_{\bar{a}_{1}}},  \nonumber \\
&&G_{a_{2}}^{(0)}(%
\omega)=\frac{P_{2}^{a_{2}}}{\omega-\Delta_{\bar{a}_{2}}}\approx \frac{1}{%
\omega-\Delta_{\bar{a}_{2}}},
\label{stand1:eq}
\end{eqnarray}
and the Fermion GF for the filled orbitals is 
\begin{equation}
F_{\nu _{0}}\simeq \frac{1}{\omega -\Delta _{\bar{a}_{1}}},
\label{stand2:eq}
\end{equation}
i.e.\ we are dealing with the situation, considered in the example in 
Appendix~\ref{3-level:app}. 
Thus, let us switch on the hopping and mixing. We have to distinguish the 
Hubbard 
sub bands originating from empty and filled orbitals since hopping for them 
are different. 

In the case of empty orbitals the hopping $t_{\nu _{e}}\neq 0$ since they
have energies above Fermi energy, where the potential barrier between atoms
either is absent or is small. However, as we have seen in the previous section,
the dispersion of the lower Hubbard sub band is negligible due to the small
weight of these orbitals in this region of energies. In the case of filled
orbitals the dispersion in the lower Hubbard bands is also negligible since $%
t_{\nu _{f}}$ is small due to the large width and height of the energetic
barriers between atoms; near the upper transition their spectral weight is
small. Thus, the only dispersion which is physically important is the 
dispersion
of the empty orbitals near the upper transitions. Above we have shown that
the switching of the mixing and hopping leads to the appearance of non-zero
spectral weights, above the Fermi level for the filled orbitals and for the
empty orbitals in the region of lower transitions. But the dispersion in the
latter region is negligible for both type of orbitals. Therefore, the role
of delocalized electrons is played by delocalized upper transitions of the 
empty
orbitals and we can use this fact in order to derive some recipe for 
how to 
correct the exchange-correlation potential. On the other hand, we can
in the general formulas use the approximation $t^{\nu _{0}}\simeq 0$ as well as
the overlap matrix with neighbors should be zero. This leads to a
disperse-less lower Hubbard band. As we see, 
Eq.(\ref{stand1:eq}) and (\ref{stand2:eq}) correspond to
the standard model for the lanthanides if the upper transitions will not
contribute to the cohesive energy and this situation arises when the 
energy of the upper transition $\Delta _{a_{2}}\gg\varepsilon_{F}$. As known
from electron spectroscopy, this is not the case even for rare earths~
\cite{Gschneider_book}. However, as we have seen above, if the mixing
interaction of the conduction bands with the upper transition is not equal
to zero, the spectral weight at both, upper and lower energies, deviates
from unity and, therefore, it is desirable to correct
the localized model at least in this place. Let us introduce this correction.

The exchange-correlation potential is a function of total charge density.
Within the real-space representation of atomic-like wave functions 
this charge density is given by Eq.(\ref{charge_dens100:eq}), where the
Fermion GF $\langle Ta_{1}a_{2}^{\dagger }\rangle _{\omega }$ should be
represented via the many-electron GFs, $G$. Within the approximation of
dispersion-less lower transitions $\Delta _{a_{1}}$ we can write the structure
of the $XX$-GFs as follows: 
\begin{equation}
G_{ja,j^{\prime }\bar{a}^{\prime }}^{(XX)}(\omega )\simeq \delta
_{jj^{\prime }}\delta ^{aa^{\prime }}\delta ^{aa_{1}}\frac{P^{a_{1}}}{\omega
-\Delta _{\bar{a}_{1}}}+\delta ^{aa_{2}}\delta ^{a^{\prime
}a_{2}}g_{ja_{2},j^{\prime }\bar{a}_{2}^{\prime }}^{(XX)}(\omega ),
\end{equation}
while all other GFs, $G^{(Xc)},G^{(cc)}$ and $G^{(cX)}$, do not have the lower
pole due to the absence of hopping and mixing in this region of energy. Here $%
a_{1}$ denote the lower transitions $\ $and $a_{2}$ the upper ones.
Inserting this formula into the expression for the charge density,
Eq.(\ref{charge_dens100:eq}), we find in terms of retarded GFs
\begin{eqnarray}
&&\hspace*{-4mm}\rho (x) =-\frac{1}{2\pi }\int \dd\omega 
\frac{1}{e^{\beta (\omega -\mu )}+1}%
\nonumber \\
&&\hspace*{-3mm}\times 
\left \{\lim_{x^{\prime }\rightarrow x}\left[ \sum_{jL,j^{\prime }L^{\prime
}}\phi _{jL}(x)\phi _{j^{\prime }L^{\prime }}^{\ast }(x^{\prime })
Im F_{jL,j^{\prime }L^{\prime }}(\omega +i\delta )\right. \right.\nonumber \\ 
&&\hspace*{-3mm}+\left.\sum_{j\nu a,j^{\prime }L^{\prime }}
\phi _{j\nu }(x)\phi _{j^{\prime }L^{\prime }}^{\ast }
(x^{\prime })(f_{\nu })^{a}
Im G_{ja,j^{\prime }L^{\prime }}^{(Xc)}(\omega +i\delta )\right]  \nonumber \\
&&\hspace*{-3mm}+\lim_{x^{\prime }\rightarrow x}\left[ 
\sum_{jL,j^{\prime }\nu a}\phi
_{jL}(x)\phi _{j^{\prime }\nu \bar{a}^{\prime }}^{\ast }(x^{\prime })(f_{\nu
}^{\dagger })^{\bar{a}}Im G_{jL,j^{\prime }\bar{a}^{\prime
}}^{(cX)}(\omega +i\delta ) \right. \nonumber \\ 
&&\hspace*{-3mm}+\left.\left.\sum_{j\nu _{e}a,j^{\prime }\nu _{e}^{\prime
}a^{\prime }}\phi _{j\nu _{e}}(x)\phi _{j^{\prime }\nu _{e}^{\prime }}^{\ast
}(x^{\prime })(f_{\nu _{e}})^{a}(f_{\nu _{e}^{\prime }}^{\dagger })^{\bar{a}%
^{\prime }}Im g_{ja,j^{\prime }\bar{a}^{\prime }}^{(XX)}(\omega
+i\delta )\right] \right\}  \nonumber \\
&&\hspace*{-3mm}+\lim_{x^{\prime }\rightarrow x}\sum_{j\nu _{e}a,j^{\prime }\nu
_{e}^{\prime }a^{\prime }}\phi _{j\nu _{0}}(x)\phi _{j^{\prime }\nu
_{0}^{\prime }}^{\ast }(x^{\prime })(f_{\nu _{0}})^{a}P^{a} \nonumber \\
&&\hspace*{-3mm}\times \frac{1}{e^{\beta
(\Delta _{\bar{a}}-\mu )}+1}(f_{\nu _{0}^{\prime }}^{\dagger })^{\bar{a}}.
\label{charge_dens100:eq}
\end{eqnarray}
The last term represents the contribution of the core states into the spectrum
of the Fermion-like excitations. This expression differs from the standard one
used in {\it ab initio} calculations only by the numbers $P^{a}$, which are 
present in
the last term and all GFs. At zero temperature $P^{a}$ are determined mainly
by the strength of the $c$-$f$-mixing, $f$-$f$-hopping and the proximity of $%
\Delta _{2}$ to the Fermi energy. In the last term the factor $P^{a}/[e^{\beta
(\Delta _{\bar{a}}-\mu )}+1]$ for the transition $a=[\gamma ,\Gamma ]$ gives
the statistical weight with which the orbitals $\nu _{0},\nu _{0}^{\prime }$
contribute to the states $|\gamma \rangle ,$ $|\Gamma \rangle $. 

The expression for the total energy  has the same ingredients, but 
the additional integral over coordinates gives the overlap matrixes and 
and matrix elements of the kinetic energy. Therefore, we find 
for the total energy
\begin{eqnarray}
\label{tot_appendix:eq}
&&E_{tot}=-\frac{1}{2\pi }\int \dd\omega \frac{1}{e^{\beta (\omega -\mu )}+1}
\nonumber \\
&&\times\{[O_{j_{1}L_{1},j_{2}L_{2}}\omega +t_{j_{1}L_{1},j_{2}L_{2}}]Im%
F_{j_{2}L_{2},j_{1}L_{1}}(\omega +i\delta ) \nonumber \\
&&+[O_{j_{1}L_{1},j_{2}\mu }\omega +t_{j_{1}L_{1},j_{2}\mu }]f_{\mu }^{b}%
Im G_{j_{2}b,j_{1}L_{1}}^{Xc^{\dagger }}(\omega +i\delta ) \nonumber
\\
&&+[O_{j_{1}L_{1},j_{2}L_{2}}\omega +t_{j_{1}\mu ,j_{2}L_{2}}](f_{\mu
}^{\dagger })^{\bar{b}}Im G_{j_{2}L_{2},j_{1}\bar{b}}^{cX^{\dagger
}}(\omega +i\delta ) \nonumber \\
&&+[O_{j_{1}\mu _{1},j_{2}\mu _{2}}\omega +t_{j_{1}\mu _{1},j_{2}\mu
_{2}}]f_{\mu _{2}}^{b_{2}}(f_{\mu _{1}}^{\dagger })^{\bar{b}_{1}} 
\nonumber \\
&&\times
Im g_{j_{2}b_{2},j_{1}\bar{b}_{1}}^{XX^{\dagger }}(\omega +i\delta )\}
\nonumber \\
&&+\sum_j 
[ \delta_{\mu _1,\mu _{2}}\Delta_{\bar{a}} +t_{j,\mu _{1};j,\mu _{2}}]
f_{\mu _{2}}^{a}(f_{\mu _{1}}^{\dagger })^{\bar{a}}
\frac{1}{e^{\beta
(\Delta _{\bar{a}}-\mu )}+1}.
\label{etotSEC:eq}
\end{eqnarray}
Here $t$ is kinetic energy.

\section{Discussion and Conclusions}
\label{disc:sect}
The {\it ab initio} calculations based on the LDA in DFT has become a quite 
reliable method for investigating the solid state. 
The materials with strongly correlated electrons,
which forms a quite wide class (high-$T_{c}$ superconductors, doped Mott
insulators, materials with colossal magneto resistance, heavy-Fermion systems
and Kondo lattices and, different artificial structures based on
small quantum dots) and are used in more and more applications, are
traditionally difficult for \emph{ab initio} calculations. It is
well-known that these difficulties often are caused by the
localized or quasi-localized electrons originating from open electronic
shells of transition elements, lanthanides or actinides. Three methods found
for overcoming these difficulties, LDA+U, self-interaction correction (SIC),
and the complete localization model for lanthanides (where all 
$f$-electrons are
forced to be core states) provide the same physics: they localize some of
the electrons, making the potentials for conduction electrons and for localized
ones, different. On one hand, since at least LDA+U and SIC use more than one
potential, they go beyond the DFT scheme, where only one potential
corresponds to a charge density in ground state. On the other hand, there
always questions, How to avoid
double-counting? What is ''Hubbard $U$''? Why is SIC applied to certain
orbitals and not to other ones? etc., for which it is
difficult to give an answer without considering the full problem (say, in
the language of some of perturbation theories) and understanding what is
included into LDA. An answer to first question has partly been found by Kotani 
~\cite{kotani}, who has shown that 
the LDA potential corresponds to a static random phase
approximation. This simultaneously gives an answer to the question why the 
Hartree-Fock calculations give much less accurate results than the ones 
in LDA: LDA
takes into account screening, while the Hartree-Fock method does not. 
However, this is
only a part of the answer since, on the one hand, in modern calculations the
exchange-correlation potential is taken from the (parameterized) results of
Monte-Carlo simulations for homogeneous electron gas, which certainly, go
beyond RPA. On the other hand,  
these statements and investigations are based
on an analysis of either the {\em homogeneous} or 
weakly inhomogeneous electron gas.
What can be done in the case of a strongly inhomogeneous electron liquid? The
LDA+U and SIC methods are intended 
for this case and, as was mentioned above,  they go beyond the DFT ideology 
and  contains some open questions. 
As we see, the problem of understanding, in
a formal language, what is included into LDA and what should be taken into
account additionally in it to describe correlated systems are necessary
steps in the development of the theory of solids. In the 
present work we made an 
attempt to obtain the physical picture analogous to the one generated by
these three methods, but to perform such an analysis on the basis of solid
grounds. Here we have used the perturbation theory from the atomic
limit developed in a previous paper (ref.~\cite{DT}) for the case of the 
generalized multi-orbital
Hubbard-Anderson models and extended it for the case of inter-site Coulomb
interaction. We have considered the easiest case, when a part of the electrons
of the $f$-shell is well localized. The picture, generated by the perturbation
theory is the following.

\emph{First}, under the word ''correlations'' we mean that the
intra-atomic interactions are strong enough to make it necessary to calculate 
$\langle \hat{n}\hat{n}\rangle $ beyond the decoupled form 
$\langle \hat{n} \rangle \,\langle \hat{n}\rangle $. 
Here, we use the exact relation $\langle 
\hat{n}\hat{n}\rangle =\langle \hat{n}\rangle $. This immediately leads to the 
formation of a energy gap between the occupied and empty orbitals.
Therefore, taking into account local correlations automatically leads to
orbital polarization, the picture which is obtained in the  methods LDA+U and
SIC. The potentials for ''upper'' and ''lower''
orbitals differ from each other, roughly speaking,  by the Hubbard $U$.

\emph{Second}, the language of the localized and delocalized \emph{orbitals}
is not completely correct: since the orbitals are coupled by
interactions to complexes, described by many-electron wave functions, the
role of the Fermion excitations is played by the single-electron 
\emph{transitions between many-electron states}. This is seen from the fact that
switching on/off the mixing, (and/or hopping) between the upper transitions $%
|\Gamma _{n}\rangle \rightarrow |\Gamma _{n+1}\rangle $  and the delocalized
states, leads to a deviation of the upper and lower spectral weights from zero
or unity and make these weights non-integer numbers. Then, the Fermion GFs
acquire a many-pole structure, where each pole contributes with its own 
weight. 
In turn, these weights are determined by the
population numbers (of the type $N_{\Gamma _{n}}\sim \langle \hat{n}_{1}\hat{%
n}_{2...}\hat{n}_{n}(1-\hat{n}_{n+1})(1-\hat{n}_{n+2})...(1-\hat{n}%
_{N})\rangle ,$ where N is the number of orbitals in the shell) of the
many-electron states involved. 

\emph{Third}, the number of delocalized, correlated, orbitals (delocalized
renormalized Fermions) does not obligatory coincide with the number of the
single-electron transitions.  

\emph{Fourth}, a part of the many-electron states which are composed mainly of 
the
localized orbitals, often form a (quasi)degenerate subsystem. Redistribution
of spectral weights between the transitions involving these states may be
caused by small energies. 

\emph{Fifth}, the spectral weights enter as factors to mixing, hopping, 
and Coulomb interaction,
decreasing their strength and providing the correlation caused mechanism of 
band narrowing. 
Since these weights are determined by the strength of the 
interactions, the problem requires self-consistent solution which  
may reflect the competition between the Hund-rule 
interactions and the cohesive energy. 

\emph{Sixth}, the closed equations for the GFs of the Hubbard operators in
terms of functional derivatives contain two types of contributions. One type
is responsible for the description of effects 
from kinematic interactions
originating from mixing and hopping 
( an example of the physics
involved, the correlation  mechanism of delocalization, see in Ref.~\cite
{urban_model}), another type generates an analogue of the WCPT for the 
Fermion GF.
The series for the many-electron GFs obtained by neglecting the contributions
from the kinematic interactions is isomorphic to the series for the
single-electron Fermion GF in WCPT. Using this isomorphism, and comparing the
Migdal-Galitskii formula for total energy of the non-correlated and
correlated systems as well as the Sham equation for the 
exchange-correlation potential, we came to the conclusion that the form for
the exchange-correlation potential used within weak-coupling theories can be
used in the theory of the correlated systems also, if one will use the 
mixing, hopping, the inverse
of the overlap matrix and the expression for charge density renormalized 
by many-electron spectral weights. The recipe for
concrete calculations and example of such calculation is given in future works.

\emph{Seventh}. The model Hubbard U can be expressed in terms of energies
of transitions between many-electron states. Consideration of screening of
these transitions by Coulomb interactions shows that the lower transitions
are renormalized much weaker  than the upper ones. This makes calculations
within one parameter, U$^{\rm{mod}},$ somewhat confusing, since they do
not reflect the true picture. 

Let us discuss this question in a more details, and make an
attempt to elucidate what actually should be understood under Hubbard $U$
from the point of view of this many-electron theory.
A commonly used definition of the Hubbard U has risen from the following
speculation. The energy of the n-orbital atom modeled with the Hubbard
model is
\begin{equation}
E_{n}=n\varepsilon _{0}^{d}+\frac{1}{2}n(n-1)U^{mod}.
\end{equation}
From this one finds that the energies of single-electron excitations are
\begin{eqnarray}
\Delta _{1} &=&E_{n}-E_{n-1}=\varepsilon _{0}^{d}+U^{mod}(n-1), \\
\Delta _{2} &=&E_{n+1}-E_{n}=\varepsilon _{0}^{d}+U^{mod}n
\end{eqnarray}
and, therefore,
\begin{equation}
U^{mod}=\Delta _{2}-\Delta _{1}.
\end{equation}
The parameter $\varepsilon _{0}$ can be taken from the requirement that
the effective attractive potential is able to hold $n$ electrons (this is
determined by the valence of the atom) and, therefore, $\varepsilon
_{0}^{d}=-(n-1/2)U^{mod}$ and, in turn,  $\Delta _{1}=-U^{mod%
}/2$ and $\Delta _{2}=U/2$. Thus, $\Delta _{1}$ and $\Delta _{2}$ play the
role of the centers of the Hubbard sub bands. Hubbard
used the Coulomb matrix element of
density-density type for $U^{mod}$. According to the common opinion,
it has unreasonably
large values, between 20 and 30 eV for a $d$-shell,
and, therefore, is considered irrelevant.
In LDA+U calculations ''reasonable'' values of $U^{mod}$ for
transition metals are near 5-8 eV~\cite{olof}, 
while in recent calculations, based on
dynamical mean field theory, in order to obtain a reasonable 
agreement with the
spectroscopical data for iron the authors had to take a value of about 3
eV~\cite{katsnelson99}.
There are, however, no suggestions what mechanism that is responsible for
such a drastical decrease of $U$ from 20 to 6 (or even to 3)eV.  The common
believe is that  delocalization plus screening, particularly, due to
interaction with some other bands, may provide such a decrease.  It is easy
to see, that at least, the density-density interaction with other bands does
not change the picture. Indeed, let us imagine, that besides, say, a $d$-shell
the atom also contains another shell with $m$ atoms. If we add 
the term $U_{cd}mn$ to the energy, it changes $\Delta_1$ and $\Delta_2$
\begin{equation}
\Delta _{1}\Rightarrow \Delta _{1}+U_{cd}m,\;\Delta _{2}\Rightarrow \Delta
_{2}+U_{cd}m,
\end{equation}
but does not change $U^{mod}$: $\Delta _{2}-\Delta _{1}=U^{mod%
}$. Earlier~\cite{urban_model} we also investigated the influence of the
delocalization on $U^{mod}$ via renormalization by (
kinematic) mixing interaction, it gives a decrease of at most
10\% under favorable
conditions. Therefore, either other mechanisms should be involved or the
description in terms of one parameter $U^{mod}$ is
oversimplified. From our consideration given above we can conclude that both
assumptions are relevant.

First, it is important, which experiment we want to compare with. It is
well-known that for \emph{ab initio} DFT LDA calculations the excited states
with large energy of excitations  are difficult  since they usually involve
non-single-electron excitations and the latter are not included into
the effective single-electron Kohn-Sham scheme (even with different
corrections).
However, in low-energy experiments the lower Hubbard
band is not involved. Therefore, for the description of such experiments one
is interested only in the properties of the upper Hubbard band. The model
calculation, which are trying to work with energies separated by a large
scale, usually
does not take into account self-consistent changes of the 
charge density. This involves the strongest interactions existing
in the system. Therefore, it looks like only methods which combine the ab
initio methods with the many-body theory might be able to treat such
problems. Within our many-electron approach the important parameters of the
system are $\Delta _{1}$ and $\Delta _{2}$. They are connected with the
model parameter Hubbard $U$, however, they have, obviously, many-electron
nature.  

Second, there are two reasons why it is very important,
where the levels $\Delta _{1}$ and $\Delta _{2}$ are situated with respect to
all 
other bands and the chemical potential: a) this influences the
distribution of spectral weights; b) the efficiency of screening depends on
it. Actually, the lower the transition is
under the Fermi level, the less
the Coulomb interaction, via which it interacts with other electrons, is
screened
(see the speculations about the one loop contribution to the polarization
operator, Eq.(\ref{rpa_looop:eq})).

Third, the way of calculation, obviously, depends on the reference point 
which interactions are included into  $\Delta _{2}^{(0)},\Delta _{1}^{(0)}.$
If we start with the LDA picture with a certain number of localized
electrons, we
have self-consistent $f$-wave functions, from them we can
construct of the
wave functions of the many-electron states, and calculate the energies $%
E_{\Gamma _{n-1}}^{(0)},E_{\Gamma _{n}}^{(0)},E_{\Gamma _{n+1}}^{(0)}$
without any relaxation of charge. The differences of these
energies provides $\Delta _{2}^{(0)},\Delta _{1}^{(0)}$. It should be
emphasized that in this approach the LDA based calculation is used only as a
generator of a convenient minimal set of wave functions for calculating $%
E_{\Gamma _{n-1}}^{(0)},E_{\Gamma _{n}}^{(0)},E_{\Gamma _{n+1}}^{(0)}$.
The energies themselves should be calculated on the 
basis of the many-electron states, $|\Gamma\rangle$.
Then, one can switch on mixing and hopping and take into account the
screening and the renormalization of $\Delta _{2}^{(0)},\Delta _{1}^{(0)}$
via the self-energies of the many-electron GFs. Let us start with the lower
transition. If we denote the Coulomb interaction between the localized and
delocalized electrons as $U_{ld},$ delocalized-delocalized as $U_{dd},$ it
is clear that the largest contribution to self-energy for the
renormalization of $\Delta _{1}$ comes from the graphs which contain the
lowest possible number of interactions $U_{ld}$, because, as we have seen in
previous section, each loop constructed on $U_{ld}$ involves the loop with
large denominator of the order of $\Delta _{2}-\Delta _{1}=U^{mod}$.
Therefore, the largest contribution to the self-energy with density-density
matrix elements $U_{ld}$, renormalizing the lower transition, contains  $%
U_{ld}$ only twice (see the graph in Fig.~\ref{loop:fig} and
Fig.~\ref{loop2:fig}).
Strictly speaking, the
equations for the lower (and for upper as well) GFs always form a system of
equations,
\begin{eqnarray}
\label{gfupper:eq}
&&\hspace{-5mm}
G^{a_{1}\bar{a}_{1^{\prime }}}(i\omega ) =G_{0}^{a_{1}\bar{a}_{1^{\prime
}}}(i\omega )+G_{0}^{a_{1}\bar{a}_{2}}(i\omega )\{\Sigma _{\bar{a}%
_{2}a_{4}}^{H} \nonumber \\
&&+T\sum_{\omega _{1}}[\bar{v}_{\bar{a}_{2}69a_{5}}\Pi
_{97,68}(i\omega -i\omega _{1})\nonumber \\
&&\times\bar{v}_{\bar{a}_{3}78a_{4}}G^{a_{5}\bar{a}%
_{3}}(i\omega _{1})]\}G^{a_{4}\bar{a}_{1^{\prime }}}(i\omega ),
\end{eqnarray}
where $\Sigma _{\bar{a}_{2}a_{4}}^{H}$ is the Hartree contribution and $\Pi$
is the full polarization operator. However, if we neglect the differences
between the levels $\Delta _{1}$, we can write the equation for renormalized
$\Delta _{1}$:
\begin{eqnarray}
&&\hspace{-5mm}
\Delta _{1}(i\omega )=\Delta _{1}^{0}+\alpha T
\sum_{\omega _{1}}\bar{v}_{%
\bar{a}_{1}69a_{1}}\Pi _{97,68}(i\omega -i\omega _{1}) \nonumber \\
&&\times\bar{v}_{\bar{a}%
_{1}78a_{1}}G^{a_{1}\bar{a}_{1}}(i\omega _{1}),
\end{eqnarray}
where $\alpha$ takes into account the number of internal transitions,
$a_5$ in Eq.(\ref{gfupper:eq}).
The situation with the upper transition is more complex since in this region
the bands are formed. For this reason it can be defined via the static
dielectric permeability as a solution of the equation with Hartree and,
screened by static $\varepsilon ^{-1}$, exchange interaction, i.e in
the Eq.(\ref{gfupper:eq}) using instead of the second term
with $\varepsilon ^{-1}(0)$ and $F
$ and $v$ replaced by $G$ and $\bar{v}$. As we see, the self-consistent
magnitudes  $\Delta _{2},\Delta _{1}$ become frequency-dependent already in
the
first steps of their dressing by electron-hole excitations. This makes such
calculations quite cumbersome and practically difficult. The other possible
way would be to calculate, as described above, the non-self-consistent values
of  $\Delta _{2}^{(0)},\Delta _{1}^{(0)}$ and to treat them as the input
parameters for further LDA calculations, where they will be renormalized as
centers of bands usually do.

In conclusion, we have reported the following results.
The perturbation theory for correlated electron systems developed in 
previous paper here is extended to the case of Coulomb interaction.
It is shown in the level of closed functional equations for the 
many-electron Green 
functions that all the  contributions in SCPT can be separated to two classes.
One of them forms the approximation of renormalized fermions, ARF, 
does not include kinematic interactions, and is fully analogous to the 
WCPT for single-electron GFs.
Another one includes them and,
 therefore, contains specific for CES effects. The ARF is used for formulation
of simple recipe for extension of the LDA to DFT to the case of 
strongly correlated electrons. 
The application to rare earths is given in next paper.

\section*{Acknowledgment}
The support from the Natural Science Foundation (NFR and TFR), the Swedish 
Foundation for
Strategic research (SSF) and G\"oran Gustafsson foundation, is acknowledged.


\appendix
\section{The Coulomb Terms in Equations of Motion for Green Functions}
\label{coul_eq_of_mot:app}
\emph{Class 0}.
This class belongs to the ones considered above. Nevertheless, for completeness
we give the detailed formulas. The correction to the EM for the GF $\langle
Tc_{1}(t)\eta ^{\dagger }(t^{\prime })\rangle_{u}$ ($\eta$ is either a 
$c$ or a $X$ operator)
from the interaction
between the conduction electrons is  
\begin{eqnarray}
&&\hspace{-5mm}
V_{1_{c}3_{c}4_{c}5_{c}}\langle {\cal T}c_{3}^{\dagger }(t)
c^{\vphantom{\dagger}}_{4}(t)c^{\vphantom{\dagger}}_{5}(t)\eta
^{\dagger }(t^{\prime })\rangle _{u}  \nonumber \\
&&=V_{1_{c}3_{c}4_{c}5_{c}} \lim_{t_{1}\rightarrow t+0}[(\langle
{\cal T}c_{3}^{\dagger }(t_{1})c^{\vphantom{\dagger}}_{4}(t_{1})\rangle 
+i\frac{\delta }{\delta
U_{34}^{cc}(t_{1})}) \nonumber \\
&&\times\langle {\cal T}c^{\vphantom{\dagger}}_{5}(t)
\eta ^{\dagger }
(t^{\prime })\rangle
_{u}],
\end{eqnarray}
while the GFs generated by the commutator $[X,V_{0}]$ can be expressed as
follows  
\begin{eqnarray}
&&\hspace*{-5mm}\langle {\cal T}Z_{j_{1}}^{\xi _{1}}(t)c_{3}^{\dagger }
(t)c^{\vphantom{\dagger}}_{4}(t)c^{\vphantom{\dagger}}_{5}(t)\eta
^{\dagger }(t^{\prime })\rangle _{u} \nonumber \\
&&=\lim_{t_{1}\rightarrow t+0} [(\langle {\cal T}
Z_{j_{1}}^{\xi _{1}}(t_{1})\rangle
_{u}+i\frac{\delta }{\delta U_{1}^{Z}(t_{1})})  \nonumber \\
&& \times\langle {\cal T}c_{3}^{\dagger
}(t)c^{\vphantom{\dagger}}_{4}(t)c^{\vphantom{\dagger}}_{5}(t)
\eta ^{\dagger }(t^{\prime })\rangle _{u}], \\
&&=\lim_{t_{1}\rightarrow t+0} [(\langle {\cal T}
Z_{j_{1}}^{\xi _{1}}(t_{1})\rangle
_{u}+i\frac{\delta }{\delta U_{1}^{Z}(t_{1})}) \nonumber \\
&&\times\lim_{t_{2}\rightarrow t_{1}+0}(\langle {\cal T}c_{3}^{\dagger
}(t_{2})c^{\vphantom{\dagger}}_{4}(t_{2})\rangle _{u}+
i\frac{\delta }{\delta U_{34}^{cc}(t_{2})})  \nonumber \\
&&\times\langle {\cal T}c^{\vphantom{\dagger}}_{5}(t)
\eta ^{\dagger }(t^{\prime })\rangle _{u}].
\end{eqnarray}
Below, for brevity, we shall denote the time limits by upper index ''plus'': 
\begin{equation}
\lim_{t_{2}\rightarrow t_{1}+0}R(t_{2})\equiv R(t_{1}^{+}).
\end{equation}
Further, 
\begin{eqnarray}
&&\hspace*{-5mm}\langle {\cal T}
X_{j_{1}}^{\bar{d}}(t)c^{\vphantom{\dagger}}_{4}(t)
c^{\vphantom{\dagger}}_{5}(t)\eta ^{\dagger }(t^{\prime
})\rangle _{u}  \nonumber \\
&&=(\langle {\cal T}
X_{j_{1}}^{\bar{d}}(t^{+})c^{\vphantom{\dagger}}_{4}(t^{+})\rangle _{u}+
i\frac{\delta 
}{\delta U_{j_{1}\bar{d},4_{c}}^{Xc}(t^{+})}) \nonumber \\
&&\times\langle {\cal T}
c^{\vphantom{\dagger}}_{5}(t)\eta ^{\dagger
}(t^{\prime })\rangle _{u}.
\end{eqnarray}
Then in the Hartree approximation we have to neglect the contribution 
from this 
derivative. We shall use the notation $\delta ^{(0)}G_{1_{c}}^{(c\eta )}$
for the right-hand side contribution to the EM for corresponding GF, not for
the GF itself. Thus, for the GF $(1/i)\langle {\cal T} c_{1}(t)\eta ^{\dagger
}(t^{\prime })\rangle _{u}=G_{1_{c}}^{(c\eta )}$ we have the standard
expression 
\begin{equation}
\delta ^{(0)}G_{1_{c}}^{(c\eta )}=V_{1_{c}3_{c}4_{c}5_{c}}\langle
{\cal T}c_{3}^{\dagger }(t_{1})
        c^{\vphantom{\dagger}}_{4}(t_{1})\rangle \langle 
{\cal T}c^{\vphantom{\dagger}}_{5}(t)\eta ^{\dagger
}(t^{\prime })\rangle _{u},
\end{equation}
where $V_{1_{c}3_{c}4_{c}5_{c}}\equiv
O_{1_{c}2_{c}}^{-1}v_{2_{c}3_{c}4_{c}5_{c}}$. The index $(0)$ means that the
correction comes from the class 0. This is, obviously, the standard Hartree
correction. The exchange-correction comes from the derivative. Next, 
\begin{eqnarray}
&&\hspace*{-5mm}\delta ^{(0)}G_{j_{1}a_{1}}^{(X\eta )}
=v^{\vphantom{\dagger}}_{[2_{c}3_{c}]4_{c}5_{c}}
O_{j_{1}\mu ,2_{c}}^{-1}(f_{\mu }^{\dagger })^{%
\bar{a}_{2}}\varepsilon _{\xi }^{a_{1}\bar{a}_{2}}\langle {\cal T}
Z_{j_{1}}^{\xi
_{1}}(t^{+})\rangle _{u}\nonumber \\ 
&&\times\langle {\cal T}
c_{3}^{\dagger }(t^{+})c^{\vphantom{\dagger}}_{4}(t^{+})\rangle
_{u}\langle {\cal T}c^{\vphantom{\dagger}}_{5}(t)
\eta^{\dagger }(t^{\prime })\rangle _{u} \nonumber \\
&&+\frac{1}{2}v^{\vphantom{\dagger}}_{2_{c}3_{c}4_{c}5_{c}}O_{j_{1}\mu
_{2},2_{c}}^{-1}O_{j_{1}\mu _{3},3_{c}}^{-1}(f_{\mu _{2}}^{\dagger })^{\bar{a%
}_{2}}(f_{\mu _{3}}^{\dagger })^{\bar{a}_{3}}\varepsilon _{\xi }^{a_{1}\bar{a%
}_{2}}\varepsilon _{\bar{d}}^{\xi \bar{a}_{3}}  \nonumber \\
&&\times\langle {\cal T}X_{j_{1}}^{\bar{d}}(t^{+})
c^{\vphantom{\dagger}}_{4}(t^{+})\rangle _{u}\langle
{\cal T}c^{\vphantom{\dagger}}_{5}(t)\eta ^{\dagger}(t^{\prime})\rangle_{u}.
\end{eqnarray}

\emph{Class 1}. The 
first GF arising from the $c$-operator is 
$\langle {\cal T}Zc^{\dagger }cc\eta^{\dagger}\rangle$ 
which is considered above. The other functions coming
from $[c_{1},\hat{V}_{1}]$ are 
\begin{eqnarray}
&&\hspace*{-5mm}
\langle {\cal T}X_{j_{2}}^{\bar{a}_{2}}(t)c^{\vphantom{\dagger}}_{4}(t)
c^{\vphantom{\dagger}}_{5}(t)\eta ^{\dagger
}(t^{\prime })\rangle _{u}  \nonumber \\
&&=[\langle {\cal T}X_{j_{2}}^{\bar{a}_{2}}(t^{+})
c^{\vphantom{\dagger}}_{4}(t^{+})\rangle_{u}+
i\frac{%
\delta }{\delta U_{24}^{Xc}(t^{+})}] \nonumber \\
&&\times\langle {\cal T}
c^{\vphantom{\dagger}}_{5}(t)\eta ^{\dagger
}(t^{\prime })\rangle _{u},
\end{eqnarray}
and 
\begin{eqnarray}
&&\hspace*{-5mm}
\langle {\cal T}c_{3}^{\dagger }(t)X_{j_{4}}^{a_{4}}(t)
c^{\vphantom{\dagger}}_{5}(t)\eta ^{\dagger
}(t^{\prime })\rangle _{u}  \nonumber \\
&&=[\langle {\cal T}c_{3}^{\dagger }(t^{+})X_{j_{4}}^{a_{4}}(t^{+})
\rangle_{u}+i
\frac{\delta }{\delta U_{34}^{cX}(t^{+})}] \nonumber \\
&&\times\langle {\cal T}
c^{\vphantom{\dagger}}_{5}(t)\eta ^{\dagger
}(t^{\prime })\rangle _{u}.
\end{eqnarray}
The functions which come from $[c_{1},\hat{V}_{1}]$ we rewrite as 
\begin{eqnarray}
&&\hspace*{-5mm}\delta ^{(1)}G_{1_{c}}^{(c\eta )}
=v^{\vphantom{\dagger}}_{[2_{c}3_{c}]4_{f}5_{c}}(f_{2}^{\dagger })^{\bar{a}%
_{2}}O_{1_{c}2_{f}}^{-1}f_{2}^{a_{1}}\varepsilon _{\xi }^{a_{1}\bar{a}%
_{2}}[\langle {\cal T}Z_{j_{1}}^{\xi }(t^{+})\rangle_{u} \nonumber \\
&&+i\frac{\delta}
{\delta U_{j_{1}\xi }^{Z}(t^{+})}] \nonumber \\
&&\times [\langle {\cal T}c_{3}^{\dagger}(t^{+})
c^{\vphantom{\dagger}}_{4}(t^{+})\rangle_{u}+
i\frac{\delta }{%
\delta U_{34}^{cc}(t^{+})}]\langle {\cal T}
c^{\vphantom{\dagger}}_{5}(t)\eta ^{\dagger}(t^{\prime
})\rangle _{u}] \nonumber \\
&&-v^{\vphantom{\dagger}}_{[2_{c}3_{c}]4_{f}5_{c}}(f_{2}^{\dagger })^{\bar{a}%
_{2}}O_{1_{c}3_{c}}^{-1} \nonumber \\
&&\times[\langle {\cal T}X_{j_{2}}^{\bar{a}_{2}}(t^{+})
c^{\vphantom{\dagger}}_{4}(t^{+})
\rangle_{u}+i\frac{
\delta }{\delta U_{j_{2}\bar{a}_{2},4_{c}}^{Xc}(t^{+})}]\langle {\cal T}
c^{\vphantom{\dagger}}_{5}(t)\eta
^{\dagger }(t^{\prime })\rangle _{u} \nonumber \\
&&+v^{\vphantom{\dagger}}_{[2_{c}3_{c}4_{f}5_{c}]}f_{4}^{a_{4}}
O_{1_{c}2_{c}}^{-1} 
\nonumber \\
&&\times[\langle {\cal T}c_{3}^{\dagger}(t^{+})X_{j_{4}}^{a_{4}}(t^{+})
\rangle+i\frac{
\delta}{\delta U_{3_{c},j_{4}\bar{a}_{4}}^{cX}(t^{+})}]\langle {\cal T}
c^{\vphantom{\dagger}}_{5}(t)\eta
^{\dagger}(t^{\prime })\rangle_{u}. \nonumber \\
\end{eqnarray}
Therefore, the Hartree contribution is 
\begin{eqnarray}
&&\hspace*{-5mm}\delta ^{(1)}G_{1_{c}}^{(c\eta )}
=v^{\vphantom{\dagger}}_{[2_{c}3_{c}]4_{f}5_{c}}(f_{2}^{\dagger })^{\bar{a}%
_{2}}O_{1_{c}2_{f}}^{-1}f_{2}^{a_{1}}\varepsilon _{\xi }^{a_{1}\bar{a}%
_{2}}\langle {\cal T}Z_{j_{1}}^{\xi }(t^{+})\rangle _{u} \nonumber \\
&&\times\langle 
{\cal T}c_{3}^{\dagger
}(t^{+})c^{\vphantom{\dagger}}_{4}(t^{+})\rangle _{u}\langle {\cal T}
c^{\vphantom{\dagger}}_{5}(t)\eta ^{\dagger }(t^{\prime
})\rangle _{u}] \nonumber \\
&&-v^{\vphantom{\dagger}}_{[2_{c}3_{c}]4_{f}5_{c}}(f_{2}^{\dagger })^{\bar{a}%
_{2}}O_{1_{c}3_{c}}^{-1}[\langle {\cal T}
X_{j_{2}}^{\bar{a}_{2}}(t^{+})c^{\vphantom{\dagger}}_{4}(t^{+})%
\rangle _{u}]\langle{\cal T} c^{\vphantom{\dagger}}_{5}(t)
\eta^{\dagger}(t^{\prime})\rangle_{u} 
\nonumber \\
&&+v^{\vphantom{\dagger}}_{[2_{c}3_{c}4_{f}5_{c}]}f_{4}^{a_{4}}
O_{1_{c}2_{c}}^{-1}\langle
{\cal T}c_{3}^{\dagger }(t^{+})X_{j_{4}}^{a_{4}}(t^{+})\rangle 
\langle{\cal T} c^{\vphantom{\dagger}}_{5}(t)\eta
^{\dagger }(t^{\prime })\rangle _{u}. \nonumber \\
\end{eqnarray}
The contribution to the GF $\langle {\cal T}X\eta ^{\dagger }\rangle $ is 
\begin{eqnarray}
&&\hspace*{-5mm}\delta ^{(1)}G_{j_{1}a_{1}}^{(X\eta )}
=v^{\vphantom{\dagger}}_{[2_{f}3_{c}]4_{c}5_{c}}
\varepsilon_{\xi}^{a_{1}\bar{a}%
_{2}}(f_{2}^{\dagger })^{\bar{a}_{2}}\delta _{j_{1}j_{2}}[\langle
{\cal T}Z_{j_{1}}^{\xi }(t^{+})\rangle _{u}  \nonumber \\
&&+i\frac{\delta }{\delta U_{j_{1}\xi
}^{Z}(t^{+})}] \nonumber \\
&&\times [\langle {\cal T}c_{3}^{\dagger }(t^{+})
c^{\vphantom{\dagger}}_{4}(t^{+})\rangle _{u}+
i\frac{\delta }{%
\delta U_{3_{c}4_{c}}^{cc}(t^{+})}]\langle {\cal T}c_{5}(t)\eta ^{\dagger
}(t^{\prime })\rangle_{u}] \nonumber \\
&&+v^{\vphantom{\dagger}}_{[2_{f}3_{c}]4_{c}5_{c}}O_{1_{f}3_{c}}^{-1}
\varepsilon _{\xi }^{a_{1}%
\bar{a}_{3}}(f_{2}^{\dagger })^{\bar{a}_{2}}(f_{3}^{\dagger })^{\bar{a}_{3}}%
\bar{\delta}_{j_{1}j_{2}}[\langle {\cal T}
Z_{j_{1}}^{\xi }(t^{+})\rangle _{u}  \nonumber \\
&&+i
\frac{\delta }{\delta U_{j_{1}\xi }^{Z}(t^{+})}] \nonumber \\
&&\times [\langle {\cal T}X_{j_{2}}^{\bar{a}_{2}}(t^{+})
c^{\vphantom{\dagger}}_{4}(t^{+})
\rangle _{u}+i\frac{%
\delta }{\delta U_{j_{2}\bar{a}_{2},4_{c}}^{Xc}(t^{+})}]\langle{\cal T} 
c^{\vphantom{\dagger}}_{5}(t)\eta
^{\dagger }(t^{\prime })\rangle _{u} \nonumber \\
&&+v^{\vphantom{\dagger}}_{[2_{f}3_{c}]4_{c}5_{c}}O_{1_{f}3_{c}}^{-1}
\varepsilon _{\xi }^{a_{1}%
\bar{a}_{3}}(f_{2}^{\dagger })^{\bar{a}_{2}}(f_{3}^{\dagger })^{\bar{a}_{3}}%
\bar{\delta}_{j_{1}j_{2}}\kappa _{\bar{b}}^{\bar{a}_{2}\xi } \nonumber
\\
&&\times [\langle {\cal T}X_{j_{2}}^{\bar{b}}(t^{+})
c^{\vphantom{\dagger}}_{4}(t^{+})\rangle _{u}+
i\frac{\delta 
}{\delta U_{j_{2}\bar{b},4_{c}}^{Xc}(t^{+})}]\langle{\cal T} 
c^{\vphantom{\dagger}}_{5}(t)\eta ^{\dagger
}(t^{\prime })\rangle _{u} \nonumber \\
&&+v^{\vphantom{\dagger}}_{2_{c}3_{c}[4_{f}5_{c}]}
O_{j_{1}\mu _{2},2_{c}}^{-1}f_{4}^{a_{4}}(f_{\mu
_{2}}^{\dagger })^{\bar{a}_{2}}\varepsilon _{\xi }^{a_{1}\bar{a}%
_{2}}[\langle {\cal T}
Z_{j_{1}}^{\xi }(t^{+})\rangle _{u}+i\frac{\delta }{\delta
U_{j_{1}\xi }^{Z}(t^{+})}] \nonumber \\
&&\times [\langle {\cal T}c_{3}^{\dagger }(t^{+})X_{j_{4}}^{a_{4}}(t^{+})
\rangle _{u}+i%
\frac{\delta }{\delta U_{3_{c}j_{4}a_{4},}^{cX}(t^{+})}]\langle{\cal T} 
c^{\vphantom{\dagger}}_{5}(t)\eta
^{\dagger }(t^{\prime })\rangle _{u} \nonumber \\
&&+v^{\vphantom{\dagger}}_{2_{c}3_{c}[4_{f}5_{c}]}
O_{j_{1}\mu _{3},3_{c}}^{-1}f_{4}^{a_{4}}(f_{\mu
_{3}}^{\dagger })^{\bar{a}_{3}}\varepsilon _{\xi }^{a_{1}\bar{a}_{3}}\{\bar{%
\delta}_{j_{1}j_{4}}[\langle {\cal T}c_{2}^{\dagger
}(t^{+})X_{j_{4}}^{a_{4}}(t^{+})\rangle _{u}  \nonumber \\
&&+i\frac{\delta }{\delta
U_{2_{c}j_{4},a_{4}}^{cX}(t^{+})}] \nonumber \\
&&\times [\langle {\cal T}
Z_{j_{1}}^{\xi }(t^{+})\rangle _{u}+i\frac{\delta }{\delta
U_{j_{1}\xi }^{Z}(t^{+})}]\langle{\cal T} 
c^{\vphantom{\dagger}}_{5}(t)\eta ^{\dagger }(t^{\prime
})\rangle _{u} \nonumber \\
&&+\delta _{j_{1}j_{4}}\kappa _{b}^{\xi a_{4}}[\langle {\cal T}c_{2}^{\dagger
}(t^{+})X_{j_{1}}^{b}(t^{+})\rangle _{u}+i\frac{\delta }{\delta
U_{2_{c}j_{1}b,}^{cX}(t^{+})}] \nonumber \\
&&\times\langle{\cal T} c_{5}(t)\eta ^{\dagger }(t^{\prime
})\rangle _{u}\}.
\end{eqnarray}
The corresponding Hartree part is 
\begin{eqnarray}
&&\hspace*{-5mm}
\delta ^{(1)}G_{j_{1}a_{1}}^{(X\eta )}
=v^{\vphantom{\dagger}}_{[2_{f}3_{c}]4_{c}5_{c}}
\varepsilon _{\xi }^{a_{1}\bar{a}%
_{2}}(f_{2}^{\dagger })^{\bar{a}_{2}}\delta _{j_{1}j_{2}}\langle
{\cal T}Z_{j_{1}}^{\xi }(t^{+})\rangle _{u}\nonumber \\
&&\times\langle {\cal T}c_{3}^{\dagger
}(t^{+})c^{\vphantom{\dagger}}_{4}(t^{+})\rangle _{u}\langle {\cal T}
c^{\vphantom{\dagger}}_{5}(t)\eta ^{\dagger }
(t^{\prime})\rangle _{u} \nonumber \\
&&+v^{\vphantom{\dagger}}_{[2_{f}3_{c}]4_{c}5_{c}}
O_{1_{f}3_{c}}^{-1}\varepsilon _{\xi }^{a_{1}%
\bar{a}_{3}}(f_{2}^{\dagger })^{\bar{a}_{2}}(f_{3}^{\dagger })^{\bar{a}_{3}}%
\bar{\delta}_{j_{1}j_{2}}\langle {\cal T}
Z_{j_{1}}^{\xi }(t^{+})\rangle _{u}\nonumber \\
&&\times\langle
{\cal T}X_{j_{2}}^{\bar{a}_{2}}(t^{+})
c^{\vphantom{\dagger}}_{4}(t^{+})\rangle _{u}\langle {\cal T}
c^{\vphantom{\dagger}}_{5}(t)\eta
^{\dagger }(t^{\prime })\rangle _{u} \nonumber \\
&&+v^{\vphantom{\dagger}}_{[2_{f}3_{c}]4_{c}5_{c}}
O_{1_{f}3_{c}}^{-1}\varepsilon _{\xi }^{a_{1}%
\bar{a}_{3}}(f_{2}^{\dagger })^{\bar{a}_{2}}(f_{3}^{\dagger })^{\bar{a}_{3}}%
\bar{\delta}_{j_{1}j_{2}}\kappa _{\bar{b}}^{\bar{a}_{2}\xi }\nonumber \\
&&\times\langle
{\cal T}X_{j_{2}}^{\bar{b}}(t^{+})
c^{\vphantom{\dagger}}_{4}(t^{+})\rangle _{u}\langle{\cal T} 
c^{\vphantom{\dagger}}_{5}(t)\eta
^{\dagger }(t^{\prime })\rangle _{u} \nonumber \\
&&+v^{\vphantom{\dagger}}_{2_{c}3_{c}[4_{f}5_{c}]}O_{j_{1}\mu
_{2},2_{c}}^{-1}f_{4}^{a_{4}}(f_{\mu _{2}}^{\dagger })^{\bar{a}%
_{2}}\varepsilon _{\xi }^{a_{1}\bar{a}_{2}}\langle {\cal T}Z_{j_{1}}^{\xi
}(t^{+})\rangle _{u}\nonumber \\
&&\times\langle {\cal T}c_{3}^{\dagger
}(t^{+})X_{j_{4}}^{a_{4}}(t^{+})\rangle _{u}\langle{\cal T} 
c^{\vphantom{\dagger}}_{5}(t)\eta ^{\dagger
}(t^{\prime })\rangle _{u} \nonumber \\
&&+v^{\vphantom{\dagger}}_{2_{c}3_{c}[4_{f}5_{c}]}O_{j_{1}\mu
_{3},3_{c}}^{-1}f_{4}^{a_{4}}(f_{\mu _{3}}^{\dagger })^{\bar{a}%
_{3}}\varepsilon _{\xi }^{a_{1}\bar{a}_{3}}\{\bar{\delta}_{j_{1}j_{4}}%
\langle {\cal T}c_{2}^{\dagger }(t^{+})X_{j_{4}}^{a_{4}}
(t^{+})\rangle _{u} \nonumber \\
&&\times\langle
{\cal T}Z_{j_{1}}^{\xi }(t^{+})\rangle _{u}
\langle{\cal T} 
c^{\vphantom{\dagger}}_{5}(t)\eta ^{\dagger
}(t^{\prime })\rangle _{u} \nonumber \\
&&+\delta^{\vphantom{\dagger}}_{j_{1}j_{4}}\kappa_{b}^{\xi a_{4}}
\langle {\cal T}c_{2}^{\dagger
}(t^{+})X_{j_{1}}^{b}(t^{+})\rangle _{u}\langle{\cal T} 
c^{\vphantom{\dagger}}_{5}(t)\eta ^{\dagger
}(t^{\prime })\rangle _{u}\}.
\end{eqnarray}
\emph{Class 2}. Here the corrections to the equations of motion are  
\begin{eqnarray}
&&\hspace*{-5mm}
\delta ^{(2)}G_{1_{c}}^{(c\eta )} =\frac{1}{2}
v^{\vphantom{\dagger}}_{2_{f}3_{f}4_{c}5_{c}}\{%
\delta _{j_{2}j_{3}}(f_{2}^{\dagger }f_{3}^{\dagger })^{\bar{\eta}%
}O_{1_{c}2_{f}}^{-1}f_{1}^{a_{1}}\varepsilon _{\bar{b}}^{a_{1}\bar{\eta}} 
\nonumber \\
&&\times[\langle {\cal T}X_{j_{2}}^{\bar{b}}(t^{+})
c^{\vphantom{\dagger}}_{4}(t^{+})\rangle _{u}+
i\frac{\delta 
}{\delta U_{j_{2}\bar{b},4_{c}}^{Xc}(t^{+})}]\langle{\cal T} 
c^{\vphantom{\dagger}}_{5}(t)\eta ^{\dagger
}(t^{\prime })\rangle _{u} \nonumber \\
&&+\bar{\delta}_{j_{2}j_{3}}f_{1}^{a_{1}}(f_{2}^{\dagger })^{\bar{a}%
_{2}}(f_{3}^{\dagger })^{\bar{a}_{3}}O_{1_{c}2_{f}}^{-1}\varepsilon _{\xi
}^{a_{1}\bar{a}_{2}}[\langle {\cal T}Z_{j_{2}}^{\xi }(t^{+})\rangle_{u}+
i\frac{%
\delta }{\delta U_{j_{2}\xi }^{Z}(t^{+})}] \nonumber \\
&&\times [\langle {\cal T}X_{j_{3}}^{\bar{a}_{3}}(t^{+})
c^{\vphantom{\dagger}}_{4}(t^{+})
\rangle _{u}+i\frac{%
\delta }{\delta U_{j_{3}\bar{a}_{3},4_{c}}^{Xc}(t^{+})}]\langle{\cal T} 
c^{\vphantom{\dagger}}_{5}(t)\eta
^{\dagger }(t^{\prime })\rangle _{u} \nonumber \\
&&-\bar{\delta}_{j_{2}j_{3}}f_{1}^{a_{1}}(f_{2}^{\dagger })^{\bar{a}%
_{2}}(f_{3}^{\dagger })^{\bar{a}_{3}}O_{1_{c}3_{f}}^{-1}\varepsilon _{\xi
}^{a_{1}\bar{a}_{3}}[\langle {\cal T}Z_{j_{3}}^{\xi }(t^{+})\rangle _{u}+
i\frac{%
\delta }{\delta U_{j_{3}\xi }^{Z}(t^{+})}] \nonumber \\
&&\times [\langle {\cal T}
X_{j_{2}}^{\bar{a}_{2}}(t^{+})
c^{\vphantom{\dagger}}_{4}(t^{+})\rangle _{u}+
i\frac{%
\delta }{\delta U_{j_{2}\bar{a}_{2},4_{c}}^{Xc}(t^{+})}]\langle{\cal T} 
c^{\vphantom{\dagger}}_{5}(t)\eta
^{\dagger }(t^{\prime })\rangle _{u}\} \nonumber \\
\end{eqnarray}
\begin{eqnarray}
&&\hspace*{-5mm}
\delta ^{(2)}G_{j_{1}a_{1}}^{(X\eta )} =\frac{1}{2}%
v^{\vphantom{\dagger}}_{2_{f}3_{f}4_{c}5_{c}}\{\delta _{j_{2}j_{1}}\delta
_{j_{2}j_{3}}(f_{2}^{\dagger }f_{3}^{\dagger })^{\bar{\eta}}\varepsilon _{%
\bar{b}}^{a_{1}\bar{\eta}} \nonumber \\
&&\times [\langle {\cal T}
X_{j_{2}}^{\bar{a}_{2}}(t^{+})c^{\vphantom{\dagger}}_{4}(t^{+})\rangle _{u}+
i\frac{%
\delta }{\delta U_{j_{2}\bar{a}_{2},4_{c}}^{Xc}(t^{+})}]\langle{\cal T} 
c^{\vphantom{\dagger}}_{5}(t)\eta
^{\dagger }(t^{\prime })\rangle _{u} \nonumber \\
&&+\delta _{j_{2}j_{1}}\bar{\delta}_{j_{2}j_{3}}(f_{2}^{\dagger })^{\bar{a}%
_{2}}(f_{3}^{\dagger })^{\bar{a}_{3}}\varepsilon _{\xi }^{a_{1}\bar{a}%
_{2}}[\langle {\cal T}
Z_{j_{2}}^{\xi }(t^{+})\rangle _{u}+i\frac{\delta }{\delta
U_{j_{2}\xi }^{Z}(t^{+})}] \nonumber \\
&&\times [\langle {\cal T}
X_{j_{3}}^{\bar{a}_{3}}(t^{+})c^{\vphantom{\dagger}}_{4}(t^{+})\rangle _{u}+
i\frac{%
\delta }{\delta U_{j_{3}\bar{a}_{3},4_{c}}^{Xc}(t^{+})}]\langle {\cal T}
c^{\vphantom{\dagger}}_{5}(t)\eta
^{\dagger }(t^{\prime })\rangle _{u} \nonumber \\
&&-\delta _{j_{3}j_{1}}\bar{\delta}_{j_{2}j_{3}}(f_{2}^{\dagger })^{\bar{a}%
_{2}}(f_{3}^{\dagger })^{\bar{a}_{3}}\varepsilon _{\xi }^{a_{1}\bar{a}%
_{3}}[\langle {\cal T}
Z_{j_{3}}^{\xi }(t^{+})\rangle _{u}+i\frac{\delta }{\delta
U_{j_{3}\xi }^{Z}(t^{+})}] \nonumber \\
&&\times [\langle {\cal T}
X_{j_{2}}^{\bar{a}_{2}}(t^{+})c^{\vphantom{\dagger}}_{4}(t^{+})\rangle _{u}+
i\frac{%
\delta }{\delta U_{j_{2}\bar{a}_{2},4_{c}}^{Xc}(t^{+})}]\langle{\cal T} 
c^{\vphantom{\dagger}}_{5}(t)\eta
^{\dagger }(t^{\prime })\rangle _{u}\}.\nonumber \\
\end{eqnarray}
The remaining contributions, in this class, comes from the processes of
transforming two $c$- into $f$-electrons 
\begin{eqnarray}
&&\hspace*{-5mm}
\delta ^{(2)}G_{1_{c}}^{f^{\dagger }f^{\dagger }} =\frac{1}{2}%
v^{\vphantom{\dagger}}_{2_{f}3_{f}4_{c}5_{c}}\{
\delta_{j_{2}j_{3}}(f_{2}^{\dagger
}f_{3}^{\dagger })^{\bar{\eta}}O_{1_{c}2_{f}}^{-1}f_{1}^{a_{1}}\varepsilon_{%
\bar{b}}^{a_{1}\bar{\eta}} \nonumber \\
&&\times [\langle {\cal T}X_{j_{2}}^{\bar{b}}(t^{+})
c^{\vphantom{\dagger}}_{4}(t^{+})\rangle 
+i\frac{\delta }{%
\delta U_{j_{2}\bar{b},4_{c}}^{Xc}(t^{+})}]\langle{\cal T} 
c^{\vphantom{\dagger}}_{5}(t)\eta ^{\dagger
}(t^{\prime })\rangle _{u} \nonumber \\
&&+\bar{\delta}_{j_{2}j_{3}}f_{1}^{a_{1}}(f_{2}^{\dagger })^{\bar{a}%
_{2}}(f_{3}^{\dagger })^{\bar{a}_{3}}O_{1_{c}2_{f}}^{-1}\varepsilon _{\xi
}^{a_{1}\bar{a}_{2}} \nonumber \\
&&\times [\langle Z_{j_{2}}^{\xi }(t^{+})\rangle +i\frac{\delta }{\delta
U_{j_{2}\xi }^{Z}(t^{+})}][\langle {\cal T}X_{j_{3}}^{\bar{a}%
_{3}}(t^{+})
c^{\vphantom{\dagger}}_{4}(t^{+})\rangle \nonumber \\
&&+i\frac{\delta }{\delta U_{j_{3}\bar{a}%
_{3},4_{c}}^{Xc}(t^{+})}]\langle{\cal T} 
c^{\vphantom{\dagger}}_{5}(t)\eta ^{\dagger }(t^{\prime
})\rangle _{u} \nonumber \\
&&-\bar{\delta}_{j_{2}j_{3}}f_{1}^{a_{1}}(f_{2}^{\dagger })^{\bar{a}%
_{2}}(f_{3}^{\dagger })^{\bar{a}_{3}}O_{1_{c}3_{f}}^{-1}\varepsilon _{\xi
}^{a_{1}\bar{a}_{3}} \nonumber \\
&&\times [\langle Z_{j_{3}}^{\xi }(t^{+})\rangle +i\frac{\delta }{\delta
U_{j_{3}\xi }^{Z}(t^{+})}][\langle {\cal T}X_{j_{2}}^{\bar{a}%
_{2}}(t^{+})
c^{\vphantom{\dagger}}_{4}(t^{+})\rangle \nonumber \\
&&+i\frac{\delta }{\delta U_{j_{2}\bar{a}%
_{2},4_{c}}^{Xc}(t^{+})}]\langle{\cal T} 
c^{\vphantom{\dagger}}_{5}(t)\eta ^{\dagger }(t^{\prime
})\rangle _{u}\}.
\end{eqnarray}
and 
\begin{eqnarray}
&&\hspace*{-5mm}
\delta ^{(2)}G_{j_{1}a_{1}}^{f^{\dagger }f^{\dagger }} =\frac{1}{2}%
v^{\vphantom{\dagger}}_{2_{f}3_{f}4_{c}5_{c}}\{\delta _{j_{2}j_{1}}\delta
_{j_{2}j_{3}}(f_{2}^{\dagger }f_{3}^{\dagger })^{\bar{\eta}}\varepsilon _{%
\bar{b}}^{a_{1}\bar{\eta}} \nonumber \\
&&\times [\langle {\cal T}X_{j_{2}}^{\bar{b}}(t^{+})
c^{\vphantom{\dagger}}_{4}(t^{+})\rangle 
+i\frac{\delta }{%
\delta U_{j_{2}\bar{b},4_{c}}^{Xc}(t^{+})}]\langle {\cal T} 
c^{\vphantom{\dagger}}_{5}(t)\eta ^{\dagger
}(t^{\prime })\rangle _{u} \nonumber \\
&&+\delta _{j_{2}j_{1}}\bar{\delta}_{j_{2}j_{3}}(f_{2}^{\dagger })^{\bar{a}%
_{2}}(f_{3}^{\dagger })^{\bar{a}_{3}}\varepsilon _{\xi }^{a_{1}\bar{a}%
_{2}} \nonumber \\
&&\times [\langle Z_{j_{2}}^{\xi }(t^{+})\rangle +i\frac{\delta }{\delta
U_{j_{2}\xi }^{Z}(t^{+})}][\langle {\cal T}X_{j_{3}}^{\bar{a}%
_{3}}(t^{+})c^{\vphantom{\dagger}}_{4}(t^{+})\rangle \nonumber \\
&&+i\frac{\delta }{\delta U_{j_{3}\bar{a}%
_{3},4_{c}}^{Xc}(t^{+})}]\langle {\cal T} 
c^{\vphantom{\dagger}}_{5}(t)\eta ^{\dagger }(t^{\prime
})\rangle _{u} \nonumber \\
&&-\delta _{j_{3}j_{1}}\bar{\delta}_{j_{2}j_{3}}(f_{2}^{\dagger })^{\bar{a}%
_{2}}(f_{3}^{\dagger })^{\bar{a}_{3}}\varepsilon _{\xi }^{a_{1}\bar{a}%
_{3}} \nonumber \\
&&\times [\langle Z_{j_{3}}^{\xi }(t^{+})\rangle +i\frac{\delta }{\delta
U_{j_{3}\xi }^{Z}(t^{+})}][\langle {\cal T}X_{j_{2}}^{\bar{a}%
_{2}}(t^{+})c^{\vphantom{\dagger}}_{4}(t^{+})\rangle \nonumber \\
&&+i\frac{\delta }{\delta U_{j_{2}\bar{a}%
_{2},4_{c}}^{Xc}(t^{+})}]\langle {\cal T} 
c^{\vphantom{\dagger}}_{5}(t)\eta ^{\dagger }(t^{\prime
})\rangle _{u}\}.
\end{eqnarray}
Thus, these two corrections differ only by factors of the overlap matrixes
coming from the commutator of $c$-operator with interaction.
Class 2, 3e). Indeed, a $c$-operator, located on site $j_{1}$ ''feels'' 
strong correlations on other site $j_{2}$ only due to non-orthogonality,
i.e.\ due to the $f$-component which it contains on the site $j_{2}$.
Remaining terms in this class come from the interaction 3f, where two
\f-electrons are transformed to $c$-electrons: 
\begin{eqnarray}
&&\hspace*{-5mm}
\delta ^{(2)}G_{1_{c}}^{ff}
=v^{\vphantom{\dagger}}_{[2_{c}3_{c}]4_{f}5_{f}}O_{1_{c}2_{c}}^{-1}
(f^{\vphantom{\dagger}}_{4}f^{\vphantom{\dagger}}_{5})^{\eta }\{\delta
_{j_{4}j_{5}}\kappa _{\eta }^{a_{4}a_{5}}\nonumber \\
&&\times\langle {\cal T}c_{3}^{\dagger
}Z_{j_{4}}^{\eta }(t)\eta ^{\dagger }(t^{\prime })\rangle \nonumber \\
&&+\bar{\delta}_{j_{4}j_{5}}[\langle {\cal T}c_{3}^{\dagger
}(t^{+})X_{4}^{a_{4}}(t^{+})\rangle +i\frac{\delta }{\delta
U_{3_{c},j_{4}a_{4}}^{cX}(t^{+})}]\nonumber \\
&&\times\langle {\cal T}
X_{5}^{a_{5}}(t)\eta ^{\dagger
}(t^{\prime })\rangle \},
\end{eqnarray}
and 
\begin{eqnarray}
&&\hspace*{-5mm}
\delta ^{(2)}G_{j_{1}a_{1}}^{ff}
=v^{\vphantom{\dagger}}_{[2_{c}3_{c}]4_{f}5_{f}}
(f^{\vphantom{\dagger}}_{4}f^{\vphantom{\dagger}}_{5})^{\eta
}O_{1_{f}2_{c}}^{-1}(f_{2}^{\dagger })^{\bar{a}_{2}}\varepsilon _{\xi
}^{a_{1}\bar{a}_{2}} \nonumber \\
&&\times \{\delta _{j_{4}j_{5}}\kappa _{\eta }^{a_{4}a_{5}}[\langle 
Z_{j_{1}}^{\xi
}(t^{+})\rangle +i\frac{\delta }{\delta U_{j_{1}\xi }^{Z}(t^{+})}] 
\nonumber \\
&&\times\langle {\cal T}
(c_{3}^{\dagger }Z_{j_{4}}^{\eta }(t)\eta ^{\dagger }(t^{\prime })\rangle 
\nonumber \\
&&+\bar{\delta}_{j_{4}j_{5}}[\langle Z_{j_{1}}^{\xi }(t^{+})\rangle +i\frac{%
\delta }{\delta U_{j_{1}\xi }^{Z}(t^{+})}][\langle {\cal T}c_{3}^{\dagger
}(t^{+})X_{4}^{a_{4}}(t^{+})\rangle \nonumber \\
&&+i\frac{\delta }{\delta
U_{3_{c},j_{4}a_{4}}^{cX}(t^{+})}]\langle {\cal T}
X_{5}^{a_{5}}(t)\eta ^{\dagger }(t^{\prime })\rangle .
\end{eqnarray}
We see that new types of GFs, $\langle {\cal T}c_{3}^{\dagger }Z_{j_{4}}^{\eta
}(t)\eta ^{\dagger }(t^{\prime })\rangle $, have appeared which cannot be
decoupled in a standard way. Expectation values of the type $\langle
{\cal T}c_{3}^{\dagger }(t)\eta ^{\dagger }(t^{\prime })\rangle $ or $\langle
{\cal T}Z_{j_{4}}^{\eta }(t)\rangle $ are equal to zero in the equilibrium
non-superconducting state. In the external fields 
$\int \dd tU^{\eta }(t)Z^{\eta}(t)$, they do exist. Therefore, 
we should keep these expectation values while 
the fields are
not equal to zero and after calculation of the derivatives $\delta \langle
{\cal T}Z_{j_{4}}^{\eta }(t)\rangle /\delta U^{\bar{\eta}}(t^{\prime })$ and $%
\delta \langle {\cal T}
c_{3}^{\dagger }(t)c^{\dagger }(t^{\prime })\rangle /\delta
U^{cc}(t^{\prime })$ , $\delta \langle {\cal T}c_{3}^{\dagger }(t)X^{\bar{a}%
}(t^{\prime })\rangle /\delta U^{Xc}(t^{\prime })$, etc., We can put
them equal to zero. Usually these graphs give small
contributions since they involve the gap $\sim U$ in the denominator. However,
these contributions describe the an effective anti-ferromagnetic interaction
between (quasi)localized electrons and, therefore, without their
contributions it is difficult to obtain localized antiferromagnetism at all.
Thus, we can use, for example, the decoupling 
\begin{equation}
\langle {\cal T}c_{3}^{\dagger }Z_{j_{4}}^{\eta }(t)\eta ^{\dagger }(t^{\prime
})\rangle =[\langle Z_{j_{4}}^{\eta }(t^{-})\rangle +i\frac{\delta }{\delta
U_{j_{1}\eta }^{Z}(t^{-})}]\langle {\cal T}c_{3}^{\dagger }(t)\eta ^{\dagger
}(t^{\prime })\rangle .
\end{equation}
The 
direct derivative of the zero inverted GF, 
$\langle {\cal T}c_{3}^{\dagger }(t)\eta
^{\dagger }(t^{\prime })\rangle^{-1}$, is equal to zero. However, it is 
always connected in the equations of motion with the operators 
$cZ^{\bar{\eta}}$ 
that gives the non-zero loop correction from $\langle
{\cal T}c_{3}(t)c_{2}^{\dagger }(t^{\prime })\rangle \langle {\cal T}
Z_{j_{4}}^{\eta
}(t^{\prime })Z_{j^{\prime }}^{\bar{\eta}}(t)\rangle $.

\emph{Class 3}. The terms, containing three $f$%
-operators may have the operators, belonging to the same site, or to
different ones. If they belong to the same site, we can write them as
follows 
\begin{eqnarray}
&&v^{\vphantom{\dagger}}_{2_{f}3_{f}[4_{f}5_{c}]}\delta _{j_{1}j_{2}}\delta
_{j_{2}j_{3}}
(f_{2}^{\dagger }f_{3}^{\dagger }
f^{\vphantom{\dagger}}_{4})^{\bar{a}_{2}}X_{j_{2}}^{%
\bar{a}_{2}}c^{\vphantom{\dagger}}_{5} \nonumber \\
&&+
v^{\vphantom{\dagger}}_{[2_{c}3_{f}]4_{f}5_{f}}
\delta _{j_{3}j_{4}}\delta
_{j_{4}j_{5}}(f_{3}^{\dagger }f^{\vphantom{\dagger}}_{4}
f^{\vphantom{\dagger}}_{5})^{a}c_{2}^{\dagger }X_{j_{3}}^{a}.
\end{eqnarray}
They, obviously, also belong to mixing interaction and have been included to
it in Ref.~\cite{DT}.

Next are the terms containing two $f$-operators, one of them belonging to
one site, and another one to other site. They can be written as follows  
\begin{eqnarray}
&&\hspace*{-5mm}
\delta^{(3)}G_{1_{c}}^{ff} =
v^{\vphantom{\dagger}}_{2_{f}3_{f}[4_{f}5_{c}]}\delta _{j_{4}j_{3}}%
\bar{\delta}_{j_{2}j_{3}}(f_{2}^{\dagger })^{\bar{a}_{2}}(f_{3}^{\dagger
}f^{\vphantom{\dagger}}_{4})^{\xi _{3}} \nonumber \\
&&\times \{O_{1_{c}2_{f}}^{-1}(f^{\vphantom{\dagger}}_{2})^{a_{6}}
\varepsilon_{\xi_{2}}^{a_{6}
\bar{a}%
_{2}}[\langle Z_{j_{2}}^{\xi _{2}}(t^{+})\rangle +i\frac{\delta }{\delta
U_{j_{2}\xi _{2}}^{Z}(t^{+})}] \nonumber \\
&&\times [\langle Z_{j_{3}}^{\xi _{3}}(t^{+})\rangle +i\frac{\delta }{\delta
U_{j_{3}\xi _{3}}^{Z}(t^{+})}]\langle {\cal T} 
c^{\vphantom{\dagger}}_{5}(t)\eta ^{\dagger }(t^{\prime
})\rangle _{u} \nonumber \\
&&-O_{1_{c}2_{f}}^{-1}(f^{\vphantom{\dagger}}_{3})^{a_{6}}
\varepsilon _{b}^{a_{6}\xi
_{3}}[\langle {\cal T} 
X_{j_{2}}^{\bar{a}_{2}}(t^{+})X_{j_{3}}^{b}(t^{+})\rangle 
\nonumber \\ &&+i
\frac{\delta }{\delta U_{j_{2}\bar{a}_{2,}j_{3}b}^{XX}(t^{+})}]\langle
{\cal T}c^{\vphantom{\dagger}}_{5}(t)\eta ^{\dagger }(t^{\prime })
\rangle _{u}\} \nonumber \\
&&+v^{\vphantom{\dagger}}_{[2_{c}3_{f}]4_{f}5_{f}}
\delta_{j_{4}j_{3}}\bar{\delta}%
_{j_{5}j_{3}}(f_{3}^{\dagger }f^{\vphantom{\dagger}}_{4})^{\xi
}(f^{\vphantom{\dagger}}_{5})^{a_{5}}\{O_{1_{c}2_{c}}^{-1}
[\langle Z_{j_{3}}^{\xi
_{3}}(t^{+})\rangle\nonumber \\ 
&&+i\frac{\delta }{\delta U_{j_{3}\xi _{3}}^{Z}(t^{+})}%
]\langle {\cal T} 
X_{j_{5}}^{a_{5}}(t)\eta ^{\dagger }(t^{\prime })\rangle _{u} 
\nonumber \\
&&-O_{1_{c}3_{f}}^{-1}(f^{\vphantom{\dagger}}_{3})^{a_{6}}
\varepsilon _{b}^{a_{6}\xi
_{3}}[\langle {\cal T}c_{2}^{\dagger }(t^{+})X_{j_{3}}^{b}(t^{+})\rangle 
\nonumber \\
&&+i\frac{%
\delta }{\delta U_{2_{c},j_{3}b}^{cX}(t^{+})}]\langle {\cal T}
X_{5}^{a_{5}}(t)\eta
^{\dagger }(t^{\prime })\rangle \}.
\end{eqnarray}
Then, for the GF $\langle {\cal T}
X_{j_{1}}^{a_{1}}(t)\eta ^{\dagger }(t^{\prime
})\rangle $ we have  
\begin{eqnarray}
&&\hspace*{-5mm}
\delta ^{(3)}G_{j_{1}a_{1}}^{ff} =
v^{\vphantom{\dagger}}_{2_{f}3_{f}[4_{f}5_{c}]}\delta
_{j_{4}j_{3}}\bar{\delta}_{j_{2}j_{3}}(f_{2}^{\dagger })^{\bar{a}%
_{2}}(f_{3}^{\dagger }f^{\vphantom{\dagger}}_{4})^{\xi _{3}} \nonumber \\
&&\times \{\delta^{\vphantom{\dagger}} _{j_{1}j_{2}}
\varepsilon _{\xi _{2}}^{a_{1}\bar{a}_{2}}
[\langle
Z_{j_{2}}^{\xi _{2}}(t^{+})\rangle +i\frac{\delta }{\delta U_{j_{2}\xi
_{2}}^{Z}(t^{+})}] \nonumber \\
&&\times \lbrack \langle Z_{j_{3}}^{\xi _{3}}(t^{+})\rangle +
i\frac{\delta }{\delta
U_{j_{3}\xi _{3}}^{Z}(t^{+})}]\langle {\cal T} 
c^{\vphantom{\dagger}}_{5}(t)\eta ^{\dagger }(t^{\prime
})\rangle _{u} \nonumber \\
&&-\delta^{\vphantom{\dagger}} _{j_{1}j_{3}}\varepsilon _{b}^{a_{6}\xi _{3}}
[\langle {\cal T} X_{j_{2}}^{%
\bar{a}_{2}}(t^{+})X_{j_{1}}^{b}(t^{+})\rangle \nonumber \\
&&+i\frac{\delta }{\delta
U_{j_{2}\bar{a}_{2,}j_{1}b}^{XX}(t^{+})}]\langle {\cal T}
c^{\vphantom{\dagger}}_{5}(t)\eta ^{\dagger
}(t^{\prime })\rangle _{u}\} \nonumber \\
&&+v^{\vphantom{\dagger}}_{[2_{c}3_{f}]4_{f}5_{f}}
\delta _{j_{4}j_{3}}\bar{\delta}%
_{j_{5}j_{3}}(f_{3}^{\dagger }f^{\vphantom{\dagger}}_{4})^{\xi }
(f^{\vphantom{\dagger}}_{5})^{a_{5}} \nonumber \\
&&\times \{O_{1_{f}2_{c}}^{-1}(f_{6}^{\dagger })^{\bar{a}_{6}}\varepsilon_{\xi
_{2}}^{a_{1}\bar{a}_{6}}[\langle Z_{j_{1}}^{\xi _{2}}(t^{+})\rangle +i\frac{%
\delta }{\delta U_{j_{1}\xi _{2}}^{Z}(t^{+})}] \nonumber \\
&&\times \lbrack \langle Z_{j_{3}}^{\xi }(t^{+})\rangle +i\frac{\delta }{\delta
U_{j_{3}\xi }^{Z}(t^{+})}]\langle {\cal T}X_{j_{5}}^{a_{5}}(t)\eta ^{\dagger
}(t^{\prime })\rangle _{u} \nonumber \\
&&-\delta _{j_{1}j_{3}}\varepsilon _{b}^{a_{6}\xi }[\langle {\cal T}
c_{2}^{\dagger
}(t^{+})X_{j_{1}}^{b}(t^{+})\rangle \nonumber \\
&&+i\frac{\delta }{\delta
U_{2_{c},j_{1}b}^{cX}(t^{+})}]\langle {\cal T}
X_{j_{5}}^{a_{5}}(t)\eta ^{\dagger
}(t^{\prime })\rangle _{u}\}.
\end{eqnarray}
\emph{Class 4}. Since the terms, which contain four $f$-operators and belong
to the same site, are included into the definition of the on-site energies 
$E_{\Gamma }$ of the terms $\Gamma $, we start with the density-density
interaction between different sites  
\begin{eqnarray}
&&\hspace*{-5mm}
\delta ^{(4)}G_{1_{c}}^{ff} =\frac{1}{2}
v^{\vphantom{\dagger}}_{2_{f}3_{f}4_{f}5_{f}}\delta
_{j_{5}j_{2}}\delta _{j_{4}j_{3}}\bar{\delta}_{j_{2}j_{3}}(f_{2}^{\dagger
}f^{\vphantom{\dagger}}_{5})^{\xi _{2}}(f_{3}^{\dagger }
f^{\vphantom{\dagger}}_{4})^{\xi _{3}} \nonumber \\
&&\times \{O_{1_{c}2_{f}}^{-1}(f^{\vphantom{\dagger}}_{2})^{a_{6}}
\varepsilon _{b}^{a_{6}\xi
_{2}}[\langle Z_{j_{3}}^{\xi }(t^{+})\rangle \nonumber \\
&&+i\frac{\delta }{\delta
U_{j_{3}\xi }^{Z}(t^{+})}]\langle {\cal T}X_{j_{2}}^{b}(t)\eta ^{\dagger
}(t^{\prime })\rangle _{u} \label{cl4dgff:eq} \\
&&+O_{1_{c}3_{f}}^{-1}(f^{\vphantom{\dagger}}_{3})^{a_{6}}
\varepsilon _{b}^{a_{6}\xi
_{3}}[\langle Z_{j_{2}}^{\xi _{2}}(t^{+})\rangle \nonumber \\
&&+i\frac{\delta }{\delta
U_{j_{2}\xi _{2}}^{Z}(t^{+})}]\langle {\cal T}X_{j_{3}}^{b}(t)\eta ^{\dagger
}(t^{\prime })\rangle _{u}\},
\end{eqnarray}
and 
\begin{eqnarray}
&&\hspace*{-5mm}
\delta^{(4)}G_{j_{1}a_{1}}^{ff} =\frac{1}{2}
v^{\vphantom{\dagger}}_{2_{f}3_{f}4_{f}5_{f}}%
\delta_{j_{5}j_{2}}\delta _{j_{4}j_{3}}\bar{\delta}_{j_{2}j_{3}}(f_{2}^{%
\dagger}f^{\vphantom{\dagger}}_{5})^{\xi_{2}}(f_{3}^{\dagger }
f^{\vphantom{\dagger}}_{4})^{\xi _{3}} \nonumber \\
&&\times\{\delta^{\vphantom{\dagger}}_{j_{1}j_{2}}
\varepsilon _{b}^{a_{1}\xi _{2}}[\langle
Z_{j_{3}}^{\xi_{3}}(t^{+})\rangle \nonumber \\
&&+i\frac{\delta }{\delta U_{j_{3}\xi
_{3}}^{Z}(t^{+})}]\langle {\cal T}X_{j_{2}}^{b}(t)\eta ^{\dagger }(t^{\prime
})\rangle_{u} \\
&&+\delta_{j_{1}j_{3}}\varepsilon _{b}^{a_{1}\xi _{3}}[\langle
Z_{j_{2}}^{\xi _{2}}(t^{+})\rangle \nonumber \\
&&+i\frac{\delta }{\delta U_{j_{2}\xi
_{2}}^{Z}(t^{+})}]\langle {\cal T}X_{j_{3}}^{b}(t)\eta ^{\dagger }(t^{\prime
})\rangle _{u}\}. \label{cl4dgff2:eq}
\end{eqnarray}
At last, the corrections to the EM from \emph{single-particle} potential are
considered in details in Ref.~\cite{DT}.

\section{Spectral Weight Transfer: An example of Two Localized Electrons in 
Three-Orbital Atoms.}
\label{3-level:app}
The LDA+U and self-interaction correction methods are used in DFT for creating 
an orbital polarization, which allows to describe localized and 
delocalized electrons with different potentials.
In this section we illustrate the mechanism of orbital polarization generated 
by strong intra-atomic interactions considering the
example of an ion with 2 localized f-electrons. In order to make
consideration more transparent we consider a hypothetical 3-orbital atom
with spin less electrons. This electron shell can be described by the
Hamiltonian: 
\begin{equation}
\HH^{d}=\HH_{d}^{0}+\HH_{dd}^{U}=\sum_{m}\varepsilon _{0}\hat{n}_{m}+
U\sum_{m\neq
m^{\prime }}\hat{n}_{m}\hat{n}_{m^{\prime }},
\end{equation}
where $m,m^{\prime }=1,2,3$ and $\hat{n}_{m}\equiv d_{m}^{\dagger }d_{m}$.
The parameter $\varepsilon _{0}$ describes the attraction to 
the nuclear potential which provides the ground state with two bonded 
electrons 
\begin{equation}
E_{n}=\varepsilon _{0}n+\frac{1}{2}Un(n-1),\;\left. \frac{dE_{n}}{dn}\right|
_{n=2}=0\;\Rightarrow \varepsilon _{0}=-\frac{3}{2}U.
\end{equation}
The Hamiltonian has the following eigenstates and eigenvalues  \\ \vspace*{5mm}
\begin{center}
\begin{tabular}{lclcc}
\hline \hline 
\hspace*{5mm}n& & state & & $E_{n}^{(i)}$ \hspace*{5mm}\\ 
\hline \hline 
\hspace*{5mm}0& & 
$|0\rangle =|000\rangle$  & & 0 \hspace*{5mm}\\ 
\hspace*{5mm}1& & 
$d_{1}^{\dagger }|0\rangle =|100\rangle$  & & $-\frac{3}{2}U$ \hspace*{5mm}\\ 
\hspace*{5mm}1& & 
$d_{2}^{\dagger }|0\rangle =|010\rangle$  & & $-\frac{3}{2}U$ \hspace*{5mm}\\ 
\hspace*{5mm}1& & 
$d_{3}^{\dagger }|0\rangle =|001\rangle$  & & $-\frac{3}{2}U$ \hspace*{5mm}\\ 
\hspace*{5mm}2& & 
$d_{1}^{\dagger }d_{2}^{\dagger }|0\rangle =|110\rangle$  & & $-2U$ \\ 
\hspace*{5mm}2& & 
$d_{2}^{\dagger }d_{3}^{\dagger }|0\rangle =|011\rangle$  & & $-2U$ \\ 
\hspace*{5mm}2& & 
$d_{1}^{\dagger }d_{3}^{\dagger }|0\rangle =|101\rangle$  & & $-2U$ \\ 
\hspace*{5mm}3& & 
$d_{1}^{\dagger }d_{2}^{\dagger }d_{3}^{\dagger }|0\rangle =|111\rangle $
             & & $-\frac{3}{2}U$ \\ \hline \hline 
\end{tabular} \\ \vspace*{5mm}
\end{center}
Introducing the Hubbard operators $|i\rangle \langle j|=(1-\delta
_{ij})X^{ij}+\delta _{ij}h^{i}$ in terms of these states we diagonalize the
Hamiltonian  
\begin{eqnarray}
\HH&=&
-\frac{3}{2}U[h^{100}+h^{010}+h^{001}+h^{111}] \nonumber \\
&-&2U[h^{110}+h^{011}+h^{101}].
\end{eqnarray}
The Fermion operators in terms of Hubbard ones are 
\begin{eqnarray}
d_{1} &=&\langle 000|d_{1}|100\rangle X^{000,100}+\langle
010|d_{1}|110\rangle X^{010,110}  \nonumber \\
&&+\langle 001|d_{1}|101\rangle X^{001,101}+\langle 011|d_{1}|111\rangle
X^{011,111}  \nonumber \\
&=&X^{000,100}+X^{010,110}+X^{001,101}+X^{011,111},
\end{eqnarray}
\begin{equation}
d_{2}=X^{000,010}-X^{100,110}+X^{001,011}-X^{101,111},
\end{equation}
\begin{equation}
d_{3}=X^{000,001}-X^{100,101}-X^{010,011}+X^{110,111}.
\end{equation}
Correspondently, the Fermion GFs are  
\begin{eqnarray}
&&\hspace*{-5mm}
F_{11}(i\omega ) =\langle {\cal T}d^{\vphantom{\dagger}}_{1}(\tau )
d_{1}^{\dagger }(\tau
^{\prime })\rangle_{i\omega }  \nonumber \\
&=&\frac{1}{i}\langle
{\cal T}
(X^{000,100}+X^{010,110}+X^{001,101}+X^{011,111})_{\tau }  \nonumber \\
&&\times(X^{100,000}+X^{110,010}+X^{101,001}+X^{111,011})_{\tau^{\prime}}
\rangle
_{i\omega }  \nonumber \\
&=&\frac{1}{i}\langle {\cal T}
X^{000,100}(\tau )X^{100,000}(\tau ^{\prime })\rangle
_{i\omega }\nonumber \\
&&+\frac{1}{i}\langle {\cal T}
(X^{010,110}(\tau )X^{110,010}(\tau ^{\prime
})\rangle _{i\omega }  \nonumber \\
&&+\frac{1}{i}\langle {\cal T}
X^{001,101}(\tau )X^{101,001}(\tau ^{\prime })\rangle
_{i\omega }\nonumber \\
&&+\frac{1}{i}\langle {\cal T}
X^{011,111}(\tau )X^{111,011}(\tau ^{\prime
})\rangle _{i\omega }  \nonumber \\
&=&-i\frac{N_{000}+N_{100}}{i\omega +\frac{3}{2}U+\mu }-i\frac{%
N_{010}+N_{110}}{i\omega +\frac{1}{2}U+\mu }\nonumber \\
&& -i\frac{N_{001}+N_{101}}{i\omega
+\frac{1}{2}U+\mu }-i\frac{N_{011}+N_{111}}{i\omega -\frac{1}{2}U+\mu },
\end{eqnarray}
The non-diagonal GF like $\frac{1}{i}\langle {\cal T}(X^{0,I_{1}}(\tau
)X^{II_{1},I_{2}}(\tau ^{\prime })\rangle_{i\omega }=0$ due to the
diagonal form of the Hamiltonian in this representation. In the same fashion 
\begin{eqnarray}
F_{22}(i\omega ) &=& \langle {\cal T}d^{\vphantom{\dagger}}_{2}(\tau )
d_{2}^{\dagger }(\tau
^{\prime })\rangle_{i\omega }  \nonumber \\
&=&-i\frac{N_{000}+N_{010}}{i\omega +\frac{3}{2}U+\mu }
   -i\frac{%
N_{100}+N_{110}}{i\omega +\frac{1}{2}U+\mu } \nonumber \\
&-&i\frac{N_{001}+N_{011}}{i\omega
+\frac{1}{2}U+\mu }-
  i\frac{N_{101}+N_{111}}{i\omega -\frac{1}{2}U+\mu }, 
\end{eqnarray}
\begin{eqnarray}
F_{33}(i\omega ) &=&\langle {\cal T}d^{\vphantom{\dagger}}_{3}(\tau )
d_{3}^{\dagger }(\tau ^{\prime
})\rangle _{i\omega }  \nonumber \\
&=&-i\frac{N_{000}+N_{001}}{i\omega +\frac{3}{2}U+\mu }-i\frac{%
N_{100}+N_{101}}{i\omega +\frac{1}{2}U+\mu } \nonumber \\
&-&i\frac{N_{010}+N_{011}}{i\omega
+\frac{1}{2}U+\mu }-i\frac{N_{110}+N_{111}}{i\omega -\frac{1}{2}U+\mu }. 
\end{eqnarray}
The population numbers in the numerators for the non-interacting atoms are just
the Gibbs weights, 
\begin{eqnarray}
N_{000} &=&1/Z_{0},\;N_{100}=N_{010}=N_{010}=e^{\beta (\mu +\frac{3}{2}%
U)}/Z_{0},  \nonumber \\
N_{110} &=&N_{011}=N_{101}=e^{\beta (2\mu +2U)}/Z_{0},\nonumber \\
N_{111}&=&e^{\beta
(3\mu +\frac{3}{2}U)}/Z_{0};  \nonumber \\
Z_{0} &=&1+3e^{\beta (\mu +\frac{3}{2}U)}+3e^{\beta (2\mu +2U)}+e^{\beta
(3\mu +\frac{3}{2}U)}.
\end{eqnarray}
The chemical potential has to provide that the total number of electrons 
in the ion equals two. The equation for the number of electrons in this ion, 
\begin{equation}
n_{d}=\sum_{i=1}^{3}\langle d_{i}^{\dagger }
d^{\vphantom{\dagger}}_{i}\rangle =2,
\end{equation}
or written in terms of the many-electron population numbers, 
\begin{eqnarray}
&&\hspace*{-5mm}
n_{d} =0\cdot N_{000}+1\cdot (N_{100}+N_{010}+N_{010}) \nonumber \\
&&\hspace*{-4mm}+2\cdot
(N_{110}+N_{011}+N_{101})+3\cdot N_{111}  \nonumber \\
&&\hspace*{-4mm}=\frac{3e^{\beta (\mu +\frac{3}{2}U)}+
6e^{\beta (2\mu +2U)}+3e^{\beta
(3\mu +\frac{3}{2}U)}}{1+3e^{\beta (\mu +\frac{3}{2}U)}+3e^{\beta (2\mu
+2U)}+e^{\beta (3\mu +\frac{3}{2}U)}}=2,
\end{eqnarray}
can be fulfilled for $\mu $ in the interval: $\Delta _{1}=E_{2}-E_{1}=-\frac{%
U}{2}\leq \mu \leq \frac{U}{2}=\Delta_{2}=E_{3}-E_{2}$. 
In this interval of $\mu $ the population numbers fulfill the condition 
\begin{equation}
N_{000}=N_{111}=N_{100}=N_{010}=N_{001}=0.
\end{equation}
Since the sum of all population numbers is equal to unity, 
\begin{equation}
1=\langle \{f^{\vphantom{\dagger}}_{\nu },f_{\nu }^{\dagger }\}\rangle =
\sum_{\Gamma }N_{\Gamma },
\end{equation}
and, also, 
\[
N_{011}=N_{101}=N_{110}
\]
all three zero Fermion GFs, 
\begin{equation}
iF_{11}^{0}(i\omega )=\frac{N_{110}}{i\omega +\frac{1}{2}U+\mu }+\frac{%
N_{101}}{i\omega +\frac{1}{2}U+\mu }+\frac{N_{011}}{i\omega -\frac{1}{2}%
U+\mu },
\end{equation}
\begin{equation}
iF_{22}^{0}(i\omega )=\frac{N_{110}}{i\omega +\frac{1}{2}U+\mu }+\frac{%
N_{011}}{i\omega +\frac{1}{2}U+\mu }+\frac{N_{101}}{i\omega -\frac{1}{2}%
U+\mu },
\end{equation}
\begin{equation}
iF_{33}^{0}(i\omega )=\frac{N_{101}}{i\omega +\frac{1}{2}U+\mu }+\frac{%
N_{011}}{i\omega +\frac{1}{2}U+\mu }+\frac{N_{110}}{i\omega -\frac{1}{2}%
U+\mu },
\end{equation}
become equal to each other within the statistical description, 
\begin{eqnarray}
iF_{11}^{0}(i\omega )&=&iF_{22}^{0}(i\omega )=iF_{33}^{0}(i\omega )\nonumber \\
&=&\frac{2/3}{%
i\omega +\frac{1}{2}U+\mu }+\frac{1/3}{i\omega -\frac{1}{2}U+\mu }.
\end{eqnarray}
As seen, when the chemical potential is within this interval of energy, $2/3$
of the spectral weight of each orbital is concentrated at the energy 
$\omega =-\frac{1}{2}U-\mu$, while the rest of this weight is placed, 
by the sum rule,  at $\omega =\frac{1}{2}U-\mu$.

From the Fermion GFs we find that the \emph{total} spectral weight in
the lower and upper poles is the same as within the single-electron picture, 
\begin{eqnarray}
&&\hspace{-5mm}
iF_{11}^{0}(i\omega )+iF_{22}^{0}(i\omega )+iF_{33}^{0}(i\omega ) 
\nonumber \\&&=3\cdot 
\left[ \frac{2/3}{i\omega +\frac{1}{2}U+\mu }+\frac{1/3}{i\omega -\frac{1}{2}%
U+\mu }\right]  \nonumber \\
&&=\frac{2}{i\omega +\frac{1}{2}U+\mu }+\frac{1}{i\omega -\frac{1}{2}U+\mu },
\end{eqnarray}
and the same number of particles for the chemical potentials in this
interval of energy  
\begin{eqnarray}
&&n_{d}=n_{1}+n_{2}+n_{3} \nonumber \\
&&=3\left[ \frac{2/3}{e^{\beta \lbrack
-2U-\mu]}+1}+\frac{1/3}{e^{\beta \lbrack -3U/2-\mu
]}+1}\right] =2.
\end{eqnarray}
This explains why the phenomenological recipe of the DFT LDA calculations
with some fixed number of the localized electrons (in the case under
consideration, two) works well: at large values of $U$ the contribution of
the upper pole is not taken into account and when the states are degenerate,
there is no difference, which of the orbitals of the non-filled shell 
that are occupied. However, it is instructive also to go further and 
consider the case when the degeneracy of these states is lifted away.

Let us imagine  that some interaction pushes the energy of one of these
orbitals below the other two (e.g.\ Hund rules, or
long-range order). Then, we have to shift down 
the energy of, say, the first orbital, $\varepsilon _{0}^{(1)}\rightarrow
\varepsilon _{0}-h_{eff},\;$and, $\varepsilon _{0}^{(2)}\rightarrow
\varepsilon _{0},\;\varepsilon _{0}^{(3)}\rightarrow \varepsilon _{0}+h_{eff}
$ in the Hamiltonian. 
The population numbers of  the levels without electrons and with three
electrons are still equal to zero because $h_{eff}\ll U$. Although the
energies of the three orbitals which share two electrons differ from each
other by the magnitude $h_{eff}\ll U$, at low enough temperatures $\beta
^{-1}\ll h_{eff}$ the population numbers become not equal to each other, 
\begin{eqnarray}
&&N_{110}=\frac{e^{\beta h_{eff}}}{1+e^{\beta h_{eff}}+e^{-\beta h_{eff}}}%
\approx 1, \nonumber \\
&&N_{101}=\frac{1}{1+e^{\beta h_{eff}}+e^{-\beta h_{eff}}}\approx
0,\nonumber \\
&&N_{011}=\frac{e^{-\beta h_{eff}}}{1+e^{\beta h_{eff}}+e^{-\beta h_{eff}}}%
\approx 0,\;
\end{eqnarray}
this leads immediately to a redistribution of the spectral weight 
\begin{eqnarray}
\label{kalle1:eq}
&&\hspace*{-5mm}
iF_{11}^{0}(i\omega ) =\frac{N_{110}}{i\omega +\frac{1}{2}U+\mu }+\frac{%
N_{101}}{i\omega +\frac{1}{2}U+h_{eff}+\mu }\nonumber \\
&&+\frac{N_{011}}{i\omega -\frac{1%
}{2}U+h_{eff}+\mu }  \nonumber \\
&\simeq &\frac{1}{i\omega +\frac{1}{2}U+\mu},
\end{eqnarray}
\begin{eqnarray}
&&\hspace*{-5mm}
iF_{22}^{0}(i\omega ) =\frac{N_{110}}{i\omega +\frac{1}{2}U+\mu }+\frac{%
N_{011}}{i\omega +\frac{1}{2}U+\mu }+\frac{N_{101}}{i\omega -\frac{1}{2}%
U+\mu }  \nonumber \\
&&\simeq \frac{1}{i\omega +\frac{1}{2}U+\mu},
\end{eqnarray}
\begin{eqnarray}
&&\hspace*{-5mm}
iF_{33}^{0}(i\omega ) =\frac{N_{101}}{i\omega +\frac{1}{2}U+\mu }+\frac{%
N_{011}}{i\omega +\frac{1}{2}U-h_{eff}+\mu }\nonumber \\
&&+\frac{N_{110}}{i\omega -\frac{1%
}{2}U+\mu }  \nonumber \\
& &\simeq\frac{1}{i\omega -\frac{1}{2}U+\mu},
\label{kalle2:eq}
\end{eqnarray}
Thus, in this case, the spectral weight of the two localized orbitals is
concentrated near the lower poles, $i\omega =E(110)-E(100)\simeq -\frac{1}{2}%
U$, while the spectral weight of the third orbital is concentrated at the 
much higher energy, $i\omega \simeq +\frac{1}{2}U$. 
This phenomenon is the \emph{%
orbital polarization}, and it is caused here by the combined effect of the
effective field and the sum rules. As we see, actually, the phenomenon of
the orbital polarization is caused by two (in our case) different
interactions which involve very different energy scales. First, the strong
Coulomb intra-atomic repulsion compensates for the strong attraction of
electrons by the nuclear potential to the extent, corresponding to the valence,
displayed by the particular ion in the particular surrounding. The latter,
should be taken into account via a Coulomb renormalization of the
energies of the single-electron transitions and it is considered in 
the discussion and conclusion. 

Second, a much smaller energy than the ''Hubbard $U$'' and Hund-rule
integrals, like an energy of magnetic ordering, or, crystalline electric field,
can provide a further polarization of the electronic shell.
The Fermion GFs have purely single-electron form with location of
orbitals at very different energies and, as we have seen, it is the
intra-atomic interaction plus the sum rules for the many-electron spectral
weights that provide their effectively single-electron form. Up to now we 
discussed only fully localized states.

Let us now discuss the following questions:

1. How does the delocalization of the non-filled orbitals arise and how 
does the the spectral weights change?

2. What approximations should be done in order to be able to separate
the calculations into core and valent ones as it is usually done in 
{\it ab initio} calculations?

To consider these questions, we have to 
add to the Hamiltonian the mixing interaction with some conduction
band and the hopping term: 
\begin{eqnarray}
\HH &=&(\HH_{d}^{0}+\HH_{dd}^{U})+\HH_{c}^{0}+\HH_{mix}+\HH_{hop}; \nonumber \\
\HH_{c}^{0} &=&\sum_{k}\varepsilon _{k}c_{k}^{\dagger
}c_{k};\;\HH_{hop}=\sum_{\langle ij\rangle \nu }t_{ij}^{\nu \nu }d_{i\nu
}^{\dagger }d_{j\nu }; \nonumber \\
\HH_{mix} &=&\sum_{k}V_{k}^{\nu }c_{k}^{\dagger }d_{\nu k}+V_{k}^{\ast \nu
}d_{\nu k}^{\dagger }c_{k}. \nonumber
\end{eqnarray}
Here we have chosen a simple form of the hopping, i.e.\ hopping is allowed
from the orbital $\nu $ in one site to the same orbital in the neighboring
ion, in a real lattice it is often not so.

The first immediate conclusion comes from the form of the Fermion GFs, $F$. 
Only a 
\emph{part} of the $f(d)$-electron shell is delocalized since, at the energies 
$\omega \sim $ $-\frac{1}{2}U$, the hopping matrix elements $%
t_{11},t_{12},t_{13},t_{23},t_{22}$ are exponentially small 
($\psi^{f} \sim \exp{(-\int {\rm d}x \sqrt{2m[U(x)-\Delta_1]})}$, 
where $U(x)$ is the 
potential separating the $f$-electrons belonging to neighboring atoms) 
and the transitions 
$\Delta _{1}=E_{2}-E_{1}=-\frac{1}{2}U$ remain localized, while the $\Delta
_{2}=E_{3}-E_{2}=+\frac{1}{2}U$ may be located not very much higher the
chemical potential and the magnitude of the hopping matrix elements $t_{33}$
may be noticeable; the mixing interaction is effective also only with the
orbital which energy crosses a band. Then it looks like if we are back to the
standard band picture, but only for the ''upper'' orbitals; therefore, the 
expected (index "exp") equation for the Fermion GF for the third orbital is  
\begin{equation}
\label{ff_exp:eq}
F_{exp;33}^{-1}(i\omega )=[F_{33}^{0}(i\omega )]^{-1}-
t^{\vphantom{\dagger}}_{33}(k)-\frac{%
|V_{k}^{(\nu =3)}|^{2}}{i\omega -\varepsilon _{k}},
\end{equation}
and  the important physical consequence expected from this form is that the
tails of this upper $f(d)$ band may go under the Fermi level and contribute to
the cohesive energy~\cite{urbansqrt}. Below we will see that 
this single-electron approach misses the important transfer of the 
spectral weight.

Let us consider the problem a bit more accurately. First we find the GFs for
the transitions involved, the equations for the population numbers and, 
make conclusions about the spectral weights and the approximation needed to
be done in order to make the problem as close as possible to the
single-electron picture. First, since only the term $t_{33}(k)d_{3}^{\dagger
}d_{3}$ is effective, the main role in the formation of the collective state 
will be
played by those transitions for which the matrix elements $%
d_{3}^{a(\gamma \Gamma )}=\langle \gamma |d_{3}|\Gamma \rangle$ is non-zero. 
It is convenient to introduce a short notation for them 
\begin{eqnarray}
2  &\equiv &[110,111],\;\bar{2}=[111,110];  \nonumber \\
1  &\equiv &[100,101],\;\bar{1}=[101,100];  \nonumber \\
1' &\equiv &[010,011],\;\bar{1}'=[011,010].
\end{eqnarray}
Then, the system of equations for the GFs within the Hubbard-I
approximation, 
\begin{eqnarray}
{\bf D}_{HIA}^{-1}{\bf G}_{HIA}={\bf P,}
\end{eqnarray}
(which is sufficient for our target here) looks as follows  
\begin{eqnarray}
&&\hspace*{-5mm}\left(
\begin{array}{ccc}
\Omega _{2}-\lambda _{2} & \lambda _{2} & \lambda _{2} \\
\lambda _{1} & \Omega _{1}-\lambda _{1} & -\lambda _{1} \\
\lambda _{1'} & -\lambda _{1'} & \Omega _{1'}-\lambda _{1'}
\end{array}
\right) \left(
\begin{array}{ccc}
G^{2\bar{2}} & G^{2\bar{1}} & G^{2\bar{1'}} \\
G^{1\bar{2}} & G^{1\bar{1}} & G^{1\bar{1'}} \\
G^{1'\bar{2}} & G^{1'\bar{2}} & G^{1'\bar{1}'}
\end{array}
\right) \nonumber \\ &&=\left(
\begin{array}{ccc}
P^{2\bar{2}} & 0 & 0 \\
0 & P^{1\bar{1}} & 0 \\
0 & 0 & P^{1'\bar{1}'}
\end{array}
\right) .
\end{eqnarray}
Here we we have used the notations for transitions $2,1,1'$ used above
(don't confuse with the orbital numbers): $G^{2\bar{2}}=-i\langle {\cal T}
X^{2}X^{\bar{2}}\rangle _{i\omega }\equiv -i\langle
{\cal T}X^{[110,111]}X^{[111,110]}\rangle _{i\omega },$ etc.   
\begin{eqnarray}
\Omega _{j}&=&i\omega -\Delta _{\bar{j}};\;\lambda _{j}\equiv 
\lambda(k,i\omega)\nonumber \\  
P_{j}&=&
\left[ t_{33}(k)+\frac{|V(k)|^{2}}{i\omega -\varepsilon _{k}}\right]
P_{j},\;j=2,1,1',
\end{eqnarray}
\begin{eqnarray}
&&P_{j} \equiv P^{j\bar{j}};\;P_{2}=N_{111}+N_{101};  \nonumber \\
&&P_{1} =N_{100}+N_{101}\simeq N_{101};\nonumber \\ 
&&P_{1'}=N_{010}+N_{011}\simeq N_{011}.
\end{eqnarray}
Then, 
\begin{equation}
{\bf G}_{HIA}=\frac{1}{||D^{-1}_{HIA}||}\left( 
\begin{array}{ccc}
\Phi _{11'}P_{1} & -\lambda _{2}\Omega _{1'}P_{1} & -\lambda _{2}\Omega
_{1}P_{1'} \\ 
-\lambda _{1}\Omega _{1'}P_{2} & \Phi _{21'}P_{1} & \lambda _{1}\Omega
_{2}P_{1'} \\ 
-\lambda _{1'}\Omega _{1}P_{2} & -\lambda _{1'}\Omega _{2}P_{1} & \Phi
_{21}P_{1'}
\end{array}
\right) ,
\end{equation}
where 
\begin{eqnarray}
||D^{-1}_{HIA}|| &=&\Omega _{2}\Omega _{1}\Omega _{1'}-\lambda _{2}\Omega _{1}
\Omega
_{1'}-\Omega _{2}\lambda _{1}\Omega _{1'}-\Omega _{2}\Omega _{1}\lambda _{1'}, 
\nonumber \\
\Phi _{ij} &=&\Omega _{i}\Omega _{j}-\Omega _{i}\lambda_{j}-
\lambda_{i}\Omega _{j}.
\end{eqnarray}
Let us write down the Fermion GFs via the Hubbard GFs. We start
with the Fermion GF of the delocalized ''orbital''  
\begin{eqnarray}
&&\hspace{-5mm}
F_{1'1'}^{HIA} =d_{1'}^{a}(d_{1'}^{\dagger})^{\bar{b}}G^{a\bar{b}%
}=G^{22}-G^{21}-G^{21'} \nonumber \\
&&-G^{12}+G^{11}+G^{11'}-G^{1'2}+G^{1'1}+G^{1'1'} \nonumber \\
&=&\frac{P_{2}\Omega _{1}\Omega_{1'}+\Omega_{2}P_{1}\Omega _{1'}+\Omega
_{2}\Omega _{1}P_{1'}}{\Omega_{2}\Omega_{1}\Omega _{1'}-\lambda _{2}\Omega
_{1}\Omega _{1'}-\Omega_{2}\lambda_{1}\Omega _{1'}-\Omega _{2}\Omega
_{1}\lambda _{1'}},
\end{eqnarray}
and inspect the form of this GF in two regions of energy, $i\omega \sim
\Delta _{\bar{2}}$ and $i\omega \sim \Delta _{\bar{1}}$. In order to see a
rough structure, let use the fact that the magnitude of the effective field, 
$h_{eff}$, 
is invisible on the scale of energies $\sim U$, therefore, 
we can put $\Omega _{1}=\Omega _{1'}$; and 
\begin{equation}
P_{2}+P_{1}+P_{1'}=(N_{110}+N_{111})+N_{101}+N_{011}=1,
\end{equation}
(with the accuracy $N_{100}=N_{010}=N_{001}\simeq 0$); then, we see that the
approximate GF for the third orbital is:
\begin{equation}
F_{1'1'}^{HIA} =\frac{P_{2}\Omega _{1}+(1-P_{2})\Omega _{2}}{\Omega
_{2}\Omega _{1}-\lambda (1-P_{2})\Omega _{2}-\lambda P_{2}\Omega _{1}}. 
\end{equation}
Then, we have a third-order equation for the spectrum. 
In order to make the further 
formulas readable we switch off 
the mixing since this does not change the principal features of the model. 
From the poles of this GF we find the band spectrum:
\begin{equation}
E_{k}^{\pm }=\frac{1}{2}(\lambda _{k}+\nu _{k});\;\nu _{k}=\sqrt{(U+\lambda
_{k})^{2}-4\lambda _{k}U(1-P_{1})}.
\end{equation}
This formally coincides with the spectrum of the s-band Hubbard model in the 
Hubbard-I approximation if we take into account that, in the form of the
spectrum, we used the choice $\varepsilon _{0}=-3U/2$ explicitly. The GF of
the delocalized orbital can be written then in the form:
\begin{equation}
F_{1'1'}^{HIA}=\frac{u_{k}^{2}}{\omega -E_{k}^{-}}+\frac{v_{k}^{2}}{\omega
-E_{k}^{+}},
\end{equation}
where
\begin{equation}
u_{k}^{2}=\frac{1}{2}\left( 1-\frac{\lambda _{k}-U(1-2P_{2})}{\nu _{k}}%
\right) ,\;v_{k}^{2}=1-u_{k}^{2}.
\end{equation}

Now we have a look at $F_{22}$ and $F_{33}$.
Since $t_{2\nu }=V_{2\nu }=0,$ and the matrix elements $d_{1}^{a}$ in the
expansion 
\begin{equation}
F_{11}^{HIA}=d_{1}^{a}(d_{1}^{\dagger })^{\bar{b}}G^{a\bar{b}}
\end{equation}
are equal to zero for the transitions $a=2\equiv[110,111];1\equiv
[100,101];1'\equiv[010,011]$. Therefore, 
\begin{eqnarray}
F_{22}^{HIA} &=&\frac{N_{110}+N_{101}}{i\omega +U/2+h_{eff}}+\frac{%
N_{011}+N_{111}}{i\omega -U/2+h_{eff}}, \\
F_{11}^{HIA} &=&\frac{N_{110}+N_{011}}{i\omega +U/2}+\frac{N_{101}+N_{111}}{%
i\omega -U/2},
\end{eqnarray}
i.e.\ the only difference with the case when the hopping and mixing are
absent is that the population number $N_{111} \neq 0$. Indeed,
we  find from from the GF $G^{1\bar{1}}$: 
\begin{eqnarray}
N_{111} &=&(N_{111}+N_{110})\sum_{k}[a_{k}^{2}f(E_{k}^{-}-\mu
)\nonumber \\ 
&&\hspace*{2.4cm}+b_{k}^{2}f(E_{k}^{+}-\mu )] \\
&\equiv &(N_{111}+N_{110})\Theta \neq 0. 
\end{eqnarray}
Here,  
\begin{equation}
a_{k}^{2}=\frac{1}{2}\left[ 1-\frac{U+\lambda _{k}-2\lambda _{k}(1-P_{2})}{%
\nu _{k}}\right] ,\;b_{k}^{2}=1-a_{k}^{2}.
\end{equation}
We consider the situation when the bottom of the conduction band is
above the energy of the lower transitions, $\lambda _{k}^{bottom}>\Delta _{%
\bar{1}},\Delta _{\bar{1}'}$, and the energy of the upper transition is above
the Fermi energy, $\Delta _{\bar{2}}>\varepsilon _{F}$. 
Since the level $\Delta _{\bar{2}}=U/2$ before switching hopping and mixing 
was empty, the
band $\lambda_k$, being transformed into $E_{k}^{-}$ 
changing is filled only slightly and, therefore, $\Theta \ll 1$. 
Within the accuracy we need this
makes it  sufficient to replace the sum $\Theta(P)$ by $\Theta (P^{(0)})$, 
i.e.\ to find it  without self-consistency, using non-perturbed
values for the population numbers. Then, $\Theta$ becomes just a given small
number (which value we do not need to know here). Therefore, although the
analytical form of the GFs, $F_{22}^{HIA},F_{11}^{HIA}$, coincides with the $%
F_{22}^{0},F_{11}^{0},$ but the population numbers in the numerators,  which
determine the spectral weights, are slightly redistributed. Since a further
analysis is possible only numerically, in order to get some impression of
what is changed let us formally write down the population numbers in terms
of $\Theta $ using the sum rule for the population numbers and the GFs $%
G^{[011,111][111,011]},G^{[101,111][111,101]},G^{[110,111][111,110]}$:  
\begin{eqnarray}
N_{111} &=&\frac{\Theta }{1+\Theta \lbrack e_{1}+e_{2}]},\;N_{110}=\frac{%
1-\Theta }{1+\Theta \lbrack e_{1}+e_{2}]},\nonumber \\ 
N_{011}&=&\frac{\Theta e_{1}}{1+\Theta [ e_{1}+e_{2}]}, 
\;N_{101}=\frac{\Theta e_{2}} {1+\Theta [e_{1}+e_{2}]} \nonumber \\ 
e_{1}&\equiv& e^{\beta (U/2-\mu -h_{eff})},e_{2}\equiv e^{\beta (U/2-\mu )}.
\end{eqnarray}
Since at $\lambda _{k}=0$ we have that 
$P_{2}=P_{2}^{0}=N_{110}^{0}+N_{111}^{0}=N_{110}^{0}$ and
equal to unity when the degeneracy is removed, it is clear that $%
(1-P_{2})\ll 1$. Using this, we can write down 
an approximate expression for the Fermion GF,
\begin{equation}
\label{F_HIA:eq}
F_{33}^{HIA}\simeq \frac{1-P_{2}}{\omega +\frac{U}{2}+(1-P_{2})\lambda _{k}}+%
\frac{P_{2}}{\omega -\frac{U}{2}+\lambda _{k}P_{2}}.
\end{equation}
We can neglect a weak and non-important dispersion $\propto (1-P_{2})\lambda
_{k}$ in the lower pole since this Hubbard sub band is fully filled and
does not contribute to the cohesive energy of the system. However, the
important thing following from this expression is that due to the
delocalization of the orbital in the region of energies $\omega \sim \frac{U%
}{2}$ the bandwidth  of the band decreases,
 the spectral weight in upper pole
is also decreasing from unity to $P_{2}<1$, and in a lower pole a small, but
non-zero weight arises.

Thus, we see that the function $F_{1'1'}$ is reduced indeed to the form of 
$F_{exp;1'1'}$
only in the case when $P_{2}=N_{110}+N_{111}=1$. Since all other population 
numbers are 
not equal to zero, $P_{2} < 1$ and the picture based on an effective 
single-electron potential should be corrected.

Switching on mixing does not change the picture in this approximation, but 
slightly 
Shifts the value of $ \Theta $ and $P_2$.

What physical consequence can be expected of those shifts of the
spectral weights? 

\emph{First}, since a non-zero spectral weight of the orbitals 1 and 2 has
appeared at the energy  $i\omega \sim U/2,$ 
and the energetic barrier between  the neighboring ions is not
that wide nothing prevents a hopping between these
orbitals belonging to different ions to develop, 
as well as  mixing of these orbitals with 
the conduction bands. Therefore, the self-consistency loops should lead to
the formation of bands due to hopping, etc. It should be noted, 
that the magnitude of the 
shifted spectral weight is small because its value is determined by the 
competition 
between the Hund-rule intra-atomic exchange energy and cohesive energy due to 
a delocalization of the empty part of the $f$-orbitals. 
The latter essentially depends on the 
degree of proximity of the upper transition $\Delta_2$ to the Fermi 
energy~\cite{urban_model}, and, as will be seen below, the Hund-rule integrals 
split the upper transitions. This also leads to a suppression of the band 
formation 
with the transitions which are much above Fermi energy.

\emph{Second}, the expectation that pure Fermion bands will be developed
from the ''upper'' orbitals happen to be valid within the approximation $%
N_{111}=0$. Actually, there is no purely single-electron bands. 

\emph{Third}, these bands are narrowed by the spectral weights. Although
this is a well known effect from the studies of the Hubbard models, this is
important in our discussion of an possible improvement of the 
{\it ab initio} 
calculations: even when SIC mimics the orbital polarization and improves
essentially the description of the matter~\cite{strange_svane}, it does not
provide a fully correct physical picture.     

\emph{Fourth}, not only the bandwidth of the band $t_{1'1'}(k)$ is decreased, $%
t_{1'1'}(k)\Rightarrow t_{1'1'}(k)[N_{110}+N_{111}]$, but, also, a part of the 
spectral weight is moved from the high-energy region $i\omega \sim \Delta _{%
\bar{2}}$ to the low-energy $i\omega \sim \Delta _{\bar{1}}$ (see the
Eqs for the Fermionic GFs $F$). Since usually the main contribution to the
value of the moment comes from the localized orbitals, one can expect a slight
deviation (decrease) of this moment from an integer value. In rare earths
this effect is weak, but it is clear that, for example, in compounds with
intermediate valence the deviation of the moment  from integer is quite strong.
Thus, this effect of the transfer of the spectral weight occurs due to
delocalization of one of the transitions (detailed consideration of the
mechanism of delocalization due to a correlation-caused transfer of spectral 
weight is given in a recent paper~\cite{urban_model}).

\emph{Fifth}, when a long-range magnetic order arises
in the system, the effective field, shifting the lower levels, as discussed
above, changes the magnitudes of all population numbers (again due to the
sum rule); the latter may change the bandwidth. Thus, if the correlated
bands are involved into the formation of magnetism, the essential physics of
the formation of the magnetic moment consists not in a rigid 
(self-consistent) shift
of the bands, but in the changes of the bandwidths involved.
From the technical point of view the lesson which one can extract from the
consideration of this model is the following. The speculations, based on
Equations 
(\ref{kalle1:eq}) and (\ref{kalle2:eq}) and which lead us to the 
Eq.(\ref{ff_exp:eq}) might be considered as a
ground for the \emph{ab initio} calculations either within the standard
model for the lanthanides and, also, within the SIC method, where SIC is 
applied to some of the orbitals. 
Some of the orbitals are fully localized and are placed
very deep in energy either ''by hands'' in first case, or by a SIC potential
in the second one, while only the orbitals remaining at higher energies are 
delocalized. However, as we see from Eq.(\ref{F_HIA:eq}) 
this picture is not fully
correct it misses the shifts of the spectral weights. As we shall see
below, a very simple recipe to correct for this exists.
In spite of the simplicity of the approximation,
Hubbard-I, used in this example (which is the same in spirit as the ARF,
discussed in the previous section), should reflect the mechanisms
of the spectral weight transfer correctly since the driving mechanism is the 
sum rules,
which comes from the exact commutation relations. Therefore, the mechanisms
discussed above are expected to remain valid in higher approximations.

 


\begin{figure}
\caption{The first contribution to the GF comes from the hopping/mixing
matrix
element. The solid line is the locator, ${\cal D}$, the circle is the
spectral
weights and the wavy line is the hopping/mixing. All lines in further
graphs can be renormalized by this insertion.} \label{A:fig}
\end{figure}
\begin{figure}
\caption{The second diagram, a, appearing in SCPT, The dashed line is
the Coulomb interaction.} \label{a:fig}
\end{figure}
\begin{figure}
\caption{The exchange graph, b, corresponding to the expression given in
Eq.(\ref{exc_contr:eq}).} \label{b:fig}
\end{figure}
\begin{figure}
\caption{A loop diagram in SCPT, graph c. The zig-zagged line represents
the end in
$\epsilon$, the constants of the algebra. The thick
line represents the incoming and outgoing indices in $\epsilon$,
i.e.\ $1''$ and $\bar{2}$ in $\epsilon^{1''\bar{2}}_{4_b}$. The
corresponding analytical expression is given by Eq.(\ref{c-graph:eq}).}
\label{c:fig}
\end{figure}
\begin{figure}
\caption{Contribution d in Eq.(\ref{d-graph:eq}).
The curly line represent the Bosonic Green function $\langle{\cal T}
Z^{6_b}Z^{4_b}\rangle$.}
\label{d:fig}
\end{figure}
\begin{figure}
\caption{Contribution e in Eq.(\ref{e-graph:eq}).} \label{e:fig}
\end{figure}
\begin{figure}
\caption{Contribution f in Eq.(\ref{f-graph:eq}).} \label{f:fig}
\end{figure}
\begin{figure}
\caption{Graph which has an exact analogy in WCPT. A two-exchange
process. Contribution g.}\label{g:fig}
\end{figure}
\begin{figure}
\caption{Graphs which has an exact analogy in WCPT. Contribution h.}
\label{h:fig}
\end{figure}
\begin{figure}
\caption{Graphs which has an exact analogy in WCPT. Contribution k.}
\label{k:fig}
\end{figure}
\begin{figure}
\caption{Graph which has an exact analogy in WCPT. An exchange diagram
with a loop insertion. Contribution l.}\label{l:fig}
\end{figure}
\begin{figure}
\caption{Graph which has an exact analogy in WCPT. An exchange diagram
with two loop insertion. Contribution m.} \label{m:fig}
\end{figure}
\begin{figure}
\caption{Graph with one line renormalized by the first
correction, Fig.~\ref{a:fig}. Any locator line can
be dressed by the graph in Fig.~\ref{a:fig}.
Contribution n.} \label{n:fig}
\end{figure}
\begin{figure}
\caption{The screening process, the localized transitions are screened
by the delocalized transitions. The screening of the delocalized
transitions by localized is {\it very} weak due to the large energy gap
involved.}
\label{loop:fig}
\end{figure}
\begin{figure}
\caption{The screening process can be continued, however, only delocalized 
transitions screen the Coulomb interaction.}
\label{loop2:fig}
\end{figure}

\end{document}